\documentclass[twocolumn]{aastex701}

\usepackage{xcolor}
\usepackage{soul}

\usepackage{mathtools}
\usepackage{graphicx}
\usepackage[caption=false]{subfig}

\newcommand{\gaia}{\textit{Gaia}}
\newcommand{\rst}{\textit{Roman}}
\newcommand{\jwst}{\textit{JWST}}
\newcommand{\euclid}{\textit{Euclid}}
\newcommand{\lsst}{\textit{Rubin/LSST}}
\newcommand{\hst}{\textit{HST}}

\newcommand{\bpm}{\texttt{BP3M}}

\sethlcolor{white}
\newcommand{\kam}[0]{}

\date{30 January 2026}
\received{\kam{30 January 2026}}
\revised{\kam{27 March 2026}}
\accepted{FUTURE DATE}
\submitjournal{Publications of the Astronomical Society of the Pacific}

\shorttitle{\rst+\gaia\, Astrometry}
\shortauthors{McKinnon \& van der Marel}

\begin{document}

\title{Simulating \rst +\gaia\ Combined Astrometry, Parallaxes, and Proper Motions}

\correspondingauthor{Kevin McKinnon}
\email{kevinmckinnon95@gmail.com}
\author[0000-0001-7494-5910]{Kevin A. McKinnon}
\affiliation{David A. Dunlap Department of Astronomy \& Astrophysics, University of Toronto, 50 St George Street, Toronto ON M5S 3H4, Canada}
\email{kevin.mckinnon@utoronto.ca}

\author[0000-0001-7827-7825]{Roeland P. van der Marel}
\affiliation{Space Telescope Science Institute, 3700 San Martin Drive, Baltimore, MD 21218, USA}
\affiliation{Center for Astrophysical Sciences, The William H. Miller III Department of Physics \& Astronomy, Johns Hopkins University, Baltimore, MD 21218, USA}
\email{marel@stsci.edu}

\begin{abstract}
The next generation of high-precision astrometry is rapidly approaching thanks to ongoing and upcoming missions like \euclid, \lsst, and \rst. We present a new tool (available at \url{https://github.com/KevinMcK95/gaia_roman_astrometry}) to simulate the astrometric precision that will be achieved when combining \gaia\, data with \rst\, images. \kam{The statistics that underpin this method generalize to combinations of astrometric datasets from any telescope.} We construct realistic \rst\, position uncertainties as a function of filter, magnitude, and exposure time, which are combined with \gaia\, precisions and user-defined \rst\, observing strategies to predict the expected uncertainty in position, parallax, and proper motion (PM). We also simulate the core \rst\, surveys to assess their end-of-mission astrometric capabilities, finding that the High Latitude and Galactic Bulge Time Domain Surveys will deliver \gaia\,-DR3-quality PMs down to $G=26.5$~mag and $G=29.0$~mag, respectively. Due to its modest number of repeat observations, we find that the astrometry of the High Latitude Wide Area Survey (HLWAS) is very sensitive to particular choices in observing strategies. We compare possible HLWAS strategies to highlight the impact of parallax effects and conclude that a multi-year \rst-only baseline is required for useful PM uncertainties ($<100$~mas/yr). This simulation tool is actively being used for ongoing \rst\, proposal writing to ensure astrometric requirements for science goals will be met. Subsequent work will expand this tool to include simulated observations from other telescopes to plan for a future where all surveys and datasets are harnessed together.
\end{abstract}

\keywords{Proper motions (1295), Astrostatistics (1882), Astrostatistics tools (1887), \newline Stellar kinematics (1608)}

\section{Introduction} \label{sec:intro}

The astrometry provided by the \gaia\, mission \citep{Gaia_2016,Gaia_2018,Gaia_2022_ref_frame,Gaia_2023_dr3} continues to facilitate incredible scientific revolutions, with additional improvements and data releases scheduled for the next half-decade. However, the moderately high \gaia\, limiting magnitude -- that is, $G=21.5$~mag at the faint limit and $G\sim19$~mag for well-measured proper motions -- means that a large amount of Local Group science has been restricted to either brighter evolved stars, which are not very numerous in a given population, or stars at relatively small distances when pushing down to the main sequence or fainter. In the Milky Way (MW), this manifests as limited angular and distance resolution in the stellar halo. For nearby dwarf galaxies, the majority of stars lie below the main-sequence turn-off and are currently missing from \gaia-based analyses.

Fortunately, the next generation of astrometry -- largely based on active data collection from \euclid\, \citep{EuclidCollaboration_2025} and \lsst, \citep{Ivezi_2019} and the soon-to-launch \rst\, \textit{Space Telescope} \citep{Spergel_2015,Akeson_2019} -- will complement \gaia's successes by extending to much fainter magnitudes with similar precision in proper motion (PM) and parallax. While high precision stellar astrometry is not the primary goal of many of these missions/telescopes, they will nonetheless deliver incredible stellar localizations down to very faint sources \citep[e.g., $\sim26.5$~mag for $5\sigma$ source detections for many \rst\, surveys;][]{Roman_ROTAC_2025}. These new faint stars will increase the number of stars in a given population by more than an order of magnitude, unlocking a wealth of previously hidden information. A far-from-exhaustive list of potential science applications includes: high spatial and distance resolution views of MW dynamics to understand its merger history \citep[e.g.,][]{Bullock_2005,Antoja_2018,Cunningham_2019b,Apfel_2025}, the discovery and characterization of distant MW stellar halo kinematic substructure \citep[e.g.,][]{Belokurov_2018,Helmi_2018,Naidu_2020}, detection of faint or dark companions in low-inclination binary systems \citep[e.g.,][]{GaiaCollaboration_2024}, bulk motions of nearby galaxies to understand the Local Group interaction history \citep[e.g.,][]{vanderMarel_2012,McConnachie_2020,Pace_2022,Bennet_2024}, and resolved internal kinematics of MW stellar streams \citep[e.g.,][]{Ibata_2020,Bonaca_2025,Nibauer_2025} and  nearby dwarf galaxies \citep[e.g.,][]{Read_2021,Massari_2018,Vitral_2024,Vitral_2025} for constraints on the nature of dark matter. In this paper, we specifically focus on the capabilities of \rst, so we point to the work of \citet{WFIRST_Astrometry_2019}, \citet{Han_2023}, and \citet{Dey_2023} for a more-complete list of the Local Group science applications that pristine \rst\, astrometry will significantly advance. \kam{For a focused discussion on} \euclid's \kam{potential impact on Local Group astrometry, please refer to} \citet{Bedin_2025}. It is difficult to understate the inevitable scientific breakthroughs that will be enabled by next generation high precision astrometry over the next 5 to 10 years. 

In preparing for the astrometric future, it is prudent to ensure that we harness the full potential of previous datasets and surveys, namely those of \gaia. While it is often easier to measure stellar PMs from images taken with the same instrument/filter/telescope orientation because it removes many sources of difficult-to-calibrate systematics \citep[e.g.][]{Anderson_2004,Anderson_2006,Bellini_2009,Bellini_2011,Anderson_2007,Anderson_2022,Sohn_2012,Sohn_2018}, recent work has shown success with cross-observatory astrometry measurements. Various \gaia\, data releases, for example, have been combined with different telescopes and surveys -- including Pan-STARRS1 \citep{Magnier_2020}, \hst\, \citep[e.g.,][]{Massari_2017,delPino_2022,Warfield_2023,McKinnon_2024}, \jwst\, \citep{Libralato_2023,Libralato_2024a}, and \euclid\, \citep{Libralato_2024b,Bedin_2025} -- to measure significantly more precise PMs by benefiting from an increased time baseline.

The goal of this work is to provide a method and tool that users can use when planning for \rst\, end-of-lifetime astrometric results or in writing \rst\ proposals. Our tool simulates combinations of \gaia\, catalogs with future \rst\, images to estimate the resulting astrometric precision for stars (or any source moving with a relatively constant PM over decades, such as free-floating exoplanets). The most significant improvement that this tool provides over previous estimates of \rst\, astrometry \citep[e.g.,][]{WFIRST_Astrometry_2019} is the generation and use of realistic \rst\, position uncertainties as a function of magnitude, exposure time, and filter due to recent optical path simulators (i.e. \texttt{stpsf} and \texttt{pandeia}). Our tool is especially useful for planning observations with the goal of reaching a particular astrometric/proper motion precision, as well as for predicting the types of dynamical measurements that will be possible in the \rst\, era using data from the core surveys: the High Latitude Wide Area Survey (HLWAS), the High Latitude Time Domain Survey (HLTDS), the Galactic Bulge Time Domain Survey (GBTDS), and the Galactic Plane Survey (GPS). We hope that our tool will help the community answer questions like ``What proper motion uncertainty will the HLTDS reach at $G=23$~mag?'' and ``If I change the observing cadence of my proposal, what is the highest achievable parallax precision?'' and ``What number of epochs and time baselines are needed to measure useful tangential velocities in nearby dwarf galaxies?''. Our tool is actively being used for \rst\, proposal development to ensure that the proposed observing strategies meet different astrometric requirements for various science goals. 

The structure of this work is organized as follows. Section~\ref{sec:math} presents the underlying theory and statistics that enables us to predict \rst\, astrometry. Section~\ref{sec:methods} introduces the simulation tool and gives an example for its use. Section~\ref{sec:caveats} gives caveats for the tool's use as well as describes some key assumptions and limitations. Section~\ref{sec:applications} applies the simulation tool to the core \rst\, surveys to explore the expected astrometric precision and its dependence on particular observing strategies. Finally, Section~\ref{sec:conclusions} presents our conclusions and gives suggestions for the focus of future work. 

\section{Underlying Theory \& Statistics} \label{sec:math}

We are interested in making 5 astrometric measurements per star: its true position on the sky at some reference epoch (2D measurement), its parallax (1D measurements), and its proper motion vector (2D measurement). We will make this measurement by comparing \gaia\, data -- providing prior information about positions, parallaxes, and PMs -- with positions measured in different \rst\, images at different epochs. The statistics presented in this section are inspired by the math presented in the work describing the \hst+\gaia\, astrometry tool \bpm\, \citep[Bayesian Positions, Parallaxes, and Proper Motions;][]{McKinnon_2024}. \kam{While this paper focuses on} \rst\, \kam{data in combination with the \gaia\, mission, the framework we present is generally applicable to combine (rectified and aligned}\footnote{\kam{We note that the rectification and alignment of images to a global reference frame is often the most difficult step in high-precision astrometry, but we assume this has been done for us in the body of this paper.}}\kam{) positions from any current or future  missions.}

\begin{figure}
    \centering
    \includegraphics[width=1.0\linewidth]{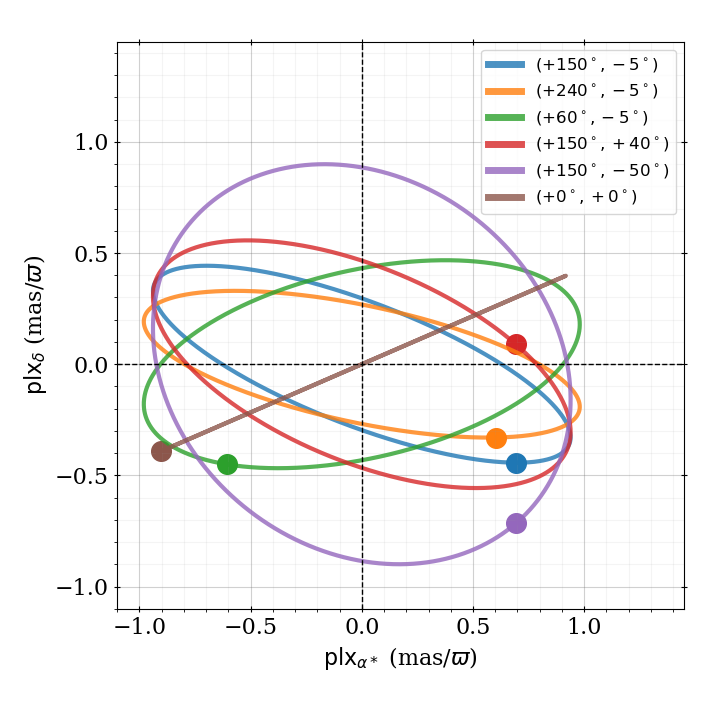}
    \caption{Examples of parallax offsets swept out over the course of a year for different RA, Dec coordinates (colored lines). The large points on each ellipse show the parallax offset at the start of the Julian year (e.g. J2027.0).}
    \label{fig:parallax_example}
\end{figure}

We start by defining some key terms: we set $i$ to be the index of the star, $t$ to be the time of the reference epoch (i.e. J2016.0 for \gaia\, DR2 and DR3), and $\vec \theta_{T,i}$ to be the true position of the star at time $t$. Here, the uppercase $T$ refers to ``True'' while the lowercase $t$ refers to ``time''. \kam{The true position of the star consists of its right ascension, $\alpha_{T,i}$, and declination, $\delta_{T,i}$.} We note that $\vec \theta_{T,i}$ is the parallax-independent position of the star, meaning it is the position of the star if it had zero parallax. Assuming the star's distance from the Sun is not changing much over time -- that is, the radial velocity of the star does not significantly change the heliocentric distance over the course of a few decades\footnote{As estimated in Appendix~\ref{sec:los_appendix}, we require $D > 50$~pc for a 100~km/s line-of-sight velocity over a 10 year time baseline to ensure the PM impact from changing distance is smaller than 0.01~mas/yr.} -- we have the following time evolution definition to describe the true position at time $t_j$:
\begin{equation} 
\begin{split}
        &\vec \theta_{T,i,j} = \vec \theta_{T,i} \\&+ \left(\begin{matrix}
            1/\cos\delta_{T,i} & 0\\
            0 & 1\\
    \end{matrix}\right) \cdot \left(\begin{matrix}
            \Delta t_j & 0 & \mathrm{plx}_{\alpha*}(t_j,\vec \theta_{T,i})\\
            0 & \Delta t_j & \mathrm{plx}_{\delta}(t_j,\vec \theta_{T,i})\\
    \end{matrix}\right) \cdot \left(\begin{matrix}
    \mu_{\alpha*,i}\\
    \mu_{\delta,i}\\
    \varpi_i\\
    \end{matrix}\right)
\end{split}
\end{equation}
where $\Delta t_j = t_j - t$ is the time baseline from the reference epoch. The parallax offset term $\mathrm{plx}_{\alpha*}(t_j,\vec \theta_{T,i})$ is the change in $\alpha*$ a star would experience at a parallax of 1~mas at time $t_j$ given the parallax-independent position $\vec\theta_{T,i}$; this parallax offset can be calculated using the Earth's obliquity and the Sun's geocentric longitude (e.g., using \texttt{astropy}). Examples of the parallax offsets over the course of a year for different RA, Dec coordinates are shown in Figure~\ref{fig:parallax_example}.

Using \gaia\, information, we can redefine the true position as:
\begin{equation} \label{eq:sky_evolution}
\begin{split}
        \vec \theta_{T,i,j} = \vec \theta_{G,i} + \pmb J_i \cdot \pmb U_{i,j} \cdot \vec v_{T,i}
\end{split}
\end{equation}
with 
$$\pmb J_{i} = \left(\begin{matrix}
            1/\cos\delta_{G,i} & 0\\
            0 & 1\\
    \end{matrix}\right)$$
and
$$\pmb U_{i,j} = \left(\begin{matrix}
            1 & 0 & \Delta t_j & 0 & \mathrm{plx}_{\alpha*}(t_j,\vec \theta_{G,i})\\
            0 & 1 & 0 & \Delta t_j & \mathrm{plx}_{\delta}(t_j,\vec \theta_{G,i})\\
    \end{matrix}\right)$$
and 
$$\vec v_{T,i} = \left(\begin{matrix}
    \Delta \alpha_i* \\
    \Delta \delta_i \\
    \mu_{\alpha*,i} \\
    \mu_{\delta,i} \\
    \varpi_i \\
\end{matrix}\right)$$
where the $G$ subscript refers to \gaia\, (which is defined at reference epoch $t_G$), $\vec \theta_{G,i}$ is the \gaia-measured parallax-independent position\footnote{One could argue that $1/\cos\delta_{G,i}$, $\mathrm{plx}_{\alpha*}(t_j,\vec \theta_{G,i})$, and $\mathrm{plx}_{\delta}(t_j,\vec \theta_{G,i})$ in $\pmb J_i$ and $\pmb U_{i,j}$ should refer to the true parallax-independent position $\vec\theta_{T,i}$ instead of the \gaia-measured values, but this would prevent the problem from having a clean linear solution. In any case, the difference is unlikely to matter significantly given the small \gaia\, uncertainties in position (i.e. $<< 10$~mas). Given $\vec \theta_{G,i}$ and $\vec v_{T,i}$, we can define $\vec \theta_{T,i}$ while also measuring the PM and parallax for a star.}, and we have now chosen to set $t=t_G$. The $\pmb U_{i,j}$ and $\vec v_{T,i}$ definitions must contain terms that account for updates to the position measured by \gaia; that is, we expect that the \gaia\, position is slightly different from the truth, and we would like to update it to a more accurate value when all our measurements are combined. We note that the $\pmb U_{i,j}\cdot \vec v_{T,i}$ term could also be adjusted to include a constant acceleration vector (e.g., $\Delta t_{j}^2/2$ terms in $\pmb U_{i,j}$).

Using the previous definitions, we can define a hierarchical model that describes the true motion vector of star $i$:
\begin{equation} \label{eq:vec_heirarchy}
    \begin{split} 
        \vec v_{T,i} &\sim \mathcal{N}\left(\vec v_{\mathrm{pop},i},\pmb C_{v,\mathrm{pop},i}\right)\\
        \vec v_{G,i} &\sim \mathcal{N}\left(v_{T,i}, \pmb C_{G,i}\right)\\
        \vec{\Delta \theta}_{i,j} = \vec \theta_{G,i}-\vec \theta_{i,j} &\sim \mathcal{N}\left(\pmb U_{i,j} \cdot \vec v_{T,i},\,\, \pmb C_{i,j} \right)
    \end{split}
\end{equation}
where $\vec \theta_{i,j}$ is the position of star $i$ measured using \rst\, image $j$, and $\pmb C_{i,j}$ is the associated position covariance matrix. For convenience, we have introduced the observed change in position from \gaia, $\vec{\Delta\theta}_{i,j}$, which we assume has already converted into units of mas and scaled $\Delta\alpha$ by $\cos \delta_{G,i}$. We have also introduced the \gaia-measured astrometric vector for star $i$, $\vec v_{G,i}$ -- with an associated covariance matrix $\pmb C_{G,i}$ -- at reference epoch $t_G$, which looks like:
$$\vec v_{G,i} = \left(\begin{matrix}
    0\\
    0\\
    \mu_{\alpha*,G,i} \\
    \mu_{\delta,G,i} \\
    \varpi_{G,i} \\
\end{matrix}\right).$$ The first two elements of this vector are 0 because \gaia\, has no ($\Delta \alpha_i*,\Delta \delta_i$) offset from its reported position $\vec \theta_{G,i}$. That is, the \gaia\, measurements of position have, on average, no expected difference (i.e. bias) from the true positions. For stars that are too faint for \gaia\, measurements ($G > 21.5$~mag), the inverse of the $\pmb C_{G,i}$ matrix is set to all zeros. In this case, only \rst\, information will be providing the astrometric constraints, which requires a minimum of three \rst\, observations to measure the 5-dimensional $\vec v_{T,i}$ when no \gaia\, priors are available. 

Finally, the $\vec v_{\mathrm{pop},i}$ and $\pmb C_{v,\mathrm{pop},i}$ parameters in Equation~\ref{eq:vec_heirarchy} define a population distribution for star $i$, which could be a function of magnitude/distance, color, membership to different PM groups, etc., and could be different for each star. In this work, we simply use it to place extremely diffuse-but-reasonable constraints on the PM and parallax with the following choices:
\begin{equation} \label{eq:diffuse_priors}
    \begin{split}
        \Delta \theta_{\mathrm{pop},i}/\mathrm{mas} &\sim\mathcal{N}(\mu=0,\sigma=\infty),\\
        \mu_{\mathrm{pop},i}/(\mathrm{mas\cdot yr^{-1})} &\sim\mathcal{N}(\mu=0,\sigma=10^5),\\
        \varpi_{\mathrm{pop},i}/\mathrm{mas} &\sim\mathcal{N}(\mu=0,\sigma=10^4).
    \end{split}
\end{equation}
These population priors ensure numerical stability when inverting the posterior astrometric matrices, especially for stars that are too faint for \gaia\, to provide information. The PM and parallax distribution widths are $\sim 10$ times larger than the largest possible stellar PM or parallax seen from Earth. Depending on the particular choices of simulated \rst\, observing strategies, the final astrometric precision outputs may contain these same uncertainties (i.e., $10^5$~mas/yr for PM and $10^4$~mas for parallax), and these results should be considered unconstrained measurements. We discuss the implications of including a population prior when building astrometric catalogs later in Section~\ref{sec:caveats}.

Using the model in Equation~\ref{eq:vec_heirarchy} allows us to define a posterior distribution on the astrometry vector for star $i$ after $n_{\mathrm{obs}}$ \rst\, observations:
\begin{equation} \label{eq:posterior_astrometry}
    \vec v_{T,i} = \mathcal{N}\left(\widehat{v_{T,i}}, \pmb \Sigma_{v,i} \right)
\end{equation}
with 
$$
\pmb \Sigma_{v,i} = \left[ \pmb C_{v,\mathrm{pop},i}^{-1} + \pmb C_{G,i}^{-1} + \sum_j^{n_{\mathrm{obs}}} \pmb U_{i,j}^T \cdot \pmb C_{i,j}^{-1} \cdot \pmb U_{i,j} \right]^{-1}
$$
and 
\begin{equation*}
    \begin{split}
        \widehat{v_{T,i}} = \pmb \Sigma_{v,i} \cdot \Bigl[ & \pmb C_{v,\mathrm{pop},i}^{-1} \cdot \vec v_{\mathrm{pop},i} + \pmb C_{G,i}^{-1} \cdot \vec v_{G,i} . \\
        &  + \sum_j^{n_{\mathrm{obs}}} \pmb U_{i,j}^T \cdot \pmb C_{i,j}^{-1} \cdot \vec{\Delta \theta_{i,j}} \Bigr].
    \end{split}
\end{equation*}
For this work, we are interested in measuring $\pmb \Sigma_{v,i}$ because it contains the desired \gaia+\rst\, astrometric uncertainties; this is the key result we estimate with our simulation tool discussed in the next section. We focus only on $\pmb \Sigma_{v,i}$ because it provides general results that depend solely on the sizes of \gaia\, astrometric uncertainties and \rst\, position uncertainties, which we can estimate fairly easily. Predictions and simulations using $\widehat{ v_{T,i}}$ can be very useful -- such as in determining how well a co-moving stellar population can be separated from a background distribution -- but they depend on the individual motion measurements of particular stars. We leave this class of analyses for future work. 

For a purely diagonal \gaia\, astrometry covariance matrix and a diagonal matrix for the uncertainty in \rst\, positions, we find that the final position, PM, and parallax uncertainties from $\pmb \Sigma_{v,i}$ are approximated by:
$$\sigma_{\mathrm{pos},i} \approx \left(\frac{1}{\sigma_{\mathrm{pop},\mathrm{pos},i}^2} +\frac{1}{\sigma_{G,\mathrm{pos},i}^2} + \sum_j \frac{1}{\sigma_{i,j}^2}\right)^{-1/2},$$
$$\sigma_{\mathrm{PM},i} \approx \left(\frac{1}{\sigma_{\mathrm{pop,PM},i}^2} +\frac{1}{\sigma_{G,\mathrm{PM},i}^2} + \sum_j \frac{\Delta t_j^2}{\sigma_{i,j}^2}\right)^{-1/2},$$
\begin{equation*}
    \begin{split}
        \sigma_{\mathrm{plx},i} \approx  & \left( \frac{1}{\sigma_{\mathrm{pop,plx},i}^2} +\frac{1}{\sigma_{G,\mathrm{plx},i}^2} \right. \\
        & \hspace{1cm}+ \left. \sum_j \frac{\mathrm{plx}_{\alpha*,i,j}^2+\mathrm{plx}_{\delta,i,j}^2}{\sigma_{i,j}^2}\right)^{-1/2},
    \end{split}
\end{equation*}
which can build trust in our model given some expected behavior. For example, the final PM uncertainty decreases by approximately $1/\Delta t$, and the position uncertainty improves by a factor of $\sqrt{n_{\mathrm{obs}}}$ after $n_{\mathrm{obs}}$ \rst\, images. These approximations also show that larger parallax offset factors will lead to better parallax constraints. To be clear, however, the astrometric precision outputs from our simulation tool and the results in this work are derived by explicitly calculating the $\pmb \Sigma_{v,i}$ matrix defined in Equation~\ref{eq:posterior_astrometry}. 

\begin{figure*}[t]
    \centering
    \includegraphics[width=1.0\linewidth,clip,trim={0cm 0cm 0cm 4cm}]{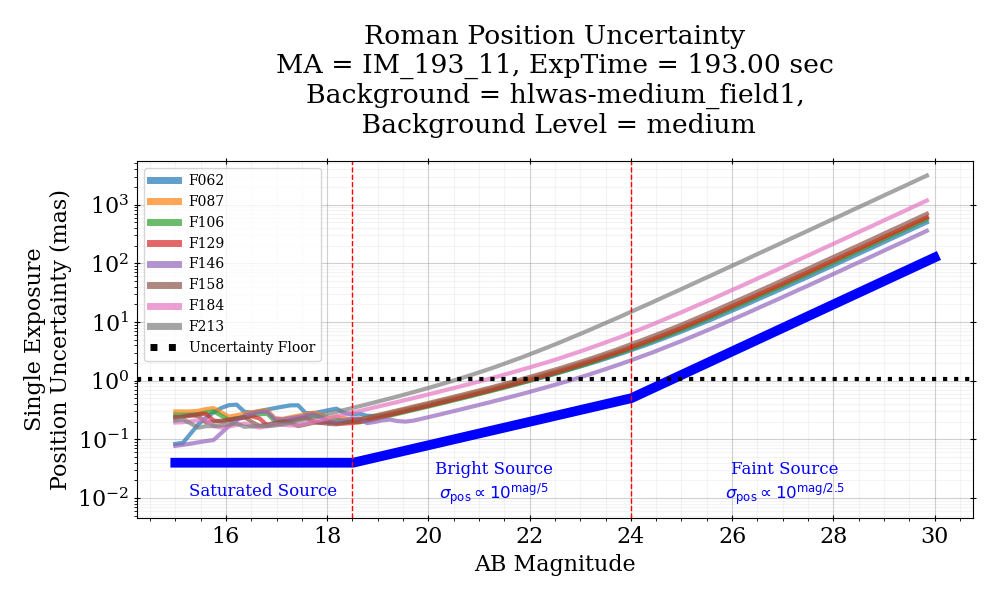}
    \caption{Position uncertainty (per coordinate) as a function of magnitude for different \rst\, filters for a $\sim193$ second exposure with a ``medium'' background level in the HLWAS Medium Field 1. A guiding thick blue line (with arbitrary y values) breaks down different expected trends for different source brightness regimes. Before using the curves in this figure, we add (in quadrature), a $1\%~\mathrm{pixelwidth}\approx 1.1$~mas noise floor to all of the position uncertainties (black dotted line), as motivated in the text. We have computed these curves for all \rst\, filters, all exposure times, and all available \texttt{pandeia} background choices at the ``medium'' background level.}
    \label{fig:roman_pos_errs}
\end{figure*}


\section{Astrometry Simulation Tool} \label{sec:methods}

From the statistics presented in the previous section -- Equation~\ref{eq:posterior_astrometry}, in particular, and the definition of the combined astrometry precision matrix $\pmb \Sigma_{v,i}$ -- we only require three pieces of information to make \gaia+\rst\, combined-astrometry precision predictions: (1) an estimate of the \gaia\,-measured astrometric uncertainties, (2) the position uncertainty in \rst\, images, and (3) the (RA, Dec) pointing and time of the \rst\, image. We have created a publicly available tool\footnote{Our simulation tool is published on Zenodo at \href{https://doi.org/10.5281/zenodo.15852051}{https://doi.org/10.5281/zenodo.15852051} and can be found at this GitHub repository: \url{https://github.com/KevinMcK95/gaia_roman_astrometry}} that is able to build the required inputs, incorporate different user-defined \rst\, observing strategies, and then calculate the final expected position, parallax, and PM uncertainties. 

The first step is to estimate the \gaia\, astrometric uncertainty, which is provided by the \gaia\, collaboration forecasting different precisions as a function of magnitude for DR3 through DR5 \footnote{\gaia\, astrometric precision relationships originally found at this webpage: \url{https://www.cosmos.esa.int/web/gaia/science-performance}}. For posterity, we state those relationships here:
\begin{equation}\label{eq:gaia_precisions}
\begin{split}
    G_\mathrm{in} &= \max(G,13)\\
    z &= 10^{0.4 \cdot (G_\mathrm{in} - 15)}\\
    \sigma_\mathrm{\varpi}/\mathrm{mas} &= \frac{T}{1000}  \sqrt{40 + 800 z + 30 z^2}
\end{split}
\end{equation}
where $T \in [1.0, 0.749, 0.527]$ for DR3, DR4, and DR5, respectively. The uncertainties for position and proper motion are simply a multiplicative factor times the $\sigma_\mathrm{\varpi}$ definition. For $(\Delta\alpha*,\Delta\delta)$ position uncertainties (measured in mas), these multiplicative factors are $(0.8,0.7)$ for all data releases. For $(\mu_{\alpha*},\mu_\delta)$ PM uncertainties (measured in mas/yr), these multiplicative factors change from $(1.03,0.89)$ for DR3, to $(0.58,0.50)$ for DR4, to $(0.29,0.25)$ for DR5. For $G>20.7$~mag, the \gaia\, parallax and PM uncertainties are undefined (i.e. uncertainties are effectively infinite), but the \gaia\, position uncertainties are still defined by this relationship for $20.7<G<21.5$~mag.  

Next, we use the \texttt{pandeia}\footnote{see \texttt{pandeia} documentation at \url{https://roman-docs.stsci.edu/simulation-tools-handbook-home/roman-wfi-exposure-time-calculator/pandeia-for-roman/overview-of-pandeia}} \citep{Pontoppidan_2016} and \texttt{stpsf}\footnote{see \texttt{stpsf} documentation at \url{https://stpsf.readthedocs.io/en/latest/intro.html}} \citep{Perrin_2012,Perrin_2025} tools to simulate how well we can expect to measure the position of a source in a \rst\, image as a function of filter, magnitude, and exposure time. We start by obtaining a super-sampled version of the \rst\, PSF for a given filter using \texttt{stpsf}. We then generate a simulated \rst\, observation of a star at a given flux level, including Poisson noise. Using the super-sampled PSF, we iteratively fit for the $(x,y)$ position of the star as well as its flux. In particular, we take numerical derivatives of the PSF with respect to the position to linearize the fitting problem, which allows us to quickly update our parameter guesses and converge on the final answer. These numerical derivatives, when combined with arguments about the shape of the $\chi^2$ fitting surface, also allow us to estimate the full covariance matrix that relates the position and flux uncertainties. Agreeing with the expectation from the Cram\'er–Rao bound \citep{Cramer_1946,Rao_1947}, we have:
\begin{equation} \label{eq:position_and_flux_errs}
\begin{split}
    \sigma_{x} &= \frac{1}{f}\cdot\left[ \sum_{p}\frac{1}{\sigma_p^2}\cdot\left(\frac{\partial~\mathrm{PSF}_p}{\partial x}\right)^2 \right]^{-1/2},\\
    \sigma_{y} &= \frac{1}{f}\cdot\left[ \sum_{p}\frac{1}{\sigma_p^2}\cdot\left(\frac{\partial~\mathrm{PSF}_p}{\partial y}\right)^2 \right]^{-1/2},\\
    \sigma_{f} &= \left[ \sum_{p}\frac{\mathrm{PSF}_p^2}{\sigma_p^2} \right]^{-1/2},
\end{split}
\end{equation}
where $f$ is the total flux of the source, $p$ is the index of pixels which we sum over, $\sigma_p$ is the flux uncertainty in each pixel, and $\mathrm{PSF}_p$ is the height of the PSF model in pixel $p$. Ultimately, it is the position uncertainty relative to the flux uncertainty that we are interested in because it allows us to easily estimate position uncertainty given the more-commonly-calculated flux signal-to-noise ($\mathrm{SNR} = f/\sigma_f$). Then, we can use the \texttt{pandeia}-calculated relationships between flux SNR and magnitude to build functions of position uncertainty versus magnitude. 

We save time during our PSF-fitting procedure by not fitting across a range of simulated fluxes for each filter. Instead, we leverage the fact that we expect there to be a constant (i.e. flux/magnitude independent) multiplier between the flux SNR and the position uncertainty (in pixels) for a given PSF -- as suggested by Equation~\ref{eq:position_and_flux_errs} -- such that \begin{equation} \label{eq:SNR_to_pos_err}
    \sigma_{\mathrm{pos}} = \frac{k}{\mathrm{SNR}_\mathrm{flux}}= \kappa \cdot \frac{\mathrm{FWHM}}{\mathrm{SNR}_\mathrm{flux}},
\end{equation}
where $k$ (in pixels) is the constant we need to measure, FWHM (in pixels) is the full width at half maximum of a given PSF, and $\kappa = k/ \mathrm{FWHM}$ is a unitless version of $k$. We have added a second version of the position uncertainty relationship with $\kappa$ because it is often straightforward to estimate the FWHM of a PSF directly from observations. For a 1D Gaussian, it is easy to show that $k$ is exactly the $\sigma$ width of the input Gaussian, with $\kappa = \sigma/\mathrm{FWHM} = 0.425$. Taking advantage of the relationship in Equation~\ref{eq:SNR_to_pos_err}, we only needed to fit a single PSF (typically with high SNR) for each filter to measure the $k$ factor between flux SNR and position uncertainty; as a sanity check, we verified that changing the input flux returned the same $k$ value per filter. Table~\ref{tab:snr_to_pos_errs} lists the $k$ values we measure for different filters when using the \texttt{stpsf}-defined PSFs, in addition to columns showing the FWHM (in \rst\, pixels) in different coordinates. One way to interpret the $k$ column is that it gives the width of an equivalent 1D Gaussian PSF with $\sigma = k$, or the width-per-coordinate (i.e. the diagonal elements of a covariance matrix) of a multivariate Gaussian PSF. Here, we must be careful to point out that the $k$ and $\sigma_\mathrm{pos}$ we report are the per-coordinate uncertainty in $x$ or $y$ (or RA or Dec) because we are interested in the values that go into the diagonal elements of $\pmb C_{i,j}$ (e.g. see Equation~\ref{eq:vec_heirarchy}). This means that our $\sigma_{\mathrm{pos}}$ values are smaller (by a factor of approximately $\sqrt{2}$) than what is sometimes referred to as the ``centroid uncertainty'', which is the quadrature sum of the $x$ and $y$ uncertainties. 

With the Table~\ref{tab:snr_to_pos_errs} results in hand, we can then quickly convert the \rst\, SNR versus magnitude outputs from \texttt{pandeia} into position uncertainty as a function of magnitude. As this step is the slowest in the tool, we save these position uncertainty functions for different configurations of the exposure time, filter, and \texttt{pandeia} backgrounds. With these saved outputs, the slowest step should not need to be run on a user's computer, though we include the code used to create these results for transparency. 

\begin{table}[h]
    \centering
    \begin{tabular}{|cccccc|}
        \hline
Filter & $k$ & FWHM$_x$ & FWHM$_y$ & FWHM$_\mathrm{ave}$ & $\kappa_\mathrm{ave}$ \\ \hline
F062 & 0.858 & 0.585 & 0.605 & 0.595 & 1.441 \\
F087 & 0.763 & 0.728 & 0.757 & 0.742 & 1.028 \\
F106 & 0.709 & 0.874 & 0.905 & 0.889 & 0.797 \\
F129 & 0.713 & 1.031 & 1.068 & 1.050 & 0.680 \\
F146 & 0.792 & 1.024 & 1.063 & 1.044 & 0.758 \\
F158 & 0.756 & 1.249 & 1.295 & 1.272 & 0.594 \\
F184 & 0.784 & 1.444 & 1.508 & 1.476 & 0.531 \\
F213 & 0.873 & 1.664 & 1.741 & 1.702 & 0.513 \\
\hline
        \end{tabular}
    \caption{Constant factor $k$ per \rst\ filter in pixel units that relates the position uncertainty to the flux SNR, as defined by Equation~\ref{eq:SNR_to_pos_err}. Additional columns show the $x$ and $y$ coordinate FWHM in units of \rst\, pixels as well as their mean. The final column provides the unitless $\kappa_\mathrm{ave} = k/\mathrm{FWHM}_\mathrm{ave}$. These values were calculated using the \texttt{stpsf}-defined PSFs, taking a $101\times101$ \rst\, pixel cutout.}
    \label{tab:snr_to_pos_errs}
\end{table}

An example of these outputs are shown in Figure~\ref{fig:roman_pos_errs} for a $\sim193$~second exposure. Increasing the exposure time typically has the effect of sliding these curves to the right, such that position uncertainties for fainter stars \kam{decrease} compared to shorter exposures, up to the limit for saturated sources. To be clear, the position uncertainty used in the final astrometry precision calculation is defined by profiles similar to the ones in Figure~\ref{fig:roman_pos_errs}, which are then added in quadrature with an adjustable floor position uncertainty, nominally set to 1\% of the \rst\, pixel width. From this figure, we see that F146 gives the most precise position per exposure for most magnitudes, likely because this wide filter is able to collect more flux SNR per exposure than its narrower counterparts. However, we caution that the large wavelength width of F146 may allow difficult-to-handle chromatic variations in the PSF to impact position measurements. While we may be able to quantify this effect once we start collecting real \rst\, data, chromatic PSF variation across F146 might ultimately lead to large systematic uncertainties that we do not account for in our simulations. 

It is important to point out that these realistic position uncertainty functions in Figure~\ref{fig:roman_pos_errs} are a key development provided by this work; before the creation of \rst\, optics/observing simulators, previous astrometric estimates necessarily made simplifying assumptions about the expected \rst\, position uncertainty (e.g., magnitude- and filter-independent uncertainties of $\sim1$\% pixel). Because we all often share a 1\% pixel floor uncertainty assumption, we will generally produce results that agree with those previous estimates for moderate to bright sources. For instance, our predictions for \rst-only HLTDS PMs in $15<G<22$~mag are very similar to the ``WFIRST HLS'' predictions shown in Figure~1 of \citet{WFIRST_Astrometry_2019}. 

Under the curves for the various \rst\, filters, we have drawn a thick blue line to show the expected trend in position uncertainty across different magnitude regimes, which is general to all telescopes/filters/exposure times. At the brightest end, pixels become saturated and cannot constrain better positions for more input flux; in fact, for many telescopes there is a magnitude limit where the position uncertainty is effectively infinite/undefined because localization becomes impossible when too many neighboring pixels saturate. Here, our saturation regime is defined by the \texttt{pandeia} flux SNR behavior. 

\rst\,\kam{ detectors, like those of \jwst \,, are different from classical CCDs in that they collect time series counts in each pixel. The flux in a pixel is then calculated using the best-fit slope of the observed counts, sometimes referred to as ``fitting a ramp'' }\citep[e.g.,][]{Fixsen_2000,Offenberg_2001,Rauscher_2007,Brandt_2024,Li_2026}\kam{. One key benefit of these detectors is that they enable higher dynamic range images because any pixel that does not saturate within the first two readouts (e.g., $\lesssim9.5$~seconds for most} \rst\,\kam{readout strategies) is able to measure a scientifically useful flux}\footnote{\kam{In theory, the first readout alone can be used to measure the flux. In practice, the resetting/flushing of the detector between exposures is imperfect such that counts from the previous exposure can bias the flux estimate.}}\kam{. However, the particular flux levels will determine which reads are unsaturated and useful for ramp fitting, which causes the flux S/N and position uncertainty to change as a non-monotonic function of magnitude. This likely explains the oscillating/saw-tooth behavior of the \texttt{pandeia} S/N outputs at the bright end of Figure}~\ref{fig:roman_pos_errs}\kam{.}

\kam{We can split} \rst\, \kam{pixels into unsaturated, ``soft'' saturated, and ``hard'' saturated regimes. The unsaturated regime is where all readouts remain unsaturated and can be used for measuring a flux; this threshold is filter and readout-pattern dependent, but most} \rst\, \kam{filters appear completely unsaturated for $\gtrsim 18.5$~mag in our 193 second exposure example (i.e. at the transition from ``Saturated Source'' to ``Bright Source''). We define hard saturation as the point when fewer than two readouts are unsaturated such that ramp fitting becomes biased or impossible; again, this threshold is filter dependent but is $\lesssim 15.5$~mag for most} \rst\, \kam{filters and exposure times}\footnote{\kam{Single pixel saturation times for different} \rst\,\kam{ filters can be found here:} \url{https://science.nasa.gov/mission/roman-space-telescope/time-to-saturation-for-point-source-imaging/}}\kam{. Soft saturation occurs at intermediate magnitude ranges where one or more of the later readouts are saturated, but there are enough unsaturated readouts to constrain a useful flux measurement. It is the soft saturation regime that is highlighted by the ``Saturated Source'' label in Figure}~\ref{fig:roman_pos_errs}\kam{, and we do not attempt to implement a hard saturation threshold. Stars that produce hard saturated pixels will likely need to mask their cores when PSF fitting such that only the lower S/N pixels in the PSF wings will constrain positions. For} \rst\, \kam{filters with very under-sampled PSFs (i.e., the bluer filters), masking the central pixel(s) will lead to a dramatic increase in position uncertainty because the wings contain relatively small amounts of position information. We leave a detailed analysis of the relationship between position uncertainty and the effects of changing numbers of soft and hard saturated pixels as future work that will be useful to understand the number of well-measured} \gaia\, \kam{stars in each} \rst\,\kam{ image.} 

In the \kam{unsaturated} intermediate magnitudes \kam{-- that is, the $18.5 \lesssim \mathrm{ABMAG}\lesssim 24$ range in Figure}~\ref{fig:roman_pos_errs}\kam{--} bright sources are dominated by Poisson noise from the source, meaning the following relationship is expected:
\begin{equation} \label{eq:photon_limit_bright_errs}
\begin{split}
    \mathrm{SNR}_\mathrm{flux} &= \sqrt{\mathrm{counts}}\\
    \mathrm{mag}&\propto-2.5\cdot\log_{10}{\mathrm{counts}}\\
    &\propto-5\cdot \log_{10}{\mathrm{SNR}_\mathrm{flux}}\\
    &\propto5\cdot \log_{10}{\sigma_{\mathrm{pos}}},
\end{split}
\end{equation} where the final line in the above equation comes from Equation~\ref{eq:SNR_to_pos_err}. For faint sources (\kam{$\mathrm{ABSMAG} \gtrsim 24$~mag in Figure}~\ref{fig:roman_pos_errs}), the background begins to dominate the noise and leads to faster increases in the position uncertainty, producing the following expected relationship:
\begin{equation} \label{eq:background_limit_pos_errs}
\begin{split}
    \mathrm{SNR}_\mathrm{flux} &\propto \mathrm{counts}\\
    \mathrm{mag}&\propto-2.5\cdot\log_{10}{\mathrm{counts}}\\
    &\propto-2.5\cdot \log_{10}{\mathrm{SNR}_\mathrm{flux}}\\
    &\propto2.5\cdot \log_{10}{\sigma_{\mathrm{pos}}}.
\end{split}
\end{equation}
Indeed, we find that the curves in Figure~\ref{fig:roman_pos_errs} asymptote towards the Equation~\ref{eq:background_limit_pos_errs} for all filters and exposure times at the faint end.

\begin{figure*}
    \centering
    \includegraphics[width=1.0\linewidth]{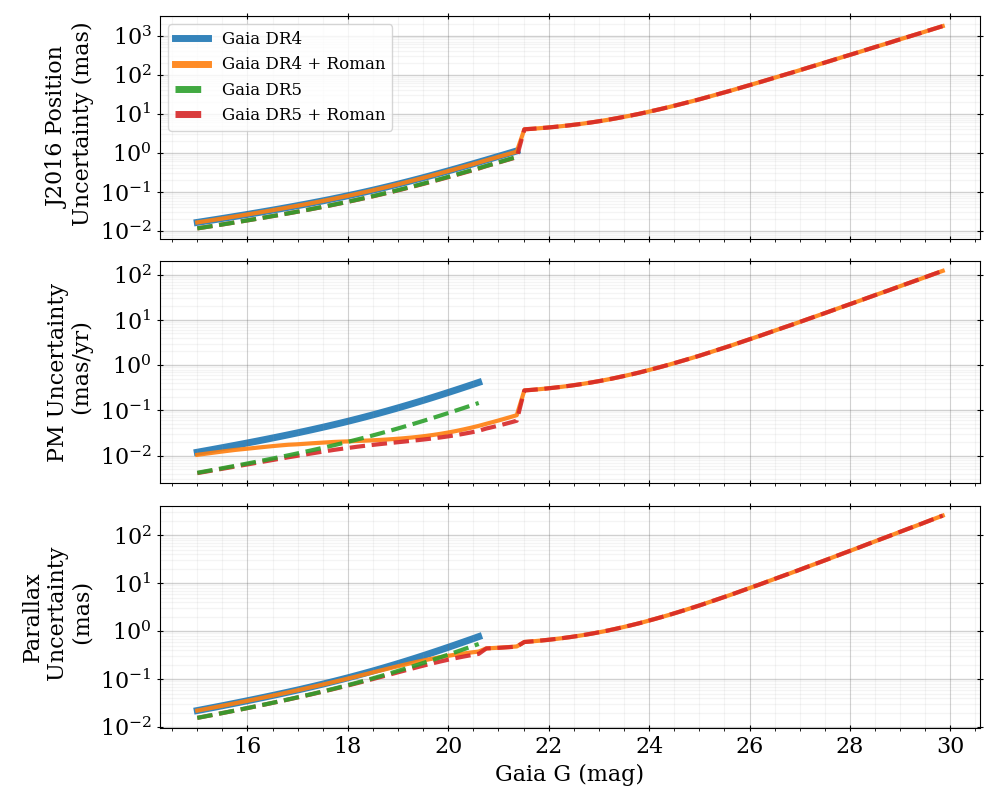}
    \caption{Example of astrometric precision after combining \gaia\, with \rst\, observations given the strategy described in the text. The solid lines show the results for \gaia\, DR4 while the dashed lines show the results for \gaia\, DR5. The \gaia\, positions extend to $G=21.5$~mag, but the \gaia\, parallaxes and PMs only go as deep as $G=20.7$~mag, which explains why there are some discontinuities at these magnitudes. In all cases, we find that the \gaia+\rst\, results are at least as precise as the corresponding \gaia\, data release alone. As we push to fainter magnitudes ($G>18$~mag), the combined \gaia+\rst\, astrometry is a significant improvement over \gaia-alone. As expected, the difference between the \rst+\gaia\, DR4 versus DR5 is negligible for $G>21.5$~mag, because the same \rst\, information is providing all the constraints to these too-faint-for-\gaia\, sources.}
    \label{fig:example_gaia+roman_astrometry}
\end{figure*}


\begin{figure*}
    \centering
    \subfloat{
        \includegraphics[width=\textwidth]{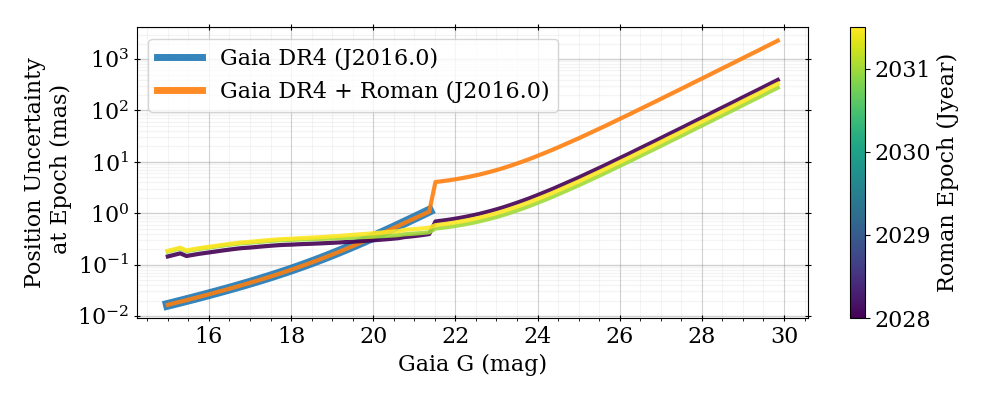}
    }\\ \vspace{-0.75cm}
    \subfloat{
        \includegraphics[width=\textwidth]{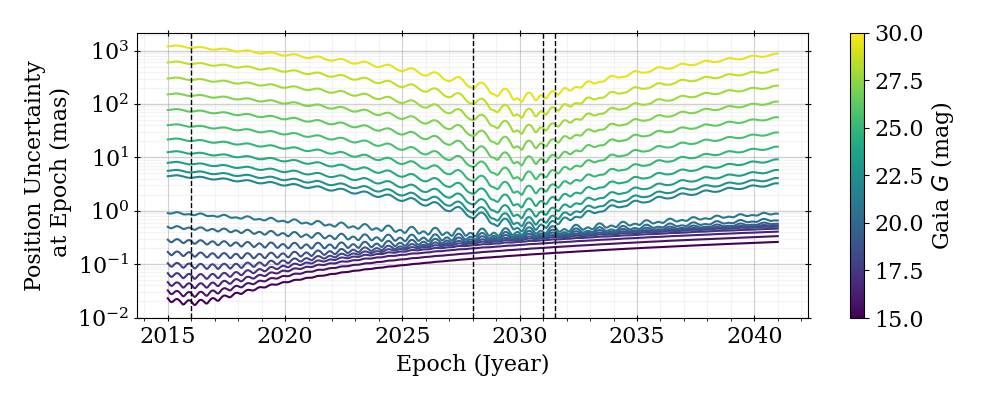}
    }
    \caption{Position uncertainty at different epochs for the \rst+\gaia\, DR4 data shown in Figure~\ref{fig:example_gaia+roman_astrometry} as a function of magnitude (top) and time (bottom). The blue and orange lines in the top panel show the results at J2016.0, while the colored lines show the results at J2028.0, J2031.0, and J2031.5. \kam{For $G<21.5$~mag, we find that J2016.0 position uncertainties from \gaia+\rst\, are almost identical to \gaia\, alone. The position uncertainties for brighter stars ($G\lesssim20$~mag) at the different }\rst\,\kam{ observation times are larger than the \gaia\, J2016.0 position uncertainties.} As expected, the position uncertainty for $G>21.5$~mag stars is much smaller at the \rst\, epochs than at J2016.0. The bottom panel shows the position uncertainty as a function of time for different magnitude sources, with vertical dashed lines showing the epochs of the different position measurements used in the simulation (either \gaia\, at J2016 or \rst\, for later times). The oscillating wiggles in the bottom panel within each year come from parallax uncertainty propagating to position uncertainty, while the general V-shape is due to PM uncertainties affecting past and future position extrapolations. The large leap in position uncertainty in the bottom panel is the transition to $G>21.5$~mag, where sources are too faint to have \gaia-measured positions at J2016.0. \label{fig:example_gaia+roman_position_uncert_vs_mag_for_epochs}} 
\end{figure*}

\begin{figure*}
    \centering
    \includegraphics[width=1.0\linewidth]{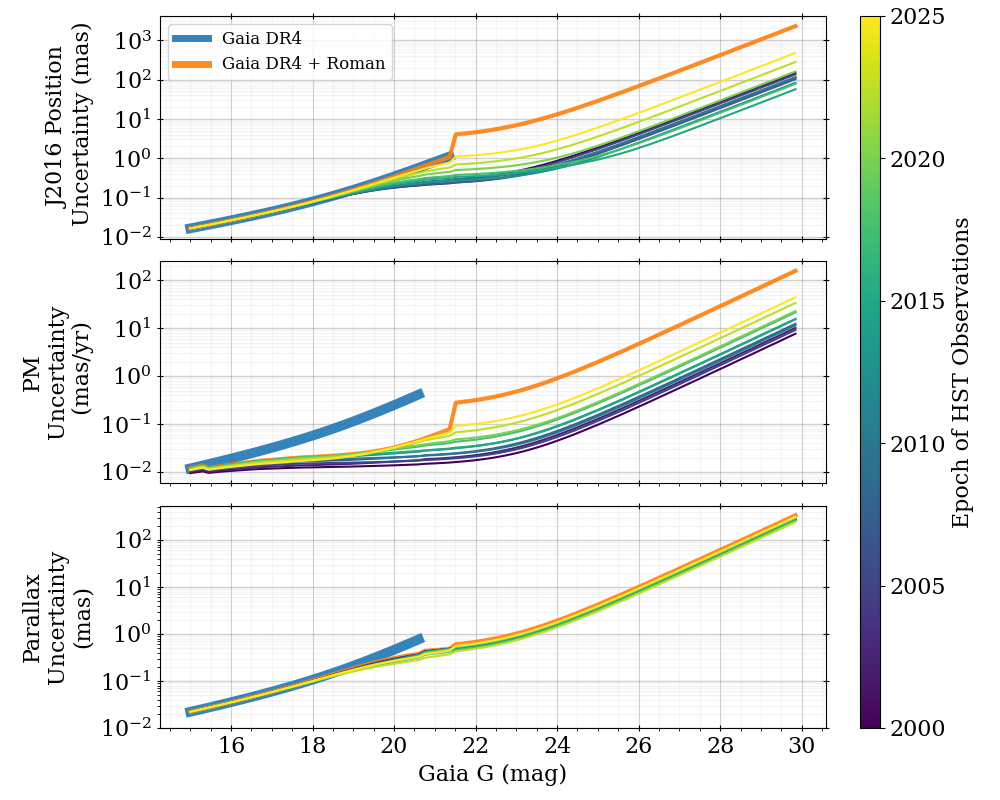}
    \caption{Example of astrometric precision after combining the \rst+\gaia\, DR4 results in Figure~\ref{fig:example_gaia+roman_astrometry} with additional \hst\, observations at different epochs. The colored lines show the result of adding a single epoch of \hst\, data with 4 dithers, assuming all magnitudes reach a per-exposure position uncertainty of $1\%$ of an \hst\, \textit{ACS/WFC} pixel (0.5~mas). Adding a single additional \hst\, epoch does not significantly improve the parallax precision, but it does lead to large improvements for the position and PM uncertainties.}
    \label{fig:example_gaia+roman+hst_astrometry}
\end{figure*}

The final required piece of information for our simulation tool is the specific \rst\, observation plan, and this is left up to the user. First, the user must specify the magnitudes of targets across all the filters that will be used, such as \gaia\, $G$, as well as the relevant \rst\, filters. Next, the user provides their desired \rst\, observing plan: the time of each \rst\, epoch, the filters used in those epochs, the number of dithers (i.e. back-to-back exposures) at each epoch, and the exposure times for each observation. Finally, the user is able to provide a (RA,Dec) pointing for the \rst\, images, which determines whether parallax effects are included in the astrometry calculations; this choice is discussed in more detail in Section~\ref{sec:caveats}. The simulation tool is then able to convert the input magnitudes into relevant \gaia\, astrometric precisions and \rst\, position uncertainties as a function of time, which are combined using Equation~\ref{eq:posterior_astrometry} to measure final position, parallax, and PM uncertainties. Using this approach, it is straightforward to test arbitrary observing strategies as well as the specific ones listed for the core surveys. Our hope is that this tool will guide conversations about optimal observing plans, both for the core mission as well as for proposal writing.

\subsection{Example Application} \label{sec:example_app}

We now use our tool to simulate the \gaia+\rst\, astrometry we predict for an arbitrary set of \rst\, images. A helpful tutorial page can be found on this \href{https://roman.ipac.caltech.edu/page/roman-gaia-astrometry-tool}{Roman Science Support Center at IPAC} webpage. Example \texttt{python}-based \texttt{Jupyter} notebooks for using the code and producing figures like those shown in this work are also found in \href{https://github.com/KevinMcK95/gaia_roman_astrometry}{this GitHub repository}.

We choose an arbitrary configuration as an illustrative example, setting the target coordinates to $RA = 150^\circ$ and $Dec=-5^\circ$, which is the approximate center of the HLWAS Medium Field 1. We choose to observe at three \rst\, epochs: J2028.0, J2031.0, and J2031.5. The first and final epochs are both set to use the F087 filter with 4 dithers per epoch, while the middle epoch is set to use the F106 filter with 5 dithers. For all epochs, each individual observation is set to use a $\sim193$~second exposure defined by the multi-accumulation (MA) strategy \texttt{IM\_193\_11}\footnote{Please refer to the following webpage for a detailed description of the allowed \rst\, multi-accumulation stategies: \url{https://roman-docs.stsci.edu/roman-instruments/the-wide-field-instrument/observing-with-the-wfi/wfi-multiaccum-ma-tables}}. We note that this MA is the one used to produce the position uncertainty functions shown in Figure~\ref{fig:roman_pos_errs}. Finally, we assume that all stars have zero color in all pairwise filter combinations, including between the \rst\, and \gaia\, filters, and simulate stars with magnitudes in the range of $15<G<30$~mag. Users interested in a particular type of star (e.g., KIII giants or MSTO stars) could use stellar isochrones to determine more accurate colors and magnitudes in each filter. 

The results of combining these \rst\, observations with \gaia\, DR4 and DR5 are shown in Figure~\ref{fig:example_gaia+roman_astrometry}. In the position and PM panels, the uncertainty we plot is a geometric mean calculated from the relevant covariance matrix using $\bar \sigma = | \pmb V_{x,y}|^{1/4}$, which accounts for the correlations between $(\alpha,\delta)$ and $(\mu_\alpha,\mu_\delta)$, respectively. As expected from the astrometry uncertainty definitions in Section~\ref{sec:math}, the final combined astrometry is always at least as precise as the corresponding \gaia\, data release. At fainter \gaia\, magnitudes such as $G > 18$~mag, the combined astrometry is a significant improvement over the \gaia-alone results for PM and parallax. For $20.7<G<21.5$~mag, \gaia\, is able to provide an additional epoch of position measurements at J2016.0 -- but no parallax or PM prior information -- which is why the combined \gaia+\rst\, parallaxes and PMs are also significantly improved in this range. For $G>21.5$~mag, there is no longer any \gaia\, measurements, so the \rst\, results are using the same \rst-only information to return identical final astrometry. 

We note that the position uncertainty in Figure~\ref{fig:example_gaia+roman_astrometry} is at the \gaia\, reference epoch of J2016.0. The position covariance matrix at some different epoch can be calculated using $\pmb U_{i,j}\cdot \pmb \Sigma_{v,i}\cdot \pmb U_{i,j}^T$, where $\pmb U_{i,j}$ contains the appropriate time and parallax offsets at a desired new time $t_j$. The position uncertainty as a function of magnitude at different epochs for our example data are shown in Figure~\ref{fig:example_gaia+roman_position_uncert_vs_mag_for_epochs}. The top panel shows the position uncertainty at each of the observed \rst\, epochs, while the bottom panel shows the position uncertainty as a function of time for different magnitude sources. In the bottom panel, the vertical dashed lines show the time of the different input position measurements. Oscillations on the timescale of a year are apparent in the bottom panel because of the propagation of parallax uncertainty into the epoch position uncertainty. There is a large increase in the position uncertainty at $G>21.5$~mag because \gaia\, is no longer able to provide a J2016.0 position measurement for these faint sources. 

It is also trivial to simulate combining additional epochs from other telescopes. The new astrometric precision matrix simply becomes 
\begin{equation} \label{eq:astrometry_add_hst}
    \left[\pmb \Sigma_{v,i}^{-1} + \sum_j^{n_{\mathrm{epochs}}} \pmb U_{i,j}^T \cdot \pmb C_{i,j}^{-1} \cdot \pmb U_{i,j} \right]^{-1}
\end{equation} with the new $\pmb C_{i,j}$ containing the relevant position uncertainty information. Here, we simulate adding in a single epoch of \hst\, observations with 4 dithers. We assume that we can measure the \hst\, position to 1\% of an \hst\, pixel width (0.5~mas for \textit{ACS/WFC}). The result of adding \hst\, observations is shown in Figure~\ref{fig:example_gaia+roman+hst_astrometry} where the different colored lines show the impact of changing the \hst\, time baseline. As expected, early \hst\, observations provide long time baselines that yield the smallest PM uncertainties, which also helps to improve position uncertainties at J2016.0. The parallax precision does not change much when only a single \hst\, epoch is used. 

\section{Caveats \& Assumptions} \label{sec:caveats}

As with many simulation tools, there are many, often reasonable, assumptions that went into making this work possible. In the following subsections, we describe our key assumptions and caveats about the use of our tool. In some cases, we suggest potential future improvements as well as provide general insight into making high-precision astrometric measurements. 

\subsection{Position Uncertainties}

One fundamental assumption of this work is that the \rst\, PSFs per filter are perfectly known as defined by \texttt{stpsf}. Often, optimal astrometric precision is achieved when a per-image PSF is constructed or adjusted from a template \citep[i.e. effective PSFs: ePSFs;][]{Anderson_2000} using the observed point sources. ePSFs are effectively a convolution of the optical PSF with the detector pixel response function (approximately a 2D boxcar), which can only be directly observed from images. One reason to use sub-pixel dithering -- in addition to improving SNR of position uncertainties at a given epoch -- is to better sample the ePSF. While \texttt{stpsf} does indeed estimate realistic ePSFs, they might not agree perfectly with true observations in every exposure, meaning that our results will likely vary at the level of a few percent. Similarly, the widest filter of F146 may experience difficulties constructing ePSFs due to color-dependent PSF variations introducing biases that affect astrometry. We have not included the difficulties or systematics that might be introduced through building ePSFs. 

Another key assumption behind our \rst\, position uncertainties is that there is no variation across the focal plane. In our work, we present position uncertainties measured at the center position of the central \rst\, detector, but the PSF is known to vary across the plane of each detector and between detectors. Future work could focus on quantifying centroiding precision across the field of view to produce more accurate astrometry simulations. We have also not included the some of the complicated effects that saturation -- or near-saturation, such as the brighter-fatter effect changing the effective PSF -- can have on centroiding. It is likely that some of the brightest magnitudes in Figure~\ref{fig:roman_pos_errs} will measure significantly worse position uncertainties than the ones we report. As a result, this means that the bright end of our simulations is likely overly optimistic. 

\begin{figure}
    \centering
    \includegraphics[width=1.0\linewidth]{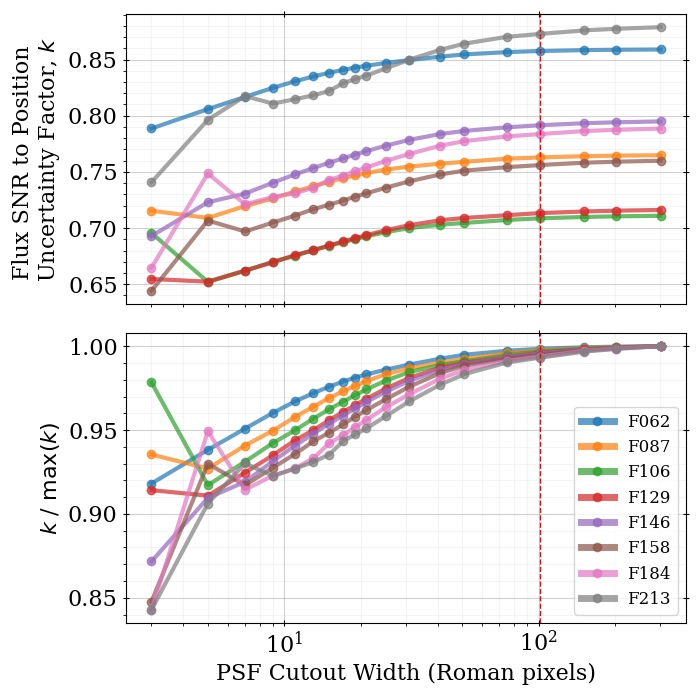}
    \caption{\textbf{Top:} Flux SNR to position uncertainty factor $k$, as defined by Equation~\ref{eq:SNR_to_pos_err}, for different cutout sizes and \rst\, filters using \texttt{stpsf}-defined PSFs. The vertical red dashed line at 101~pixels correspond to the values used in this work and presented in Table~\ref{tab:snr_to_pos_errs}. As the cutout size increase, more of the wings are included in the fit, yielding a higher position precision at a given flux SNR. \textbf{Bottom:} Comparison of each filter's $k$ to its approximate asymptotic value using a 301 pixel cutout. Even for small cutouts, the $k$ value only changes by a few percent (up to $\sim16\%$).}
    \label{fig:flux_snr_to_pos_err_vs_cutout}
\end{figure}

\begin{figure*}[t]
    \centering
    \includegraphics[width=1.0\linewidth]{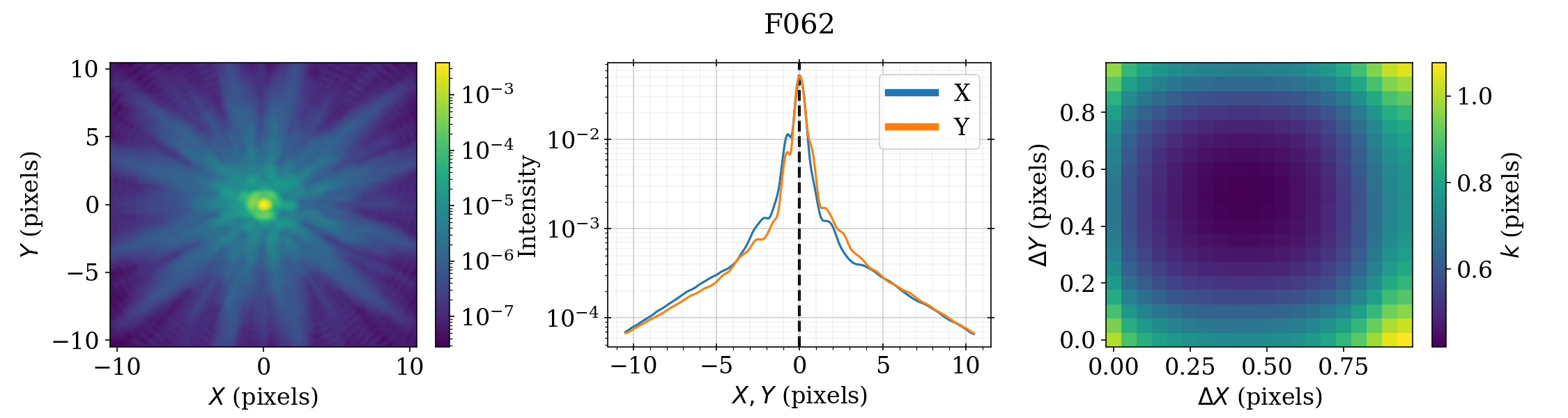}\\
    \includegraphics[width=1.0\linewidth]{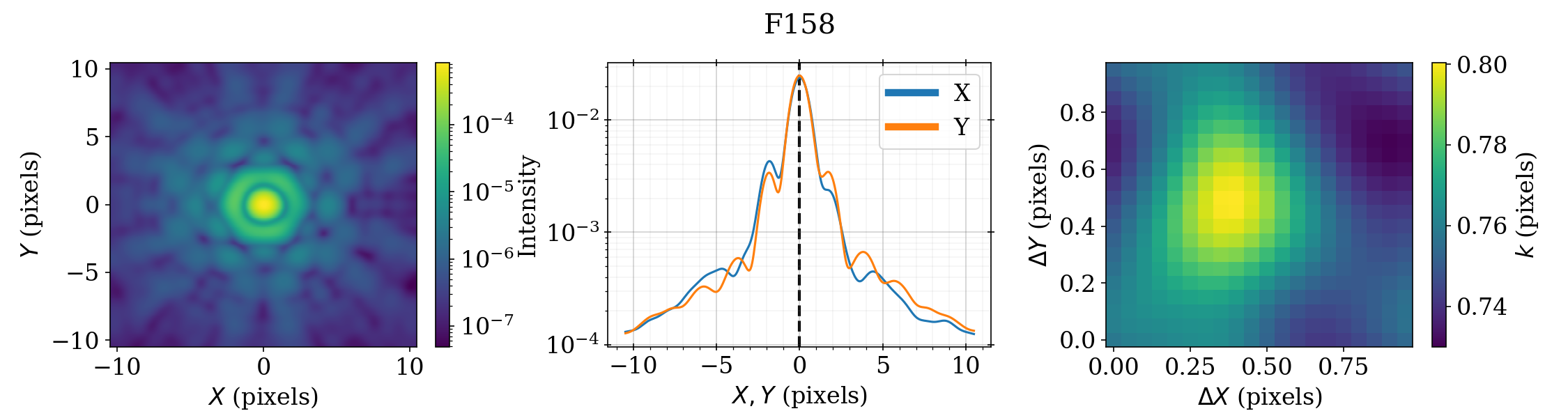}\\
    \includegraphics[width=1.0\linewidth]{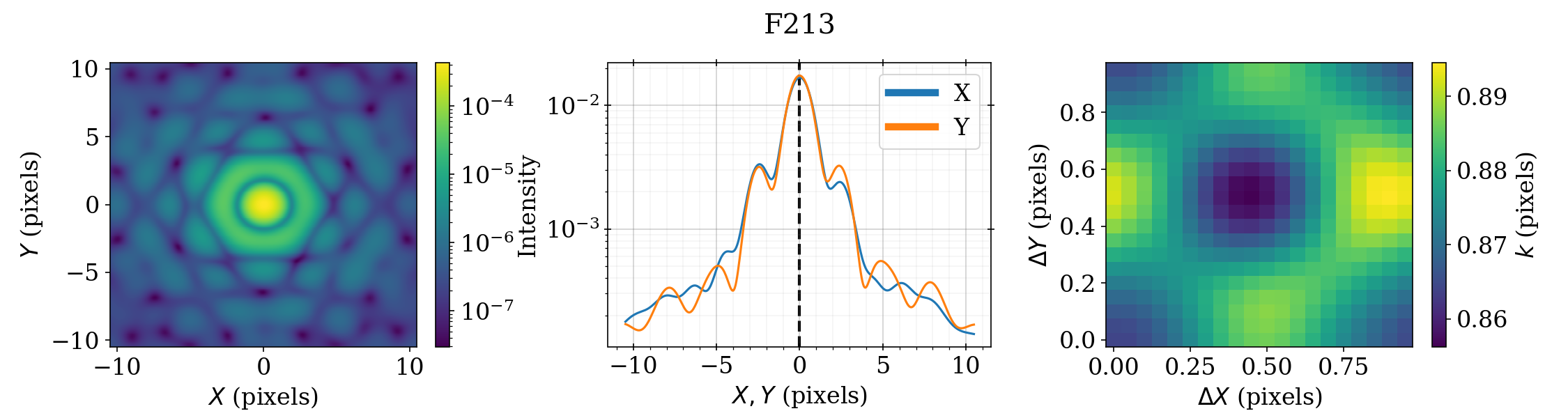}
    \caption{\kam{Comparison of the PSF distribution and effect of subpixel offsets on position uncertainty measurements for a few representative filters (top to bottom: F062, F158, F213). The left panels show the super-sampled PSF in $X,Y$ coordinates (zoomed in to the central $21\times21$ pixels), while the middle columns show 1D projections of the PSFs; these panels show the the default \texttt{stpsf} output PSFs without any subpixel offsets. The right panels demonstrate how the $k$ factor changes as a function of a subpixel offset. For very under-sampled PSFs (top), the change in $k$ is larger than for better-sampled PSFs (middle and bottom).}}
    \label{fig:subpixel_shift_psf_vs_filter}
\end{figure*}

\begin{figure}[t]
    \centering
    \includegraphics[width=1.0\linewidth]{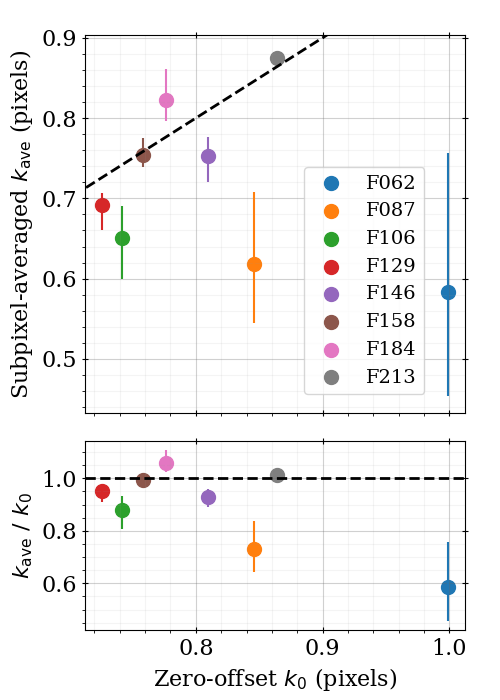}
    \caption{\kam{\textbf{Top:} Comparison of the average $k$ across subpixel offsets versus the $k$ value with no offset (i.e., the native \texttt{stpsf} output). Errorbars show a 68\% credible interval. The most under-sampled PSFs -- that is, F062, F087, and F106 -- show the largest variation in $k$ in response to subpixel offsets. \textbf{Bottom:} subpixel-averaged $k$ divided by the zero-offset $k$ to quantify the relative difference. The average $k$ value is typically within 10\% of the zero-shift $k$ value, but some under-sampled PSFs show differences at the 40\% level.}}
    \label{fig:subpixel_k_vs_k0}
\end{figure}

While measuring the relationship between position uncertainty and flux SNR (Equation~\ref{eq:SNR_to_pos_err}) using \texttt{stpsf}, we used a cutout of $101\times101$~pixel$^{2}$ to include the PSF wings. However, actual fitting of PSFs to extract positions will likely use much smaller cutouts (e.g., $9\times9$). We estimate the impact of this choice in Figure~\ref{fig:flux_snr_to_pos_err_vs_cutout} for the different filters by varying the cutout width and its impact on the conversion factor $k$ between flux SNR and position uncertainty. In summary, the $k$ values -- and therefore our uncertainty predictions -- change by $<8\%$ between the $101\times101$ and $9\times9$ cutouts. Similarly, when estimating flux SNR versus magnitude functions, we have used \texttt{pandeia}'s default 0.2" aperture in our simulations, but this aperture does not necessarily result in the optimal SNR given the FWHM of the \rst\ PSFs \citep[e.g., Figure 5 of ][]{Savino_2024}. Similarly, the perfect mapping of photometric SNR to astrometric SNR that we have assumed in this work is likely not exactly one-to-one: for instance, the relationship in Equation~\ref{eq:SNR_to_pos_err} with a magnitude-independent $k$ is likely less true at the faintest magnitudes (e.g. noise dominated by the background). However, we expect that all of the effects discussed here are likely to matter at the level of, at maximum, a few tens of percents and not orders of magnitude.

\kam{The PSF derivatives in Equation}~\ref{eq:position_and_flux_errs}\kam{ suggest that subpixel shifts of the PSF distribution will impact the position uncertainties. We explore this effect in Figure}~\ref{fig:subpixel_shift_psf_vs_filter}\kam{ for three different filters: F062, F158, and F213. The widths of these PSFs change from very under-sampled (F062) to more complete sampling (F158 and F213). In all cases, we notice that the 2D PSF -- and the 1D projections in X and Y -- are asymmetric. The right-most column summarizes the impact that subpixel offsets have on the position uncertainty measurements, quantified by changes in $k$. Given our pixel grid choices, \texttt{stpsf} places the center of the PSF in the middle of the central pixel. This explains why the bluest, most under-sampled filter shows a relatively large $k$ change from the zero-offset to (0.5,0.5) pixel offsets: the former has the PSF core falling in a single pixel while the latter splits the core light across fours pixels, enabling better centroiding. In this work, we report results using the zero-offset $k$ value (i.e. the bottom left corner of each right panel), as this is the default \texttt{stpsf} output. Future versions of our tool will likely let users select $k$ values for different subpixel offsets or use an average across all offsets.}

\kam{We compare the zero-offset $k$ to the mean $k$ after averaging over all subpixel offsets in Figure}~\ref{fig:subpixel_k_vs_k0}\kam{, finding that the average $k$ is within a few percent of the zero-shift $k$ value for most filters. However, the most under-sampled filters -- namely F062, F087, and F106 -- show that the average $k$ value is up to 40\% smaller than the zero-shift $k$ value. For simulated observations using the bluest filters, our current predictions for position uncertainties may be pessimistic at the tens of percent level, but how that effect propagates into PM and parallax precision depends on the particular observing strategy.}

\kam{After accounting for subpixel dithers, it is not surprising that smaller width PSFs could enable more precise position measurements compared to wider PSFs. Of course, this ``narrower PSF gives more precision'' trend is only true up to a certain point: for a delta function PSF, localizing a source is only possible to within approximately $\pm0.5$~pixel. Using similar logic, under-sampled PSFs necessarily rely more on the information in their relatively low S/N wings when constraining subpixel positions. This likely explains why previous empirical calibrations with real} \hst\,, \jwst\,, and \euclid\, \kam{observations have found that under-sampled PSFs do not localize stellar centroids as well as well-sampled ones at a given S/N} \citep[e.g.,][]{Bellini_2011,Libralato_2023,Libralato_2024b}\kam{. Similarly, we expect that future real} \rst\, \kam{data will reveal that the very under-sampled filters will measure less precise positions for the same flux SNR compared to better-sampled filters. While it is difficult to predict the magnitude of this effect, it further justifies our choice to use the more pessimistic zero-shift $k$ values instead of the subpixel-offset averaged ones.}

\kam{The calculations presented in this work assume that there} will not be significant uncertainties associated with aligning \rst\, images onto the \gaia\, reference frame, either through systematics or by simple propagation that leads to inflated localization uncertainties. That is, we assume the alignment uncertainty is a component of the floor uncertainty in our positions. Because of the large \rst\, field of view, this is likely a very reasonable choice: most pointings will contain enough well-measured \gaia\, stars (e.g., $>50$) to align each image. However, there may be sparse regions -- for example, in the HLWAS -- where there are not many \gaia\, stars per \rst\, pointing. When properly propagated, this alignment uncertainty may increase the uncertainty of the \rst\, positions above what we have assumed. 

Currently, our tool uses the floor error added to the \rst\, position uncertainties as a way to capture both a sense of a systematic floor as well as typical impact from alignment uncertainty. Our choice of a 1\% pixel width floor is consistent with findings from other telescopes like \hst\ \citep{Anderson_2004,Anderson_2006}, \jwst\ \citep{Libralato_2023,Libralato_2024a}, and \euclid\ \citep{Libralato_2024b}, as well as previous \rst\, predictions \citep[e.g.,][]{WFIRST_Astrometry_2019}. Users can explore the effect of changing the position floor (e.g. as a method for testing an optimistic extreme-calibration future or pessimistic bad alignment scenarios) if they so desire. 

We note that systematic effects could include, for example, centroiding issues -- either from problems with estimated empirical PSFs or detector effects like pixel sensitivity variations -- and global alignment issues (e.g., from improperly-defined distortion corrections). In this work, we have assumed that all errors are improved by $\sqrt{n_\mathrm{exposures}}$, which is likely true for many types of centroiding systematics. However, some systematics such as ill-measured distortion corrections could lead to correlated position measurements between different images that do not improve as $\sqrt{n}$. We have ignored the latter type of systematic with our simulations, and it is not clear how best to incorporate these effects without specific knowledge about the true \rst\, performance. Ideally, variations in \rst\, geometric distortions will be very stable in time at L2, and can be well-characterized or removed, in particular by harnessing the extreme number of repeat visits used by the GBTDS, with some estimates predicting improved systematic floors down to 0.1\% or 0.01\% of a pixel. 

\subsection{Population Priors} \label{sec:pop_prior_caveats}

As mentioned in Section~\ref{sec:math}, we have implemented an extremely diffuse global prior on the expected astrometry (Equation~\ref{eq:diffuse_priors}). These priors offer numerical stability when inverting the astrometry covariance matrices of $\pmb \Sigma_{v,i}$, especially for cases where fewer than 3 epochs are observed, leading to singular astrometry covariance matrices. When exploring the final astrometry precision outputs, cases where the position, parallax, and PM uncertainties are extremely large (i.e. $\sim 1\times10^{4}$~mas or $\sim 1\times10^{5}$~mas/yr) should be treated as unconstrained measurements. In principle, teams building astrometric catalogs should be careful when incorporating population-level priors. While it often is good practice to incorporate prior knowledge about a group of stars into your analysis -- such as using literature membership probabilities with distances and motions for a nearby galaxy -- these choices can make resulting catalogs less future-proof than otherwise. Any subsequent measurements that build off of the original work will necessarily also incorporate the original prior assumptions. Unless end users are provided with the exact way to remove the original prior, it becomes statistically incorrect to incorporate new prior constraints with the catalog (e.g. based on different literature choices). Conversely, forgoing a population prior makes it easy to daisy-chain new measurements into the catalog likelihood, as illustrated with Equation~\ref{eq:astrometry_add_hst}. Not including a population prior also allows for catalog users to make their own decisions about relevant background information for boutique analyses. \gaia\, is a wonderful survey in this sense because its measurements describe likelihood distributions that allow us to combine it with additional information while maintaining robust statistics.

\subsection{Astrometry Calculations}

\kam{When combining data from multiple observatories, systematic errors from all telescopes can and will manifest in the final combined astrometric solutions. While we have already discussed systematics in the} \rst\, \kam{position measurements, previous work has demonstrated biases, correlations, and underestimated uncertainties in \gaia\, PMs and parallaxes} \citep[e.g.,][]{Vasiliev_2019,Lindegren_2021}. \kam{We have not included \gaia\,-level systematics effects in this work because it is difficult to anticipate their strength or importance in future data releases. Ideally given enough time, the \gaia\, systematics may be fully characterized or removed entirely. One could include \gaia\, systematic effects by adding a floor astrometric covariance matrix to $\pmb C_{G,i}$ in Equation}~\ref{eq:vec_heirarchy}\kam{, although we note that this is not a feature of the current version of our simulation tool.}

For real analyses of \rst\, images, we suggest that the most statistically rigorous approach is to simultaneously align the \rst\, images with \gaia\, while also updating the astrometry of all the sources in the image, following the Bayesian approach of \bpm\, \citep{McKinnon_2024}. We note that the previous \bpm\, reference uses a slow-to-converge MCMC method when aligning images and fitting astrometric solutions, but we have recently developed a much faster process that is able to directly draw from posterior distributions. This new approach will be discussed in upcoming work (McKinnon et al. in prep.), and was recently successfully applied in an analysis of Draco II's bulk motion \citep{Warfield_2025}. As presented in the \bpm\, paper, the effect of the number of \gaia\, stars per image as well as the impact of the precise location of stars across the detector can be explored by generating realistic synthetic data, but we leave that for future work. It is also possible that some properties and associated uncertainties of the alignment may be constrained, for example, using self-calibration to interpolate between well-measured pointings; if this approach is indeed taken, then it is important to remember that self-calibration can be made easier or harder based on the choice of observation/tiling strategies \cite[e.g.,][]{Holmes_2012}. 

If the user does not provide an RA, Dec coordinate during the fitting, then parallax is not included in the final astrometry calculations (i.e., only position and PM are used to invert a $4\times4$ matrix). For many applications, this is not recommended, but there are situations where fitting without parallax is likely to give more accurate predictions. For example, co-moving stars in moderately distant (e.g. $D > 10$~kpc) dwarf galaxies experience effectively the same parallax effects over the \rst\, field of view: if there are enough co-moving and equidistant stars in an image to align using those sources alone, then parallax offsets will be removed during the image alignment to the \gaia\, reference frame. Similarly, ignoring parallax effects during fitting is almost equivalent to putting a very strong prior on the parallax, which we might know a priori from the literature for confirmed member stars of a nearby galaxy. 

We also note that the parallax offsets used in our calculations -- like those shown in Figure~\ref{fig:parallax_example} -- are calculated using the Earth's orbit. However, the parallax effects that \rst\, will see as a function of time will be dictated by its precise L2 orbit. True analyses of real \rst\, data will require taking the ecliptic of the telescope into account at a given time; this is also true for other L2 telescopes like \jwst\, and \euclid. We expect that using Earth's orbit instead of L2 in our simulations should only have a minor impact on the final astrometry. For instance, L2 is relatively near the Earth, simply $\sim 1\%$ further from the Sun, so the parallax effects in this work are likely within $\sim 1\%$ of the truth. 

\begin{figure*}[t]
    \centering
    \includegraphics[trim={0cm 0cm 0cm 1.5cm},clip,width=1.0\linewidth]{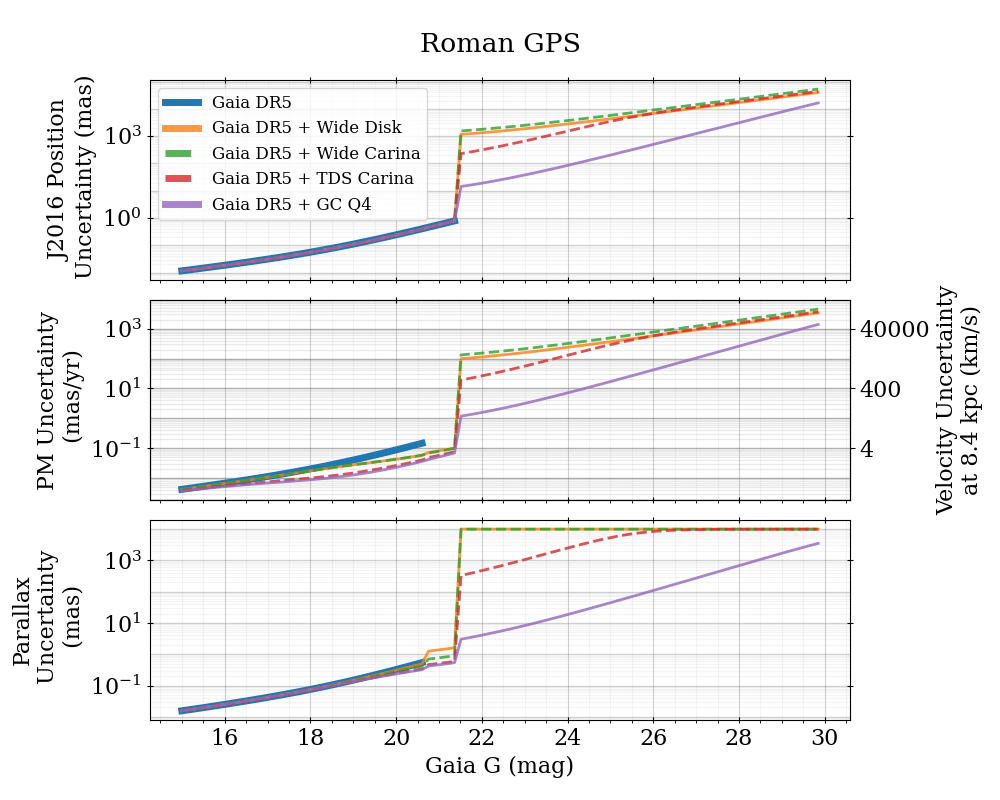}
    \caption{Expected astrometric precision for the GPS for the fields listed in Table~\ref{tab:gps_coordinates}. We convert the PM uncertainties to tangential velocity uncertainties at an approximate distance of the Galactic center. We do not expect to measure precise PMs and parallaxes for faint stars ($G>21.5$) in most GPS fields.}
    \label{fig:GPS_astrometry_errs}
\end{figure*}

Our model of stellar motion assumes that the true change in position of a star on the sky is entirely described by the PM and parallax; it does not account for position changes due to radial velocity changing the distance to a star. This means that our calculations do not apply to some of the nearest stars where radial velocity effects could be important. Appendix~\ref{sec:los_appendix} estimates the strength of this assumption, suggesting that $D>50$~pc is likely a good threshold for typical velocities, uncertainties, and time baselines. Similarly, the PM for each star is assumed to be constant, meaning our precision calculations do not strictly apply to stars experiencing significant changes in PM over human timescales; such cases include stars near the Galactic Center as well as stars in binary systems. However, this tool may still be useful for these non-constant-PM cases because it can set a threshold for determining PM uncertainties required to detect divergence from a constant-PM null hypothesis.

Given the degeneracies between parallax and PM as well as the relationship between parallax effects, pointing, and time, it behooves us to think about the precise timing of a survey. If high-precision PMs are the goal, then the following lessons are important:
\begin{enumerate}
    \item If you only have 2 epochs, then it is best to observe at the exact same phase of year such that parallax offsets  -- and therefore degeneracy with PM -- will be removed;
    \item If you only have 2 epochs, but require them to be 6 months out of phase with each other, then it is best to observe at the time that minimizes the parallax offset factors;
    \item If you have 3 or more epochs, it is best to start observing at the time that maximizes the parallax offset factors, then have subsequent epochs be 6 months out of phase with the previous epoch. This will maximize parallax constraining power and break degeneracies with PM. 
\end{enumerate}
Many of the \rst\, core surveys expect to have at least 2 epochs with a phase separation of 6 months such that the return visit is observed with a different telescope roll angle. This consideration means that item 2 in the above list is particularly relevant. These lessons are most impactful when considering small to moderate numbers of epochs across a short time baseline (e.g. HLWAS). If you instead have an extremely large number of epochs over a longer baseline (e.g. GBTDS), then the parallaxes and PMs will likely be constrained by sheer number of repeat observations, and it is less important to fuss over the initial and followup epoch times.

\section{Application to Core Roman Surveys} \label{sec:applications}

We next use our simulation tool to predict the expected astrometry from the core \rst\, surveys, whose goals and initial proposed observing strategies are described in \citet{Roman_ROTAC_2025}. The exact filters, exposure times, and cadences we use in our simulations are based off of the information provided on the \href{https://roman-docs.stsci.edu/roman-community-defined-surveys}{Roman Community Defined Surveys} webpage, which we summarize in the following subsections. As these strategies are likely to evolve in the coming months, we clarify that we have used the information available as of December 2025. In all following subsections, we compare to \gaia\, DR5 to show what we expect at the end of Roman's nominal five year lifetime. We also always assume there is exactly zero color for all pairwise comparisons of different filters.  

\subsection{Galactic Plane Survey (GPS)} \label{sec:gps}

\begin{table}[h]
    \centering
    \begin{tabular}{|ccc|}
         \hline
         Field & $l$ (deg) & $b$ (deg) \\
         \hline
         Wide Disk & 45 & 0 \\
         Wide Carina & -73.0 & -0.25 \\
         TDS Carina & -72.48 & -0.65 \\
         TDS Galactic Center Q4 & 358.49 & -0.125 \\
         \hline
    \end{tabular}
    \caption{GPS Galactocentric Coordinates used in astrometry simulations.}
    \label{tab:gps_coordinates}
\end{table}

\begin{figure*}[t]
    \centering
    \includegraphics[trim={0cm 0cm 0cm 1.5cm},clip,width=1.0\linewidth]{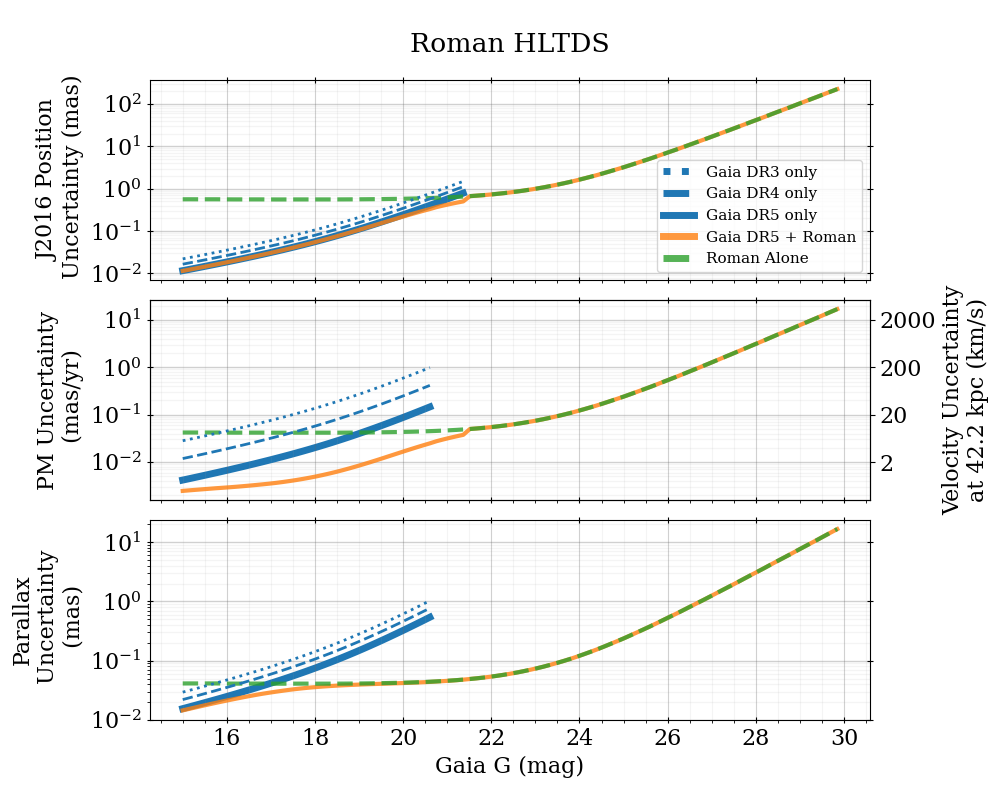}
    \caption{Expected astrometric precision for the Euclid Deep Field in the HLTDS, including a conversion to tangential velocity uncertainties at a moderate halo distance of $\sim 42$~kpc. The blue lines show the predictions for different \gaia\, data releases. The orange line shows the combined \gaia+\rst\, predictions using the observing strategy described in the text, and the green dashed line shows the predictions from the \rst\, information alone. A large number of repeat \rst\, observations with a relatively large time baseline leads to precise PMs and parallaxes for faint targets ($G>21.5$~mag). }
    \label{fig:HLTDS_astrometry_errs}
\end{figure*}

\begin{figure*}[t]
    \centering
    \includegraphics[trim={0cm 0cm 0cm 0cm},clip,width=1.0\linewidth]{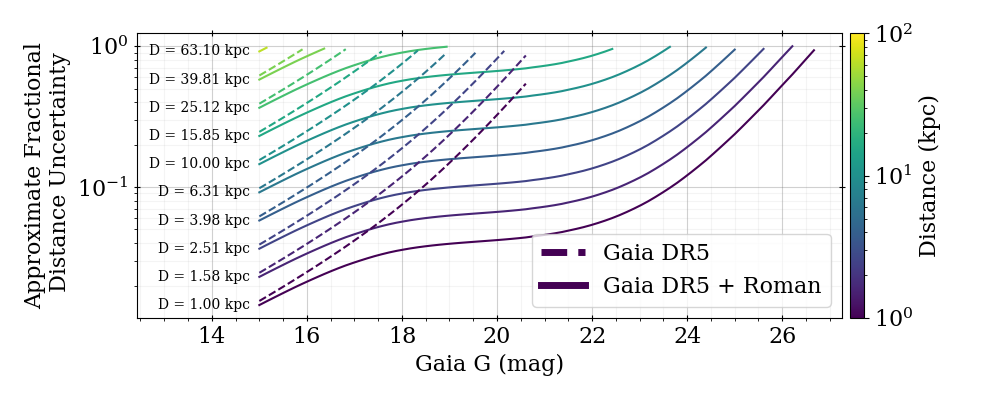}
    \caption{Approximate fractional distance uncertainty versus magnitude for the HLTDS based on the parallax uncertainty in Figure~\ref{fig:HLTDS_astrometry_errs}. Only measurements with less than 100\% distance uncertainties are included. The dashed lines show \gaia\, DR5 expectations alone, while the solid lines show the combined \gaia+\rst\, results. Lines are colored and labeled by the true input distance.}
    \label{fig:HLTDS_distance_frac_errs}
\end{figure*}

\begin{figure*}[t]
    \centering
    \includegraphics[trim={0cm 0cm 0cm 1.5cm},clip,width=1.0\linewidth]{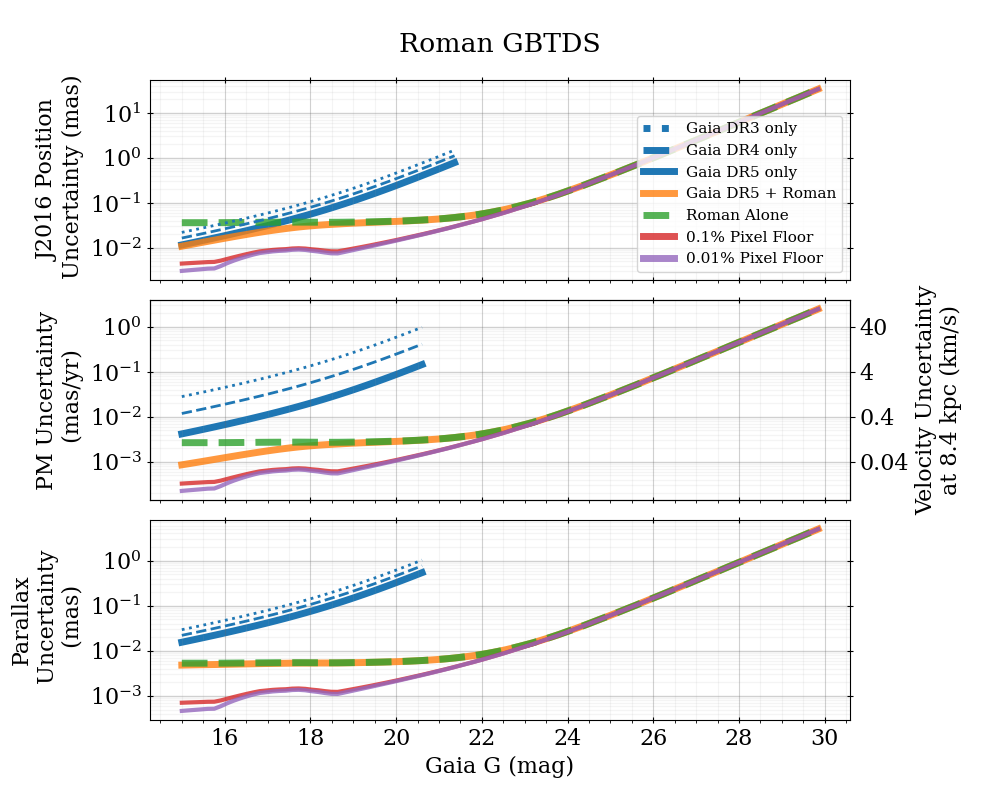}
    \caption{Same as Figure~\ref{fig:HLTDS_astrometry_errs}, but now simulated astrometric precision for the GBTDS near the Galactic center. The tangential velocity uncertainty is estimated at the approximate distance of the Galactic center. The orange line shows the result of assuming our standard 1\% \rst\, floor uncertainty in position, with red and purple lines showing the results if the systematic floor is improved down to 0.1\% and 0.01\% pixels, as are being discussed as a post-calibration possibilities. The extreme number of repeat \rst\, observations ($>54,000$) with a $\sim 5$ year time baseline leads to very precise PMs and parallaxes across a large range of magnitues, including for faint targets ($G>21.5$~mag).}
    \label{fig:GBTDS_astrometry_errs}
\end{figure*}

A summary of the GPS fields we explore are given in Table~\ref{tab:gps_coordinates}. For all fields, we set the first epoch to J2027.1 while the second epoch is set to J2028.5. All pointings use 2 dithers with an exposure time of 60 seconds. For ``Wide'' fields, the first epoch uses the F129 and F213 filters while the second epoch uses F158 and F184. For the ``TSD Carina'' and ``GC Q4'' fields, the strategy includes more observations to capture time series information. TDS Carina uses F129 and F158 at the first epoch and F184 at the second epoch as normal. Then, F213 is observed at the second epoch using the ``fast/minute'' configuration -- described on the Roman Community Defined Survey webpage -- which takes 43 exposures with an equal spacing of $11$~minutes. The TDS Galactic Center Q4 (``GC Q4'') takes F062 in the first epoch and F087 and F106 in the second epoch as normal. The first epoch then also includes F129 using the ``low/weeks'' configuration, while the second epoch includes F213 using the ``low/weeks'', ``medium/hours'', and ``fast/minutes'' configurations. The ``low/weeks'' configuration takes 5 exposures equally spaced over 60 days (i.e. every $\sim2$~weeks), while the ``medium/hours'' configuration takes 8 exposures over 80~hours with increasing spacing between each exposure (though we simulate it using equal spacing for convenience). 

The resulting astrometric predictions are shown in Figure~\ref{fig:GPS_astrometry_errs}. For the PMs, we convert to tangential velocity uncertainties (right axis of middle panel) using:
\begin{equation}\label{eq:vtan_err_from_pm_err}
    \left[\frac{\sigma_{V_T}}{\mathrm{km/s}} \right] = 4.744\cdot \left[\frac{D}{\mathrm{kpc}} \right]\cdot\left[\frac{\sigma_\mu}{\mathrm{mas/yr}} \right]
\end{equation}
where we have set $D=8.4$~kpc as an approximate distance to the Galactic center (while also having integer values at the PM axis ticks). From Equation~\ref{eq:vtan_err_from_pm_err}, $\sigma_{V_T}$ increases linearly with changing distance. As the GPS is targeting stars over a range of distances, it may be helpful to point out that the tangential velocity uncertainties can be changed to test another distance by multiplying the values on the right axis by $D/8.4$, where $D$ is in kpc. With the relatively small number of epochs in the ``Wide'' fields, we find that we do not expect to measure particularly constraining PMs or parallaxes for faint stars ($G>21.5$~mag). The GC Q4 field, with a larger number of repeat observations, should measure much more precise PMs and parallaxes compared to the other fields. All fields show a PM improvement over \gaia-alone for $G<20.7$~mag in addition to well-measured PMs and parallaxes in the $20.7<G<21.5$~mag regime.

\subsection{High Latitude Time Domain Survey (HLTDS)} \label{sec:hltds}

\begin{table}[h]
    \centering
    \begin{tabular}{|cccc|}
        \hline
        Filter  &   Exposure     & Number of &  Cadence  \\ 
         & Time (Sec) & Dithers & Group \\ \hline
        F087	&	193 &	1     &      1,2 \\
        F129	&	307 &	1     &      1   \\
        F184	& 	409 & 	4     &      1   \\
        F158	&	420 &	1     &      2   \\
        F106	&	294 &	1     &      2   \\
        \hline
    \end{tabular}
    \caption{HLTDS core cadence by filter used in astrometry simulations, which will be observed for the middle $\sim2$~years of \rst's nominal 5 year lifetime. Each cadence group will be observed 5 days after the last, alternating between groups 1 and 2.}
    \label{tab:hltds_core_cadence}
\end{table}

For simulating the HLTDS, we choose a pointing centered on the Euclid Deep Field (EDF; $\alpha,\delta = 61.241^\circ,-48.423^\circ$). The EDF will be observed in a Pilot, Early Extended, Core, and Late Extended set of configurations. The Core observing strategy is defined by Table~\ref{tab:hltds_core_cadence} and will span $\sim2$~years in the middle of \rst's nominal 5 year lifetime; for simulation purposes, we set the core observations to occur every five days from J2028.5 to J2030.5. The Core will alternate between two different cadence groups, each observed 5 days after the last, using three of the total five filters at a time. The Pilot, Early Extended, and Late Extended observations will also use the same filters, exposure times, and dithers as the Core, though all five filters will be observed at every epoch. 

Following the Roman Community Defined Survey information, we assume there will be 8 epochs in the Pilot program with a cadence of 16 days, setting the first epoch to J2027.0. The Early Extended program will end 35 days before the start of the Core program, covering 3 epochs with a 65 day cadence. The Late Extended program will start 35 days after the end of the Core program, covering 5 epochs with an 85 day cadence. 

The resulting astrometric predictions are shown in Figure~\ref{fig:HLTDS_astrometry_errs}. In this case, we have converted PM uncertainties to velocities at a moderate halo distance of $\sim 42$~kpc. For this survey, we have enough repeat \rst\, observations that we can also show the \rst-only astrometric predictions (green dashed line), which highlights the role that the \gaia-measured priors plays for the brighter magnitude sources. Compared to the GPS, now many $G>21.5$~mag stars will have very high precision PMs ($<0.1$~mas/yr) and parallaxes ($<0.1$~mas). Our results anticipate that the HLTDS will be able to measure \gaia\, DR3 quality parallaxes and PMs down to $G=26.5$~mag, including tangential velocity uncertainties smaller than $100$~km/s for $G<26$~mag to moderately large halo distances. Clearly, using a large number of repeat observations with multi-year time baselines leads to strong constraints on parallax and PM. 

We can take the parallax uncertainty estimates one step further by estimating the distance uncertainty. While the optimal conversion between distance and parallax can be quite complicated and involve incorporating carefully selected prior distance knowledge \citep[e.g.,][]{Bailer-Jones_2023}, we use the simple conversion of $D = \frac{1}{\varpi}$, leading to the following fractional distance uncertainty estimate: 
\begin{equation}
    \frac{\sigma_D}{D} = D\cdot \sigma_\varpi,
\end{equation} 
which we can evaluate across a range of distances, producing the results in Figure~\ref{fig:HLTDS_distance_frac_errs}. Here, we have only kept measurements that have less than a 100\% distance uncertainty, with dashed lines showing the \gaia\, DR5 predictions and solid lines showing the combined \gaia+\rst\, results. Evidently, stars within $10$~kpc can expect $<100$\% distance uncertainties out to $\sim23.5$~mag at the end of the HLTDS. These findings are quite exciting for many aspects of Milky Way substructure studies, where fainter magnitude stars with useful parallaxes will enable characterization of the Milky Way stellar halo to much finer distance resolution than \gaia\, alone.

\subsection{Galactic Bulge Time Domain Survey (GBTDS)} \label{sec:gbtds}

\begin{figure*}[t]
    \centering
    \includegraphics[trim={0cm 0cm 0cm 0cm},clip,width=1.0\linewidth]{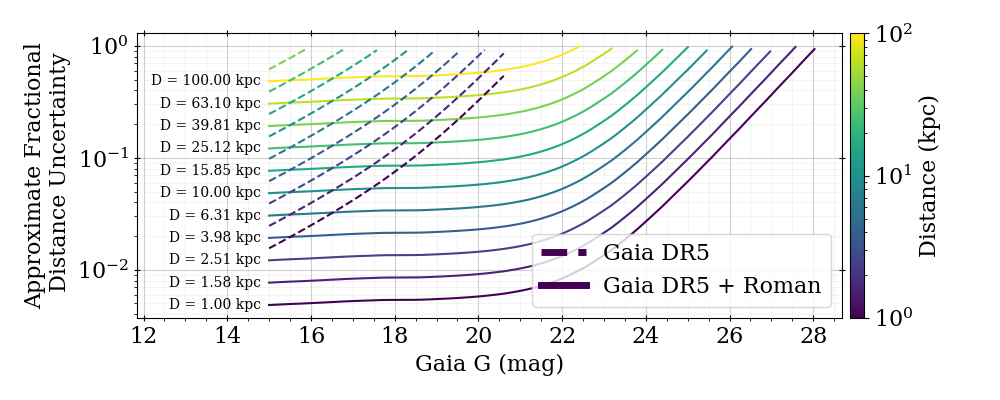}
    \caption{Same as Figure~\ref{fig:HLTDS_distance_frac_errs}, but now for the GBTDS based on the orange-line parallax uncertainties in Figure~\ref{fig:GBTDS_astrometry_errs}. Labels indicating distance align with the left edge of the \gaia+\rst\, tracks.}
    \label{fig:GBTDS_distance_frac_errs}
\end{figure*}

\begin{figure*}[t]
    \centering
    \includegraphics[trim={0cm 0cm 0cm 2cm},clip,width=1.0\linewidth]{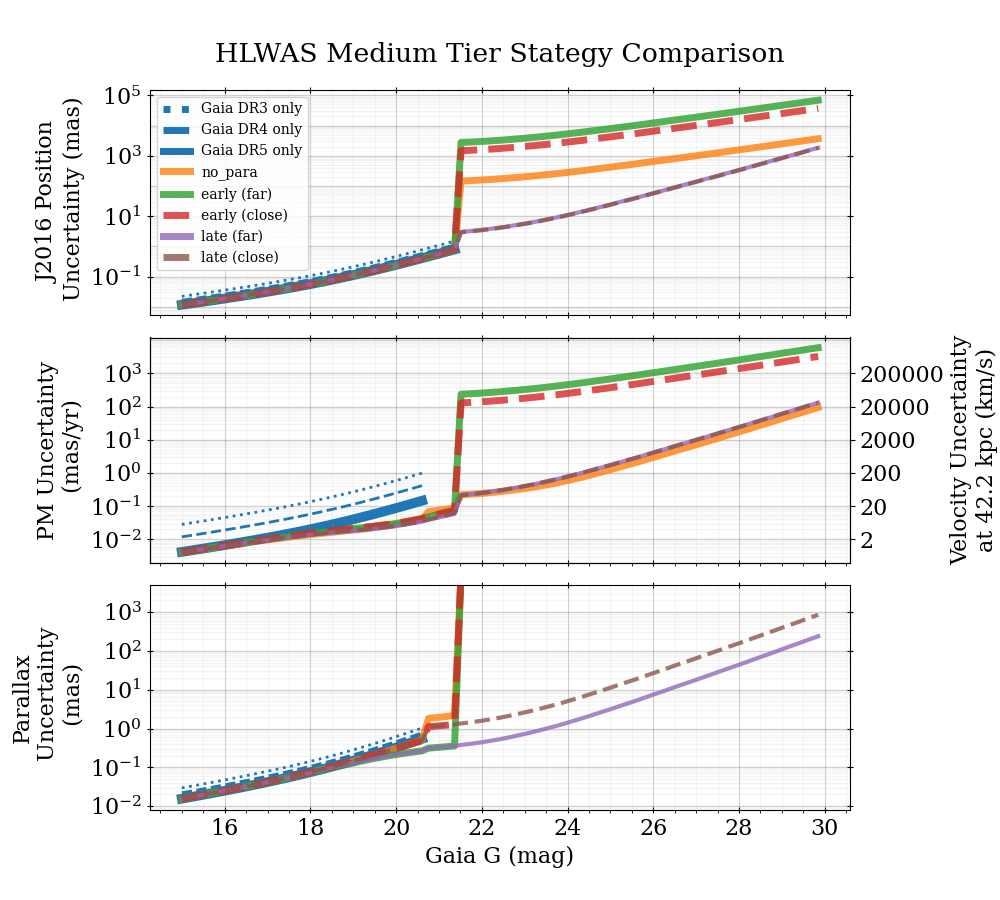}
    \caption{Expected astrometric precision for the HLWAS Medium tier as a function of magnitude when pointing in the direction of Medium Field 1, $(\alpha,\delta) = (150^\circ,-5^\circ)$. Tangential velocity uncertainties are shown for a moderate halo distance, $D\approx42$~kpc. Parallax uncertainties off of the scale of the bottom panel are effectively unconstrained. The ``no\_para'' line corresponds to observing with a 5 year time baseline at the same phase in the year. ``(far)'' versus ``(close)'' labels in the legend refer to differences in the time of the first epoch defined by parallax effects, as described in the text. ``early''-labeled lines use only 2 \rst\, epochs with a 6 month time baseline near the beginning of the survey, while ``late'' lines correspond to spacing 6 epochs over 4.5 years. Evidently, first epoch timing has important effects for the final survey astrometry. As expected, increasing the time baseline leads to more precise PMs and parallaxes.}
    \label{fig:HLWAS_medium_astrometry_errs}
\end{figure*}

The GBTDS will include $>54,000$ epochs spanning 10 ``seasons'' ($\sim72$~days each, every 6 months) over 5 years, with the first epoch taken near March 2027. The selected filters, exposure times, dithers, and epochs are summarized in Table~\ref{tab:gbtds_cadence}, as taken from the Roman Community Defined Survey webpage. The first three and last three GBTDS seasons will operate with the ``high'' cadence scheme, while the middle four seasons will operate in the ``low'' cadence scheme. For the astrometry simulations, we set the start of the observations to be half a season (36 days) before J2027.25 (approximately March 2027). The midpoints of the subsequent seasons are set to be exactly 6 months after the previous one. We set the pointing coordinate to $(l,b) = (0.5^\circ,-1.4^\circ)$, near the Galactic Center.

The resulting astrometric predictions are shown in Figure~\ref{fig:GBTDS_astrometry_errs}, with tangential velocities estimated near the Galactic center. With an extremely large number of repeat visits over 5 years, the GBTDS can expect to measure the highest precision of all the surveys: even the \rst\, alone PMs and parallaxes are expected to be better than \gaia\, alone for $G>16$~mag. Our results anticipate that the GBTDS will be able to measure \gaia\, DR3 quality parallaxes and PMs down to $G=28$~mag and $G=29$~mag, respectively. Of course, these findings assume that crowding in the Galactic Bulge regions does not impact the \rst\, centroiding too much. We have also added profiles showing the improvement gained by calibrating the systematic floor down to 0.1\% (red line) and 0.01\% (purple line) of a \rst\, pixel using the repeat GBTDS visits. In these scenarios, the most significant astrometric improvement is seen by the brightest/near-saturation sources that are most affected by the systematic floor.

As with the HLTDS, we can turn the parallax uncertainties into fractional distance uncertainties, and this is shown in Figure~\ref{fig:GBTDS_distance_frac_errs}. Now, we find that the GBTDS may be able to deliver $<10\%$ distance uncertainties -- comparable to expectations from spectro-photometric distances -- for stars within 10~kpc out to $22.5$~mag. Combined with incredible PMs, these parralaxes will make separation of stars into foreground versus Galactic Center components much easier for a vastly increased number of stars. 

\begin{table}[h]
    \centering
    \begin{tabular}{|cccccc|}
        \hline
        Season  & Filter  &   Exposure    & Dithers &  Epochs & Cadence \\ 
         Type &   &   Time (sec)    &  &   &  \\ \hline
High & F146	& 66 & 1 & 8390	& 12.1~m	\\
& F087	& 66 & 1 & 282	&     6~h \\
& F213	& 66 & 1 & 282	&     6~h	\\
& F184	& 284 &    4 & 3	   & $\sim~35$~d     \\
& F106	& 135  &   2 & 3	   & $\sim~35$~d     \\
& F129	& 85 & 2 & 3	   & $\sim~35$~d 	   \\
& F158	& 60 & 2 & 3	   & $\sim~35$~d   \\
& F062	& 60 & 2 & 3	   & $\sim~35$~d   \\
\hline
Low & F146	& 66 & 1 & 14	& 5~d	\\
& F087	& 66 & 1 & 14	&     5~d \\
& F213	& 66 & 1 & 14	&     5~d \\
& F184	& 284 &    4 & 3	   & $\sim~35$~d     \\
& F106	& 135  &   2 & 3	   & $\sim~35$~d     \\
& F129	& 85 & 2 & 3	   & $\sim~35$~d	   \\
& F158	& 60 & 2 & 3	   & $\sim~35$~d  \\
& F062	& 60 & 2 & 3	   & $\sim~35$~d	  \\
        \hline
            \end{tabular}
    \caption{GBTDS cadence and number of epochs for each filter used in the astrometry simulations. m, h, and d in the Cadence column correspond to minutes, hours, and days, respectively. The program is broken up into High and Low cadence seasons, spanning 10 seasons over 5 years. In total, there are 54252 unique exposures.}
    \label{tab:gbtds_cadence}
\end{table}

\subsection{High Latitude Wide Area Survey (HLWAS)} \label{sec:hlwas}

The exact timing of the HLWAS is not yet defined, so we explore a few different possible options here for the Medium tier. Based on the public Community Survey webpage, the observing plan will use 2 epochs each for the F158, F106, and F129 filters. Depending on the time baselines, this could correspond to as few as 2 epochs (observe all three filters back-to-back) and as many as 6 (different epochs for all filters). The proposed strategy is set to use 3 dithers per exposure and $\sim 107$~second exposure times for all filters. 

We set the pointing coordinate to  $\alpha=+150^\circ$ and $\delta=-5^\circ$, corresponding to the approximate center of the HLWAS Medium Field 1. Referring to the blue line parallax offset curve in Figure~\ref{fig:parallax_example} for this pointing, we find that the times that maximize the parallax offset occur $\sim 39.6\%$ and $\sim 89.6\%$ of the way through the Julian year. Conversely, the parallax offset minima occur $\sim 14.5\%$ and $\sim 64.5\%$ of the way through the year. Based on roll angle considerations for the grism slitless spectroscopic component of the HLWAS, it is expected that at least one followup epoch will occur 6 months out of phase with the first epoch. This means that parallax effects are guaranteed to show up in the observed position changes between the first and second epoch. As a result, only having 2 epochs means that any position change is a degenerate combination of parallax and PM, so we expect to measure large parallax and PM uncertainties in that case. Of course, having access to prior distance/parallax information could also be used to break this degeneracy, such as constraints from spectroscopy, or literature distances with membership selection derived from position in color-magnitude diagrams. For some science cases in nearby galaxies or the distant MW halo, it is fairly reasonable to force the parallaxes to be small during astrometric fits. However, in many other cases it is prudent to avoid potential biases from incorrect prior expectations (see Section~\ref{sec:pop_prior_caveats} for a discussion on the impact of population priors). 

Here, we explore the impact of the first epoch timing on the final astrometric precision by focusing on the HLWAS Medium survey using different epoch configurations. The first configuration is the ``no\_para'' case, which corresponds to observing F106, F129, and F158 at J2027.396 (i.e. starting at the time of maximum parallax offset) with a followup set of the same filters exactly 5 years later; in this way, the \rst-only data (i.e. $G>21.5$~mag stars) will have no parallax information, but we maximize the parallax information for $G<21.5$~mag. The next case is ``early (far)'' which again corresponds to starting the first epoch at J2027.396 -- the ``far'' part of the parallax ellipse -- with the second epoch occurring 0.5 years after the first. Conversely, the ``early (close)'' case begins at J2027.145 -- the time of the minimum of the parallax ellipse -- with the second epoch again 0.5 years later. These ``early'' cases correspond to one possible approach which observes all of the HLWAS pointings early in \rst's lifetime. Finally, we add ``late'' cases for both ``close'' and ``far'', which correspond to spacing the 6 Medium tier epochs over $\sim5$ years: the timing of these observations are set to $\mathrm{\texttt{first\_epoch}} + [0,0.5,3,3.5,4,4.5]$~years such that there is a 6 month phase offset between subsequent observations to alternate between the two minima/maxima of the parallax ellipse.  The results of these tests are shown in Figure~\ref{fig:HLWAS_medium_astrometry_errs}, with tangential velocity uncertainties estimated at a moderate halo distance of $\sim 42$~kpc.

While the ``no\_para'' case is likely not to be the chosen strategy -- telescope roll angle requirements suggest we require 6 month of phase between different epochs -- these results put a useful lower bound on the PM uncertainties we could reach for $G>21.5$~mag stars. We find the expected result that observing the \rst\, epochs with no phase offset leads to large final parallax uncertainties. We are still able to learn some parallax information in the $20.7<G<21.5$~mag range though, because the \gaia-measured positions provide a third epoch of position data at J2016.0. For $G>21.5$~mag, the ``no\_para'' strategy is able to measure the most-precise PMs because all position changes in the data are the result of PM alone. 

Comparing the ``early'' cases, we see that the ``close'' strategy is able to measure slightly more precise PMs than the ``far'' strategy. That is, given only 2 epochs observed with a 6 month time baseline, it is best to take those measurement at the time that minimizes parallax effects. However, the $20.7<G<21.5$~mag parallaxes of ``close'' suffer slightly compared to the ``far'' strategy. Here, we point out that there is likely legacy value in choosing the ``far'' strategy over the ``close'' one: subsequent observations -- either with the same or different telescopes -- will benefit from the maximized parallax information contained in the ``far'' strategy, leading to more precise PMs and parallaxes if a third epoch is taken later. In any case, the short \rst\, time baseline of either ``early'' strategy leads to PM uncertainties for faint stars that are very large ($>100$~mas/yr). These PMs would be effectively useless for many science cases, such as in kinematic studies of substructure in the Milky Way stellar halo or in nearby dwarf galaxies.

Instead, the ``late'' configurations provide much more precise astrometry for the exact same number of observations, benefiting from a longer total \rst\, time baseline. Now, PM precisions are almost as good as the ``no\_para'' case, and the parallaxes are also quite well-constrained. Here, we see that the ``far'' strategy is now optimal because it maximizes parallax information to provide significantly better parallax uncertainties compared to the corresponding ``close'' strategy. Vitally, we find that these ``late (far)'' PM and parallax uncertainties are more than 1000 times smaller than the ``early'' configurations. Comparing to the \gaia\, DR3 $G=20.7$~mag precisions, we find that the ``late (far)'' strategy is able to measure \gaia-like PM uncertainties out to $\sim24$~mag and parallax uncertainties out to $\sim22.5$~mag. While some subfields of astronomy would be able to do their science sooner if HLWAS collected all of its observations early using short \rst\, baselines,  our calculations suggest that this would come at a severe cost to PM science. We recommend that the observation strategy teams take this information into account when setting the final HLWAS cadence.

\section{Conclusions} \label{sec:conclusions}

We have created a realistic simulation tool that combines \gaia\, with \rst\, to predict uncertainties in position, parallax, and PM as a function of magnitude given a user-provided \rst\, observing strategy (Sections \ref{sec:math} \& \ref{sec:methods}, Figure \ref{fig:example_gaia+roman_astrometry}). \kam{The statistics at the heart of our tool can easily be generalized to include astrometric positions from other surveys and datasets.} This tool is actively being used by multiple groups writing \rst\, proposals to ensure that various astrometric thresholds for scientific goals are being met. \href{https://github.com/KevinMcK95/gaia_roman_astrometry}{Our code is publicly available} with additional \texttt{Jupyter} notebooks providing tutorials for its use. In this work, we have simulated the core surveys of the Roman mission (Section~\ref{sec:applications}): GPS, HLTDS, GBTDS, and HLWAS. We predict that the HLTDS and GBTDS will provide high quality parallaxes and PMs to much fainter magnitudes than \gaia\, alone will ever see (Figures~\ref{fig:HLTDS_astrometry_errs} to \ref{fig:GBTDS_distance_frac_errs}). We also find that the astrometric quality of the HLWAS is extremely dependent upon the final selected observing strategy (Section~\ref{sec:hlwas} and Figure~\ref{fig:HLWAS_medium_astrometry_errs}). We show that choosing a \rst\, baseline of 6 months for HLWAS leads to almost useless PM uncertainties ($>100$~mas/yr) for $G>21.5$~mag, while collecting observations over 4.5 years results in PMs that are 1000 times more precise. It is our strongest desire that the committees in charge of the \rst\, timeline considers the results of these simulations when designing the final schedule. 

Future work could explore techniques to make the \rst\, position uncertainty measurements even more realistic. This includes capturing how the position uncertainties are expected to change across the face of the detector (provided by \texttt{stpsf}), attempting to model chromatic point spread functions for wide filters, as well as generating realistic scenes to understand the impact of crowding. These realistic scenes could also be used to characterize how well an \rst\, image can be aligned to the \gaia\, reference frame as a function of the number and brightness of stars in that image. 

As demonstrated in Section~\ref{sec:example_app}, this tool can also be used to estimate the astrometry from combining \gaia+\rst\, with archival \hst\, images (Figure~\ref{fig:example_gaia+roman+hst_astrometry}). It would be worthwhile for future efforts to build complementary position uncertainty versus magnitude functions (such Figure~\ref{fig:roman_pos_errs}) for additional instruments/telescopes -- \hst, \textit{JWST}, \textit{Rubin}, and \textit{Euclid} -- to predict the total astrometry that can be leveraged from all available observations of a target. 

\begin{acknowledgments}

\kam{The authors thank the anonymous referee for comments that helped improve the clarity of this paper.} K.A.M. acknowledges supports from the University of Toronto’s Eric and Wendy Schmidt AI in Science Post-doctoral Fellowship, a program of Schmidt Sciences. K.A.M. thanks Eddie Schlafly and Robyn Sanderson for helpful conversations that improved the quality of the simulation tool. K.A.M. also thanks the teams that created and support \rst\, optical path simulation tools that make this work possible. 

\end{acknowledgments}

This work was carried out in collaboration with the  HSTPROMO (High-resolution Space Telescope PROper MOtion) Collaboration\footnote{\url{https://www.stsci.edu/~marel/hstpromo.html}}, a set of projects aimed at improving our dynamical understanding of stars, clusters, and galaxies in the nearby Universe through measurement and interpretation of proper motions from \hst\,, \gaia\,, \jwst\,, \rst\,, and other space observatories. We thank the collaboration members for sharing their ideas and software.

\vspace{5mm}
\facilities{\rst, \gaia}

\software{\texttt{astropy} \citep{astropy_2013,astropy_2018,astropy_2022}, \texttt{BP3M} \citep{McKinnon_2024}, \texttt{GaiaHub} \citep{delPino_2022}, \texttt{jupyter} \citep{jupyter_citation}, \texttt{matplotlib}  \citep{matplotlib_citation}, \texttt{numpy} \citep{numpy_citation}, \texttt{pandas} \citep{pandas_citation_2010,pandas_citation_2020}, \texttt{pandeia} \citep{Pontoppidan_2016}, \texttt{scipy} \citep{scipy_citation}, \texttt{stpsf} \citep{Perrin_2012,Perrin_2025}}

\bibliography{main}{}

@ARTICLE{numpy_citation,
 title         = {Array programming with {NumPy}},
 author        = {Charles R. Harris and K. Jarrod Millman and St{\'{e}}fan J.
                 van der Walt and Ralf Gommers and Pauli Virtanen and David
                 Cournapeau and Eric Wieser and Julian Taylor and Sebastian
                 Berg and Nathaniel J. Smith and Robert Kern and Matti Picus
                 and Stephan Hoyer and Marten H. van Kerkwijk and Matthew
                 Brett and Allan Haldane and Jaime Fern{\'{a}}ndez del
                 R{\'{i}}o and Mark Wiebe and Pearu Peterson and Pierre
                 G{\'{e}}rard-Marchant and Kevin Sheppard and Tyler Reddy and
                 Warren Weckesser and Hameer Abbasi and Christoph Gohlke and
                 Travis E. Oliphant},
 year          = {2020},
 month         = sep,
 journal       = {Nature},
 volume        = {585},
 number        = {7825},
 pages         = {357--362},
 doi           = {10.1038/s41586-020-2649-2},
 publisher     = {Springer Science and Business Media {LLC}},
 url           = {https://doi.org/10.1038/s41586-020-2649-2}
}

@ARTICLE{astropy_2013,
Adsurl = {http://adsabs.harvard.edu/abs/2013A\%26A...558A..33A},
Archiveprefix = {arXiv},
Author = {{Astropy Collaboration} and {Robitaille}, T.~P. and {Tollerud}, E.~J. and {Greenfield}, P. and {Droettboom}, M. and {Bray}, E. and {Aldcroft}, T. and {Davis}, M. and {Ginsburg}, A. and {Price-Whelan}, A.~M. and {Kerzendorf}, W.~E. and {Conley}, A. and {Crighton}, N. and {Barbary}, K. and {Muna}, D. and {Ferguson}, H. and {Grollier}, F. and {Parikh}, M.~M. and {Nair}, P.~H. and {Unther}, H.~M. and {Deil}, C. and {Woillez}, J. and {Conseil}, S. and {Kramer}, R. and {Turner}, J.~E.~H. and {Singer}, L. and {Fox}, R. and {Weaver}, B.~A. and {Zabalza}, V. and {Edwards}, Z.~I. and {Azalee Bostroem}, K. and {Burke}, D.~J. and {Casey}, A.~R. and {Crawford}, S.~M. and {Dencheva}, N. and {Ely}, J. and {Jenness}, T. and {Labrie}, K. and {Lim}, P.~L. and {Pierfederici}, F. and {Pontzen}, A. and {Ptak}, A. and {Refsdal}, B. and {Servillat}, M. and {Streicher}, O.},
Doi = {10.1051/0004-6361/201322068},
Eid = {A33},
Eprint = {1307.6212},
Journal = {\aap},
Month = oct,
Pages = {A33},
Primaryclass = {astro-ph.IM},
Title = {{Astropy: A community Python package for astronomy}},
Volume = 558,
Year = 2013}

@ARTICLE{astropy_2018,
       author = {{Astropy Collaboration} and {Price-Whelan}, A.~M. and
         {Sip{\H{o}}cz}, B.~M. and {G{\"u}nther}, H.~M. and {Lim}, P.~L. and
         {Crawford}, S.~M. and {Conseil}, S. and {Shupe}, D.~L. and
         {Craig}, M.~W. and {Dencheva}, N. and {Ginsburg}, A. and {Vand
        erPlas}, J.~T. and {Bradley}, L.~D. and {P{\'e}rez-Su{\'a}rez}, D. and
         {de Val-Borro}, M. and {Aldcroft}, T.~L. and {Cruz}, K.~L. and
         {Robitaille}, T.~P. and {Tollerud}, E.~J. and {Ardelean}, C. and
         {Babej}, T. and {Bach}, Y.~P. and {Bachetti}, M. and {Bakanov}, A.~V. and
         {Bamford}, S.~P. and {Barentsen}, G. and {Barmby}, P. and
         {Baumbach}, A. and {Berry}, K.~L. and {Biscani}, F. and {Boquien}, M. and
         {Bostroem}, K.~A. and {Bouma}, L.~G. and {Brammer}, G.~B. and
         {Bray}, E.~M. and {Breytenbach}, H. and {Buddelmeijer}, H. and
         {Burke}, D.~J. and {Calderone}, G. and {Cano Rodr{\'\i}guez}, J.~L. and
         {Cara}, M. and {Cardoso}, J.~V.~M. and {Cheedella}, S. and {Copin}, Y. and
         {Corrales}, L. and {Crichton}, D. and {D'Avella}, D. and {Deil}, C. and
         {Depagne}, {\'E}. and {Dietrich}, J.~P. and {Donath}, A. and
         {Droettboom}, M. and {Earl}, N. and {Erben}, T. and {Fabbro}, S. and
         {Ferreira}, L.~A. and {Finethy}, T. and {Fox}, R.~T. and
         {Garrison}, L.~H. and {Gibbons}, S.~L.~J. and {Goldstein}, D.~A. and
         {Gommers}, R. and {Greco}, J.~P. and {Greenfield}, P. and
         {Groener}, A.~M. and {Grollier}, F. and {Hagen}, A. and {Hirst}, P. and
         {Homeier}, D. and {Horton}, A.~J. and {Hosseinzadeh}, G. and {Hu}, L. and
         {Hunkeler}, J.~S. and {Ivezi{\'c}}, {\v{Z}}. and {Jain}, A. and
         {Jenness}, T. and {Kanarek}, G. and {Kendrew}, S. and {Kern}, N.~S. and
         {Kerzendorf}, W.~E. and {Khvalko}, A. and {King}, J. and {Kirkby}, D. and
         {Kulkarni}, A.~M. and {Kumar}, A. and {Lee}, A. and {Lenz}, D. and
         {Littlefair}, S.~P. and {Ma}, Z. and {Macleod}, D.~M. and
         {Mastropietro}, M. and {McCully}, C. and {Montagnac}, S. and
         {Morris}, B.~M. and {Mueller}, M. and {Mumford}, S.~J. and {Muna}, D. and
         {Murphy}, N.~A. and {Nelson}, S. and {Nguyen}, G.~H. and
         {Ninan}, J.~P. and {N{\"o}the}, M. and {Ogaz}, S. and {Oh}, S. and
         {Parejko}, J.~K. and {Parley}, N. and {Pascual}, S. and {Patil}, R. and
         {Patil}, A.~A. and {Plunkett}, A.~L. and {Prochaska}, J.~X. and
         {Rastogi}, T. and {Reddy Janga}, V. and {Sabater}, J. and
         {Sakurikar}, P. and {Seifert}, M. and {Sherbert}, L.~E. and
         {Sherwood-Taylor}, H. and {Shih}, A.~Y. and {Sick}, J. and
         {Silbiger}, M.~T. and {Singanamalla}, S. and {Singer}, L.~P. and
         {Sladen}, P.~H. and {Sooley}, K.~A. and {Sornarajah}, S. and
         {Streicher}, O. and {Teuben}, P. and {Thomas}, S.~W. and
         {Tremblay}, G.~R. and {Turner}, J.~E.~H. and {Terr{\'o}n}, V. and
         {van Kerkwijk}, M.~H. and {de la Vega}, A. and {Watkins}, L.~L. and
         {Weaver}, B.~A. and {Whitmore}, J.~B. and {Woillez}, J. and
         {Zabalza}, V. and {Astropy Contributors}},
        title = "{The Astropy Project: Building an Open-science Project and Status of the v2.0 Core Package}",
      journal = {\aj},
         year = 2018,
        month = sep,
       volume = {156},
       number = {3},
          eid = {123},
        pages = {123},
          doi = {10.3847/1538-3881/aabc4f},
archivePrefix = {arXiv},
       eprint = {1801.02634},
 primaryClass = {astro-ph.IM},
       adsurl = {https://ui.adsabs.harvard.edu/abs/2018AJ....156..123A}
}

@ARTICLE{astropy_2022,
       author = {{Astropy Collaboration} and {Price-Whelan}, Adrian M. and {Lim}, Pey Lian and {Earl}, Nicholas and {Starkman}, Nathaniel and {Bradley}, Larry and {Shupe}, David L. and {Patil}, Aarya A. and {Corrales}, Lia and {Brasseur}, C.~E. and {N{"o}the}, Maximilian and {Donath}, Axel and {Tollerud}, Erik and {Morris}, Brett M. and {Ginsburg}, Adam and {Vaher}, Eero and {Weaver}, Benjamin A. and {Tocknell}, James and {Jamieson}, William and {van Kerkwijk}, Marten H. and {Robitaille}, Thomas P. and {Merry}, Bruce and {Bachetti}, Matteo and {G{"u}nther}, H. Moritz and {Aldcroft}, Thomas L. and {Alvarado-Montes}, Jaime A. and {Archibald}, Anne M. and {B{'o}di}, Attila and {Bapat}, Shreyas and {Barentsen}, Geert and {Baz{'a}n}, Juanjo and {Biswas}, Manish and {Boquien}, M{'e}d{'e}ric and {Burke}, D.~J. and {Cara}, Daria and {Cara}, Mihai and {Conroy}, Kyle E. and {Conseil}, Simon and {Craig}, Matthew W. and {Cross}, Robert M. and {Cruz}, Kelle L. and {D'Eugenio}, Francesco and {Dencheva}, Nadia and {Devillepoix}, Hadrien A.~R. and {Dietrich}, J{"o}rg P. and {Eigenbrot}, Arthur Davis and {Erben}, Thomas and {Ferreira}, Leonardo and {Foreman-Mackey}, Daniel and {Fox}, Ryan and {Freij}, Nabil and {Garg}, Suyog and {Geda}, Robel and {Glattly}, Lauren and {Gondhalekar}, Yash and {Gordon}, Karl D. and {Grant}, David and {Greenfield}, Perry and {Groener}, Austen M. and {Guest}, Steve and {Gurovich}, Sebastian and {Handberg}, Rasmus and {Hart}, Akeem and {Hatfield-Dodds}, Zac and {Homeier}, Derek and {Hosseinzadeh}, Griffin and {Jenness}, Tim and {Jones}, Craig K. and {Joseph}, Prajwel and {Kalmbach}, J. Bryce and {Karamehmetoglu}, Emir and {Ka{l}uszy{'n}ski}, Miko{l}aj and {Kelley}, Michael S.~P. and {Kern}, Nicholas and {Kerzendorf}, Wolfgang E. and {Koch}, Eric W. and {Kulumani}, Shankar and {Lee}, Antony and {Ly}, Chun and {Ma}, Zhiyuan and {MacBride}, Conor and {Maljaars}, Jakob M. and {Muna}, Demitri and {Murphy}, N.~A. and {Norman}, Henrik and {O'Steen}, Richard and {Oman}, Kyle A. and {Pacifici}, Camilla and {Pascual}, Sergio and {Pascual-Granado}, J. and {Patil}, Rohit R. and {Perren}, Gabriel I. and {Pickering}, Timothy E. and {Rastogi}, Tanuj and {Roulston}, Benjamin R. and {Ryan}, Daniel F. and {Rykoff}, Eli S. and {Sabater}, Jose and {Sakurikar}, Parikshit and {Salgado}, Jes{'u}s and {Sanghi}, Aniket and {Saunders}, Nicholas and {Savchenko}, Volodymyr and {Schwardt}, Ludwig and {Seifert-Eckert}, Michael and {Shih}, Albert Y. and {Jain}, Anany Shrey and {Shukla}, Gyanendra and {Sick}, Jonathan and {Simpson}, Chris and {Singanamalla}, Sudheesh and {Singer}, Leo P. and {Singhal}, Jaladh and {Sinha}, Manodeep and {Sip{H{o}}cz}, Brigitta M. and {Spitler}, Lee R. and {Stansby}, David and {Streicher}, Ole and {{{S}}umak}, Jani and {Swinbank}, John D. and {Taranu}, Dan S. and {Tewary}, Nikita and {Tremblay}, Grant R. and {Val-Borro}, Miguel de and {Van Kooten}, Samuel J. and {Vasovi{'c}}, Zlatan and {Verma}, Shresth and {de Miranda Cardoso}, Jos{'e} Vin{'i}cius and {Williams}, Peter K.~G. and {Wilson}, Tom J. and {Winkel}, Benjamin and {Wood-Vasey}, W.~M. and {Xue}, Rui and {Yoachim}, Peter and {Zhang}, Chen and {Zonca}, Andrea and {Astropy Project Contributors}},
        title = "{The Astropy Project: Sustaining and Growing a Community-oriented Open-source Project and the Latest Major Release (v5.0) of the Core Package}",
      journal = {apj},
         year = 2022,
        month = aug,
       volume = {935},
       number = {2},
          eid = {167},
        pages = {167},
          doi = {10.3847/1538-4357/ac7c74},
archivePrefix = {arXiv},
       eprint = {2206.14220},
 primaryClass = {astro-ph.IM},
       adsurl = {https://ui.adsabs.harvard.edu/abs/2022ApJ...935..167A}
}

@ARTICLE{scipy_citation,
  author  = {Virtanen, Pauli and Gommers, Ralf and Oliphant, Travis E. and
            Haberland, Matt and Reddy, Tyler and Cournapeau, David and
            Burovski, Evgeni and Peterson, Pearu and Weckesser, Warren and
            Bright, Jonathan and {van der Walt}, St{\'e}fan J. and
            Brett, Matthew and Wilson, Joshua and Millman, K. Jarrod and
            Mayorov, Nikolay and Nelson, Andrew R. J. and Jones, Eric and
            Kern, Robert and Larson, Eric and Carey, C J and
            Polat, {\.I}lhan and Feng, Yu and Moore, Eric W. and
            {VanderPlas}, Jake and Laxalde, Denis and Perktold, Josef and
            Cimrman, Robert and Henriksen, Ian and Quintero, E. A. and
            Harris, Charles R. and Archibald, Anne M. and
            Ribeiro, Ant{\^o}nio H. and Pedregosa, Fabian and
            {van Mulbregt}, Paul and {SciPy 1.0 Contributors}},
  title   = {{{SciPy} 1.0: Fundamental Algorithms for Scientific
            Computing in Python}},
  journal = {Nature Methods},
  year    = {2020},
  volume  = {17},
  pages   = {261--272},
  adsurl  = {https://rdcu.be/b08Wh},
  doi     = {10.1038/s41592-019-0686-2},
}

@Article{matplotlib_citation,
  Author    = {Hunter, J. D.},
  Title     = {Matplotlib: A 2D graphics environment},
  Journal   = {Computing in Science \& Engineering},
  Volume    = {9},
  Number    = {3},
  Pages     = {90--95},
  publisher = {IEEE COMPUTER SOC},
  doi       = {10.1109/MCSE.2007.55},
  year      = 2007
}

@conference{jupyter_citation,
	Author = {Thomas Kluyver and Benjamin Ragan-Kelley and Fernando P{\'e}rez and Brian Granger and Matthias Bussonnier and Jonathan Frederic and Kyle Kelley and Jessica Hamrick and Jason Grout and Sylvain Corlay and Paul Ivanov and Dami{\'a}n Avila and Safia Abdalla and Carol Willing},
	Booktitle = {Positioning and Power in Academic Publishing: Players, Agents and Agendas},
	Editor = {F. Loizides and B. Schmidt},
	Organization = {IOS Press},
	Pages = {87 - 90},
	Title = {Jupyter Notebooks -- a publishing format for reproducible computational workflows},
	Year = {2016}}

@software{pandas_citation_2020,
    author       = {The pandas development team},
    title        = {pandas-dev/pandas: Pandas},
    month        = feb,
    year         = 2020,
    publisher    = {Zenodo},
    version      = {latest},
    doi          = {10.5281/zenodo.3509134},
    url          = {https://doi.org/10.5281/zenodo.3509134}
}

@InProceedings{pandas_citation_2010,
  author    = { {W}es {M}c{K}inney },
  title     = { {D}ata {S}tructures for {S}tatistical {C}omputing in {P}ython },
  booktitle = { {P}roceedings of the 9th {P}ython in {S}cience {C}onference },
  pages     = { 56 - 61 },
  year      = { 2010 },
  editor    = { {S}t\'efan van der {W}alt and {J}arrod {M}illman },
  doi       = { 10.25080/Majora-92bf1922-00a }
}

@ARTICLE{delPino_2022,
       author = {{del Pino}, Andr{\'e}s and {Libralato}, Mattia and {van der Marel}, Roeland P. and {Bennet}, Paul and {Fardal}, Mark A. and {Anderson}, Jay and {Bellini}, Andrea and {Tony Sohn}, Sangmo and {Watkins}, Laura L.},
        title = "{GaiaHub: A Method for Combining Data from the Gaia and Hubble Space Telescopes to Derive Improved Proper Motions for Faint Stars}",
      journal = {\apj},
         year = 2022,
        month = jul,
       volume = {933},
       number = {1},
          eid = {76},
        pages = {76},
          doi = {10.3847/1538-4357/ac70cf},
archivePrefix = {arXiv},
       eprint = {2205.08009},
 primaryClass = {astro-ph.GA},
       adsurl = {https://ui.adsabs.harvard.edu/abs/2022ApJ...933...76D}
}

@ARTICLE{Cunningham_2019b,
       author = {{Cunningham}, Emily C. and {Deason}, Alis J. and {Sanderson}, Robyn E. and {Sohn}, Sangmo Tony and {Anderson}, Jay and {Guhathakurta}, Puragra and {Rockosi}, Constance M. and {van der Marel}, Roeland P. and {Loebman}, Sarah R. and {Wetzel}, Andrew},
        title = "{HALO7D II: The Halo Velocity Ellipsoid and Velocity Anisotropy with Distant Main-sequence Stars}",
      journal = {\apj},
         year = 2019,
        month = jul,
       volume = {879},
       number = {2},
          eid = {120},
        pages = {120},
          doi = {10.3847/1538-4357/ab24cd},
archivePrefix = {arXiv},
       eprint = {1810.12201},
 primaryClass = {astro-ph.GA},
       adsurl = {https://ui.adsabs.harvard.edu/abs/2019ApJ...879..120C}
}

@ARTICLE{McKinnon_2024,
       author = {{McKinnon}, Kevin A. and {del Pino}, Andr{\'e}s and {Rockosi}, Constance M. and {Apfel}, Miranda and {Guhathakurta}, Puragra and {van der Marel}, Roeland P. and {Bennet}, Paul and {Fardal}, Mark A. and {Libralato}, Mattia and {Sohn}, Sangmo Tony and {Vitral}, Eduardo and {Watkins}, Laura L.},
        title = "{BP3M: Bayesian Positions, Parallaxes, and Proper Motions Derived from the Hubble Space Telescope and Gaia Data}",
      journal = {\apj},
         year = 2024,
        month = sep,
       volume = {972},
       number = {2},
          eid = {150},
        pages = {150},
          doi = {10.3847/1538-4357/ad5834},
archivePrefix = {arXiv},
       eprint = {2310.20099},
 primaryClass = {astro-ph.GA},
       adsurl = {https://ui.adsabs.harvard.edu/abs/2024ApJ...972..150M}
}

@ARTICLE{Bailer-Jones_2023,
       author = {{Bailer-Jones}, C.~A.~L.},
        title = "{Estimating Distances from Parallaxes. VI. A Method for Inferring Distances and Transverse Velocities from Parallaxes and Proper Motions Demonstrated on Gaia Data Release 3}",
      journal = {\aj},
         year = 2023,
        month = dec,
       volume = {166},
       number = {6},
          eid = {269},
        pages = {269},
          doi = {10.3847/1538-3881/ad08bb},
archivePrefix = {arXiv},
       eprint = {2311.00374},
 primaryClass = {astro-ph.GA},
       adsurl = {https://ui.adsabs.harvard.edu/abs/2023AJ....166..269B}
}

@ARTICLE{Roman_ROTAC_2025,
       author = {{Observations Time Allocation Committee}, Roman and {Community Survey Definition Committees}, Core},
        title = "{Roman Observations Time Allocation Committee: Final Report and Recommendations}",
      journal = {arXiv e-prints},
         year = 2025,
        month = may,
          eid = {arXiv:2505.10574},
        pages = {arXiv:2505.10574},
          doi = {10.48550/arXiv.2505.10574},
archivePrefix = {arXiv},
       eprint = {2505.10574},
 primaryClass = {astro-ph.IM},
       adsurl = {https://ui.adsabs.harvard.edu/abs/2025arXiv250510574O}
}

@ARTICLE{Bellini_2011,
       author = {{Bellini}, A. and {Anderson}, J. and {Bedin}, L.~R.},
        title = "{Astrometry and Photometry with HST WFC3. II. Improved Geometric-Distortion Corrections for 10 Filters of the UVIS Channel}",
      journal = {\pasp},
         year = 2011,
        month = may,
       volume = {123},
       number = {903},
        pages = {622},
          doi = {10.1086/659878},
archivePrefix = {arXiv},
       eprint = {1102.5218},
 primaryClass = {astro-ph.IM},
       adsurl = {https://ui.adsabs.harvard.edu/abs/2011PASP..123..622B}
}

@ARTICLE{Bellini_2009,
       author = {{Bellini}, A. and {Bedin}, L.~R.},
        title = "{Astrometry and Photometry with HST WFC3. I. Geometric Distortion Corrections of F225W, F275W, F336W Bands of the UVIS Channel}",
      journal = {\pasp},
         year = 2009,
        month = dec,
       volume = {121},
       number = {886},
        pages = {1419},
          doi = {10.1086/649061},
archivePrefix = {arXiv},
       eprint = {0910.3250},
 primaryClass = {astro-ph.IM},
       adsurl = {https://ui.adsabs.harvard.edu/abs/2009PASP..121.1419B}
}

@ARTICLE{McConnachie_2020,
       author = {{McConnachie}, Alan W. and {Venn}, Kim A.},
        title = "{Revised and New Proper Motions for Confirmed and Candidate Milky Way Dwarf Galaxies}",
      journal = {\aj},
         year = 2020,
        month = sep,
       volume = {160},
       number = {3},
          eid = {124},
        pages = {124},
          doi = {10.3847/1538-3881/aba4ab},
archivePrefix = {arXiv},
       eprint = {2007.05011},
 primaryClass = {astro-ph.GA},
       adsurl = {https://ui.adsabs.harvard.edu/abs/2020AJ....160..124M}
}

@ARTICLE{Read_2021,
       author = {{Read}, J.~I. and {Mamon}, G.~A. and {Vasiliev}, E. and {Watkins}, L.~L. and {Walker}, M.~G. and {Pe{\~n}arrubia}, J. and {Wilkinson}, M. and {Dehnen}, W. and {Das}, P.},
        title = "{Breaking beta: a comparison of mass modelling methods for spherical systems}",
      journal = {\mnras},
         year = 2021,
        month = feb,
       volume = {501},
       number = {1},
        pages = {978-993},
          doi = {10.1093/mnras/staa3663},
archivePrefix = {arXiv},
       eprint = {2011.09493},
 primaryClass = {astro-ph.GA},
       adsurl = {https://ui.adsabs.harvard.edu/abs/2021MNRAS.501..978R}
}

@ARTICLE{Li_2026,
       author = {{Li}, Bowen and {McKinnon}, Kevin A. and {Saydjari}, Andrew K. and {Sayres}, Conor and {Eadie}, Gwendolyn M. and {Casey}, Andrew R. and {Holtzman}, Jon A. and {Brandt}, Timothy D. and {Fernandez-Trincado}, Jose G.},
        title = "{Optimal and Unbiased Fluxes from Up-the-Ramp Detectors under Variable Illumination}",
      journal = {arXiv e-prints},
         year = 2026,
        month = jan,
          eid = {arXiv:2601.10878},
        pages = {arXiv:2601.10878},
          doi = {10.48550/arXiv.2601.10878},
archivePrefix = {arXiv},
       eprint = {2601.10878},
 primaryClass = {astro-ph.IM},
       adsurl = {https://ui.adsabs.harvard.edu/abs/2026arXiv260110878L}
}

@ARTICLE{Massari_2017,
       author = {{Massari}, D. and {Posti}, L. and {Helmi}, A. and {Fiorentino}, G. and {Tolstoy}, E.},
        title = "{The power of teaming up HST and Gaia: the first proper motion measurement of the distant cluster NGC 2419}",
      journal = {\aap},
         year = 2017,
        month = feb,
       volume = {598},
          eid = {L9},
        pages = {L9},
          doi = {10.1051/0004-6361/201630174},
archivePrefix = {arXiv},
       eprint = {1612.00183},
 primaryClass = {astro-ph.GA},
       adsurl = {https://ui.adsabs.harvard.edu/abs/2017A&A...598L...9M}
}

@ARTICLE{Pace_2022,
       author = {{Pace}, Andrew B. and {Erkal}, Denis and {Li}, Ting S.},
        title = "{Proper Motions, Orbits, and Tidal Influences of Milky Way Dwarf Spheroidal Galaxies}",
      journal = {\apj},
         year = 2022,
        month = dec,
       volume = {940},
       number = {2},
          eid = {136},
        pages = {136},
          doi = {10.3847/1538-4357/ac997b},
archivePrefix = {arXiv},
       eprint = {2205.05699},
 primaryClass = {astro-ph.GA},
       adsurl = {https://ui.adsabs.harvard.edu/abs/2022ApJ...940..136P}
}

@ARTICLE{Ibata_2020,
       author = {{Ibata}, Rodrigo and {Thomas}, Guillaume and {Famaey}, Benoit and {Malhan}, Khyati and {Martin}, Nicolas and {Monari}, Giacomo},
        title = "{Detection of Strong Epicyclic Density Spikes in the GD-1 Stellar Stream: An Absence of Evidence for the Influence of Dark Matter Subhalos?}",
      journal = {\apj},
         year = 2020,
        month = mar,
       volume = {891},
       number = {2},
          eid = {161},
        pages = {161},
          doi = {10.3847/1538-4357/ab7303},
archivePrefix = {arXiv},
       eprint = {2002.01488},
 primaryClass = {astro-ph.GA},
       adsurl = {https://ui.adsabs.harvard.edu/abs/2020ApJ...891..161I}
}

@ARTICLE{Bullock_2005,
       author = {{Bullock}, James S. and {Johnston}, Kathryn V.},
        title = "{Tracing Galaxy Formation with Stellar Halos. I. Methods}",
      journal = {\apj},
         year = 2005,
        month = dec,
       volume = {635},
       number = {2},
        pages = {931-949},
          doi = {10.1086/497422},
archivePrefix = {arXiv},
       eprint = {astro-ph/0506467},
 primaryClass = {astro-ph},
       adsurl = {https://ui.adsabs.harvard.edu/abs/2005ApJ...635..931B}
}

@ARTICLE{Bonaca_2025,
       author = {{Bonaca}, Ana and {Price-Whelan}, Adrian M.},
        title = "{Stellar streams in the Gaia era}",
      journal = {\nar},
         year = 2025,
        month = jun,
       volume = {100},
          eid = {101713},
        pages = {101713},
          doi = {10.1016/j.newar.2024.101713},
archivePrefix = {arXiv},
       eprint = {2405.19410},
 primaryClass = {astro-ph.GA},
       adsurl = {https://ui.adsabs.harvard.edu/abs/2025NewAR.10001713B}
}

@ARTICLE{Nibauer_2025,
       author = {{Nibauer}, Jacob and {Bonaca}, Ana and {Price-Whelan}, Adrian M. and {Spergel}, David N. and {Greene}, Jenny E.},
        title = "{Measurement of Dark Matter Substructure from the Kinematics of the GD-1 Stellar Stream}",
      journal = {arXiv e-prints},
         year = 2025,
        month = oct,
          eid = {arXiv:2510.02247},
        pages = {arXiv:2510.02247},
          doi = {10.48550/arXiv.2510.02247},
archivePrefix = {arXiv},
       eprint = {2510.02247},
 primaryClass = {astro-ph.GA},
       adsurl = {https://ui.adsabs.harvard.edu/abs/2025arXiv251002247N}
}

@ARTICLE{WFIRST_Astrometry_2019,
       author = {{WFIRST Astrometry Working Group} and {Sanderson}, Robyn E. and {Bellini}, Andrea and {Casertano}, Stefano and {Lu}, Jessica R. and {Melchior}, Peter and {Libralato}, Mattia and {Bennett}, David and {Shao}, Michael and {Rhodes}, Jason and {Sohn}, Sangmo Tony and {Malhotra}, Sangeeta and {Gaudi}, Scott and {Fall}, S. Michael and {Nelan}, Ed and {Guhathakurta}, Puragra and {Anderson}, Jay and {Ho}, Shirley},
        title = "{Astrometry with the Wide-Field Infrared Space Telescope}",
      journal = {Journal of Astronomical Telescopes, Instruments, and Systems},
         year = 2019,
        month = oct,
       volume = {5},
          eid = {044005},
        pages = {044005},
          doi = {10.1117/1.JATIS.5.4.044005},
archivePrefix = {arXiv},
       eprint = {1712.05420},
 primaryClass = {astro-ph.IM},
       adsurl = {https://ui.adsabs.harvard.edu/abs/2019JATIS...5d4005W}
}

@ARTICLE{GaiaCollaboration_2024,
       author = {{Gaia Collaboration} and {Panuzzo}, P. and {Mazeh}, T. and {Arenou}, F. and {Holl}, B. and {Caffau}, E. and {Jorissen}, A. and {Babusiaux}, C. and {Gavras}, P. and {Sahlmann}, J. and {Bastian}, U. and {Wyrzykowski}, {\L}. and {Eyer}, L. and {Leclerc}, N. and {Bauchet}, N. and {Bombrun}, A. and {Mowlavi}, N. and {Seabroke}, G.~M. and {Teyssier}, D. and {Balbinot}, E. and {Helmi}, A. and {Brown}, A.~G.~A. and {Vallenari}, A. and {Prusti}, T. and {de Bruijne}, J.~H.~J. and {Barbier}, A. and {Biermann}, M. and {Creevey}, O.~L. and {Ducourant}, C. and {Evans}, D.~W. and {Guerra}, R. and {Hutton}, A. and {Jordi}, C. and {Klioner}, S.~A. and {Lammers}, U. and {Lindegren}, L. and {Luri}, X. and {Mignard}, F. and {Nicolas}, C. and {Randich}, S. and {Sartoretti}, P. and {Smiljanic}, R. and {Tanga}, P. and {Walton}, N.~A. and {Aerts}, C. and {Bailer-Jones}, C.~A.~L. and {Cropper}, M. and {Drimmel}, R. and {Jansen}, F. and {Katz}, D. and {Lattanzi}, M.~G. and {Soubiran}, C. and {Th{\'e}venin}, F. and {van Leeuwen}, F. and {Andrae}, R. and {Audard}, M. and {Bakker}, J. and {Blomme}, R. and {Casta{\~n}eda}, J. and {De Angeli}, F. and {Fabricius}, C. and {Fouesneau}, M. and {Fr{\'e}mat}, Y. and {Galluccio}, L. and {Guerrier}, A. and {Heiter}, U. and {Masana}, E. and {Messineo}, R. and {Nienartowicz}, K. and {Pailler}, F. and {Riclet}, F. and {Roux}, W. and {Sordo}, R. and {Gracia-Abril}, G. and {Portell}, J. and {Altmann}, M. and {Benson}, K. and {Berthier}, J. and {Burgess}, P.~W. and {Busonero}, D. and {Busso}, G. and {Cacciari}, C. and {C{\'a}novas}, H. and {Carrasco}, J.~M. and {Carry}, B. and {Cellino}, A. and {Cheek}, N. and {Clementini}, G. and {Damerdji}, Y. and {Davidson}, M. and {de Teodoro}, P. and {Delchambre}, L. and {Dell'Oro}, A. and {Fraile Garcia}, E. and {Garabato}, D. and {Garc{\'\i}a-Lario}, P. and {Haigron}, R. and {Hambly}, N.~C. and {Harrison}, D.~L. and {Hatzidimitriou}, D. and {Hern{\'a}ndez}, J. and {Hestroffer}, D. and {Hodgkin}, S.~T. and {Jamal}, S. and {Jevardat de Fombelle}, G. and {Jordan}, S. and {Krone-Martins}, A. and {Lanzafame}, A.~C. and {L{\"o}ffler}, W. and {Lorca}, A. and {Marchal}, O. and {Marrese}, P.~M. and {Moitinho}, A. and {Muinonen}, K. and {Nu{\~n}ez Campos}, M. and {Oreshina-Slezak}, I. and {Osborne}, P. and {Pancino}, E. and {Pauwels}, T. and {Recio-Blanco}, A. and {Riello}, M. and {Rimoldini}, L. and {Robin}, A.~C. and {Roegiers}, T. and {Sarro}, L.~M. and {Schultheis}, M. and {Smith}, M. and {Sozzetti}, A. and {Utrilla}, E. and {van Leeuwen}, M. and {Weingrill}, K. and {Abbas}, U. and {{\'A}brah{\'a}m}, P. and {Abreu Aramburu}, A. and {Ahmed}, S. and {Altavilla}, G. and {{\'A}lvarez}, M.~A. and {Anders}, F. and {Anderson}, R.~I. and {Anglada Varela}, E. and {Antoja}, T. and {Baig}, S. and {Baines}, D. and {Baker}, S.~G. and {Balaguer-N{\'u}{\~n}ez}, L. and {Balog}, Z. and {Barache}, C. and {Barros}, M. and {Barstow}, M.~A. and {Bartolom{\'e}}, S. and {Bashi}, D. and {Bassilana}, J.-L. and {Baudeau}, N. and {Becciani}, U. and {Bedin}, L.~R. and {Bellas-Velidis}, I. and {Bellazzini}, M. and {Beordo}, W. and {Bernet}, M. and {Bertolotto}, C. and {Bertone}, S. and {Bianchi}, L. and {Binnenfeld}, A. and {Blanco-Cuaresma}, S. and {Bland-Hawthorn}, J. and {Blazere}, A. and {Boch}, T. and {Bossini}, D. and {Bouquillon}, S. and {Bragaglia}, A. and {Braine}, J. and {Bratsolis}, E. and {Breedt}, E. and {Bressan}, A. and {Brouillet}, N. and {Brugaletta}, E. and {Bucciarelli}, B. and {Butkevich}, A.~G. and {Buzzi}, R. and {Camut}, A. and {Cancelliere}, R. and {Cantat-Gaudin}, T. and {Capilla Guilarte}, D. and {Carballo}, R. and {Carlucci}, T. and {Carnerero}, M.~I. and {Carretero}, J. and {Carton}, S. and {Casamiquela}, L. and {Casey}, A. and {Castellani}, M. and {Castro-Ginard}, A. and {Ceraj}, L. and {Cesare}, V. and {Charlot}, P. and {Chaudet}, C. and {Chemin}, L. and {Chiavassa}, A. and {Chornay}, N. and {Chosson}, D.},
        title = "{Discovery of a dormant 33 solar-mass black hole in pre-release Gaia astrometry}",
      journal = {\aap},
         year = 2024,
        month = jun,
       volume = {686},
          eid = {L2},
        pages = {L2},
          doi = {10.1051/0004-6361/202449763},
archivePrefix = {arXiv},
       eprint = {2404.10486},
 primaryClass = {astro-ph.GA},
       adsurl = {https://ui.adsabs.harvard.edu/abs/2024A&A...686L...2G}
}

@ARTICLE{Magnier_2020,
       author = {{Magnier}, Eugene. A. and {Schlafly}, Edward. F. and {Finkbeiner}, Douglas P. and {Tonry}, J.~L. and {Goldman}, B. and {R{\"o}ser}, S. and {Schilbach}, E. and {Casertano}, S. and {Chambers}, K.~C. and {Flewelling}, H.~A. and {Huber}, M.~E. and {Price}, P.~A. and {Sweeney}, W.~E. and {Waters}, C.~Z. and {Denneau}, L. and {Draper}, P.~W. and {Hodapp}, K.~W. and {Jedicke}, R. and {Kaiser}, N. and {Kudritzki}, R.-P. and {Metcalfe}, N. and {Stubbs}, C.~W. and {Wainscoat}, R.~J.},
        title = "{Pan-STARRS Photometric and Astrometric Calibration}",
      journal = {\apjs},
         year = 2020,
        month = nov,
       volume = {251},
       number = {1},
          eid = {6},
        pages = {6},
          doi = {10.3847/1538-4365/abb82a},
archivePrefix = {arXiv},
       eprint = {1612.05242},
 primaryClass = {astro-ph.IM},
       adsurl = {https://ui.adsabs.harvard.edu/abs/2020ApJS..251....6M}
}

@ARTICLE{Holmes_2012,
       author = {{Holmes}, Rory and {Hogg}, David W. and {Rix}, Hans-Walter},
        title = "{Designing Imaging Surveys for a Retrospective Relative Photometric Calibration}",
      journal = {\pasp},
         year = 2012,
        month = nov,
       volume = {124},
       number = {921},
        pages = {1219},
          doi = {10.1086/668656},
archivePrefix = {arXiv},
       eprint = {1203.6255},
 primaryClass = {astro-ph.IM},
       adsurl = {https://ui.adsabs.harvard.edu/abs/2012PASP..124.1219H}
}

@ARTICLE{Warfield_2025,
       author = {{Warfield}, Jack T. and {McKinnon}, Kevin A. and {Sohn}, Sangmo Tony and {Kallivayalil}, Nitya and {Savino}, Alessandro and {van der Marel}, Roeland P. and {Pace}, Andrew B. and {Garling}, Christopher T. and {Ahvazi}, Niusha and {Bennet}, Paul and {Cohen}, Roger E. and {Correnti}, Matteo and {Fardal}, Mark A. and {McQuinn}, Kristen B.~W. and {Newman}, Max J.~B. and {Vitral}, Eduardo},
        title = "{The Proper Motion of Draco II with HST using Multiple Reference Frames and Methodologies}",
      journal = {arXiv e-prints},
         year = 2025,
        month = oct,
          eid = {arXiv:2510.24849},
        pages = {arXiv:2510.24849},
          doi = {10.48550/arXiv.2510.24849},
archivePrefix = {arXiv},
       eprint = {2510.24849},
 primaryClass = {astro-ph.GA},
       adsurl = {https://ui.adsabs.harvard.edu/abs/2025arXiv251024849W}
}

@ARTICLE{Vitral_2025,
       author = {{Vitral}, Eduardo and {van der Marel}, Roeland P. and {Sohn}, Sangmo Tony and {Pe{\~n}arrubia}, Jorge and {Patel}, Ekta and {Watkins}, Laura L. and {Libralato}, Mattia and {McKinnon}, Kevin and {Bellini}, Andrea and {Bennet}, Paul},
        title = "{HSTPROMO Internal Proper Motion Kinematics of Dwarf Spheroidal Galaxies: II. Velocity Anisotropy and Dark Matter Cusp Slope of Sculptor}",
      journal = {arXiv e-prints},
         year = 2025,
        month = aug,
          eid = {arXiv:2508.20711},
        pages = {arXiv:2508.20711},
          doi = {10.48550/arXiv.2508.20711},
archivePrefix = {arXiv},
       eprint = {2508.20711},
 primaryClass = {astro-ph.GA},
       adsurl = {https://ui.adsabs.harvard.edu/abs/2025arXiv250820711V}
}

@ARTICLE{Apfel_2025,
       author = {{Apfel}, Miranda and {McKinnon}, Kevin and {Rockosi}, Constance M. and {Guhathakurta}, Puragra and {Johnston}, Kathryn V.},
        title = "{Constraining the Milky Way Halo Accretion History with Simulated Stellar Halos: Designing the HALO7D-X Survey}",
      journal = {\apj},
         year = 2025,
        month = aug,
       volume = {988},
       number = {2},
          eid = {225},
        pages = {225},
          doi = {10.3847/1538-4357/ade43d},
archivePrefix = {arXiv},
       eprint = {2507.05239},
 primaryClass = {astro-ph.GA},
       adsurl = {https://ui.adsabs.harvard.edu/abs/2025ApJ...988..225A}
}

@ARTICLE{Naidu_2020,
       author = {{Naidu}, Rohan P. and {Conroy}, Charlie and {Bonaca}, Ana and {Johnson}, Benjamin D. and {Ting}, Yuan-Sen and {Caldwell}, Nelson and {Zaritsky}, Dennis and {Cargile}, Phillip A.},
        title = "{Evidence from the H3 Survey That the Stellar Halo Is Entirely Comprised of Substructure}",
      journal = {\apj},
         year = 2020,
        month = sep,
       volume = {901},
       number = {1},
          eid = {48},
        pages = {48},
          doi = {10.3847/1538-4357/abaef4},
archivePrefix = {arXiv},
       eprint = {2006.08625},
 primaryClass = {astro-ph.GA},
       adsurl = {https://ui.adsabs.harvard.edu/abs/2020ApJ...901...48N}
}

@MISC{Anderson_2007,
        author = {{Anderson}, Jay},
        title = "{Variation of the Distortion Solution}",
 howpublished = {Instrument Science Report ACS 2007-08, 12 pages},
         year = 2007,
        month = jul,
        pages = {8},
       adsurl = {https://ui.adsabs.harvard.edu/abs/2007acs..rept....8A}
}

@ARTICLE{Gaia_2016,
       author = {{Gaia Collaboration} and {Prusti}, T. and {de Bruijne}, J.~H.~J. and {Brown}, A.~G.~A. and {Vallenari}, A. and {Babusiaux}, C. and {Bailer-Jones}, C.~A.~L. and {Bastian}, U. and {Biermann}, M. and {Evans}, D.~W. and {Eyer}, L. and {Jansen}, F. and {Jordi}, C. and {Klioner}, S.~A. and {Lammers}, U. and {Lindegren}, L. and {Luri}, X. and {Mignard}, F. and {Milligan}, D.~J. and {Panem}, C. and {Poinsignon}, V. and {Pourbaix}, D. and {Randich}, S. and {Sarri}, G. and {Sartoretti}, P. and {Siddiqui}, H.~I. and {Soubiran}, C. and {Valette}, V. and {van Leeuwen}, F. and {Walton}, N.~A. and {Aerts}, C. and {Arenou}, F. and {Cropper}, M. and {Drimmel}, R. and {H{\o}g}, E. and {Katz}, D. and {Lattanzi}, M.~G. and {O'Mullane}, W. and {Grebel}, E.~K. and {Holland}, A.~D. and {Huc}, C. and {Passot}, X. and {Bramante}, L. and {Cacciari}, C. and {Casta{\~n}eda}, J. and {Chaoul}, L. and {Cheek}, N. and {De Angeli}, F. and {Fabricius}, C. and {Guerra}, R. and {Hern{\'a}ndez}, J. and {Jean-Antoine-Piccolo}, A. and {Masana}, E. and {Messineo}, R. and {Mowlavi}, N. and {Nienartowicz}, K. and {Ord{\'o}{\~n}ez-Blanco}, D. and {Panuzzo}, P. and {Portell}, J. and {Richards}, P.~J. and {Riello}, M. and {Seabroke}, G.~M. and {Tanga}, P. and {Th{\'e}venin}, F. and {Torra}, J. and {Els}, S.~G. and {Gracia-Abril}, G. and {Comoretto}, G. and {Garcia-Reinaldos}, M. and {Lock}, T. and {Mercier}, E. and {Altmann}, M. and {Andrae}, R. and {Astraatmadja}, T.~L. and {Bellas-Velidis}, I. and {Benson}, K. and {Berthier}, J. and {Blomme}, R. and {Busso}, G. and {Carry}, B. and {Cellino}, A. and {Clementini}, G. and {Cowell}, S. and {Creevey}, O. and {Cuypers}, J. and {Davidson}, M. and {De Ridder}, J. and {de Torres}, A. and {Delchambre}, L. and {Dell'Oro}, A. and {Ducourant}, C. and {Fr{\'e}mat}, Y. and {Garc{\'\i}a-Torres}, M. and {Gosset}, E. and {Halbwachs}, J. -L. and {Hambly}, N.~C. and {Harrison}, D.~L. and {Hauser}, M. and {Hestroffer}, D. and {Hodgkin}, S.~T. and {Huckle}, H.~E. and {Hutton}, A. and {Jasniewicz}, G. and {Jordan}, S. and {Kontizas}, M. and {Korn}, A.~J. and {Lanzafame}, A.~C. and {Manteiga}, M. and {Moitinho}, A. and {Muinonen}, K. and {Osinde}, J. and {Pancino}, E. and {Pauwels}, T. and {Petit}, J. -M. and {Recio-Blanco}, A. and {Robin}, A.~C. and {Sarro}, L.~M. and {Siopis}, C. and {Smith}, M. and {Smith}, K.~W. and {Sozzetti}, A. and {Thuillot}, W. and {van Reeven}, W. and {Viala}, Y. and {Abbas}, U. and {Abreu Aramburu}, A. and {Accart}, S. and {Aguado}, J.~J. and {Allan}, P.~M. and {Allasia}, W. and {Altavilla}, G. and {{\'A}lvarez}, M.~A. and {Alves}, J. and {Anderson}, R.~I. and {Andrei}, A.~H. and {Anglada Varela}, E. and {Antiche}, E. and {Antoja}, T. and {Ant{\'o}n}, S. and {Arcay}, B. and {Atzei}, A. and {Ayache}, L. and {Bach}, N. and {Baker}, S.~G. and {Balaguer-N{\'u}{\~n}ez}, L. and {Barache}, C. and {Barata}, C. and {Barbier}, A. and {Barblan}, F. and {Baroni}, M. and {Barrado y Navascu{\'e}s}, D. and {Barros}, M. and {Barstow}, M.~A. and {Becciani}, U. and {Bellazzini}, M. and {Bellei}, G. and {Bello Garc{\'\i}a}, A. and {Belokurov}, V. and {Bendjoya}, P. and {Berihuete}, A. and {Bianchi}, L. and {Bienaym{\'e}}, O. and {Billebaud}, F. and {Blagorodnova}, N. and {Blanco-Cuaresma}, S. and {Boch}, T. and {Bombrun}, A. and {Borrachero}, R. and {Bouquillon}, S. and {Bourda}, G. and {Bouy}, H. and {Bragaglia}, A. and {Breddels}, M.~A. and {Brouillet}, N. and {Br{\"u}semeister}, T. and {Bucciarelli}, B. and {Budnik}, F. and {Burgess}, P. and {Burgon}, R. and {Burlacu}, A. and {Busonero}, D. and {Buzzi}, R. and {Caffau}, E. and {Cambras}, J. and {Campbell}, H. and {Cancelliere}, R. and {Cantat-Gaudin}, T. and {Carlucci}, T. and {Carrasco}, J.~M. and {Castellani}, M. and {Charlot}, P. and {Charnas}, J. and {Charvet}, P. and {Chassat}, F. and {Chiavassa}, A. and {Clotet}, M. and {Cocozza}, G. and {Collins}, R.~S. and {Collins}, P. and {Costigan}, G. and {Crifo}, F. and {Cross}, N.~J.~G. and {Crosta}, M. and {Crowley}, C. and {Dafonte}, C. and {Damerdji}, Y. and {Dapergolas}, A. and {David}, P. and {David}, M. and {De Cat}, P. and {de Felice}, F. and {de Laverny}, P. and {De Luise}, F. and {De March}, R. and {de Martino}, D. and {de Souza}, R. and {Debosscher}, J. and {del Pozo}, E. and {Delbo}, M. and {Delgado}, A. and {Delgado}, H.~E. and {di Marco}, F. and {Di Matteo}, P. and {Diakite}, S. and {Distefano}, E. and {Dolding}, C. and {Dos Anjos}, S. and {Drazinos}, P. and {Dur{\'a}n}, J. and {Dzigan}, Y. and {Ecale}, E. and {Edvardsson}, B. and {Enke}, H. and {Erdmann}, M. and {Escolar}, D. and {Espina}, M. and {Evans}, N.~W. and {Eynard Bontemps}, G. and {Fabre}, C. and {Fabrizio}, M. and {Faigler}, S. and {Falc{\~a}o}, A.~J. and {Farr{\`a}s Casas}, M. and {Faye}, F. and {Federici}, L. and {Fedorets}, G. and {Fern{\'a}ndez-Hern{\'a}ndez}, J. and {Fernique}, P. and {Fienga}, A. and {Figueras}, F. and {Filippi}, F. and {Findeisen}, K. and {Fonti}, A. and {Fouesneau}, M. and {Fraile}, E. and {Fraser}, M. and {Fuchs}, J. and {Furnell}, R. and {Gai}, M. and {Galleti}, S. and {Galluccio}, L. and {Garabato}, D. and {Garc{\'\i}a-Sedano}, F. and {Gar{\'e}}, P. and {Garofalo}, A. and {Garralda}, N. and {Gavras}, P. and {Gerssen}, J. and {Geyer}, R. and {Gilmore}, G. and {Girona}, S. and {Giuffrida}, G. and {Gomes}, M. and {Gonz{\'a}lez-Marcos}, A. and {Gonz{\'a}lez-N{\'u}{\~n}ez}, J. and {Gonz{\'a}lez-Vidal}, J.~J. and {Granvik}, M. and {Guerrier}, A. and {Guillout}, P. and {Guiraud}, J. and {G{\'u}rpide}, A. and {Guti{\'e}rrez-S{\'a}nchez}, R. and {Guy}, L.~P. and {Haigron}, R. and {Hatzidimitriou}, D. and {Haywood}, M. and {Heiter}, U. and {Helmi}, A. and {Hobbs}, D. and {Hofmann}, W. and {Holl}, B. and {Holland}, G. and {Hunt}, J.~A.~S. and {Hypki}, A. and {Icardi}, V. and {Irwin}, M. and {Jevardat de Fombelle}, G. and {Jofr{\'e}}, P. and {Jonker}, P.~G. and {Jorissen}, A. and {Julbe}, F. and {Karampelas}, A. and {Kochoska}, A. and {Kohley}, R. and {Kolenberg}, K. and {Kontizas}, E. and {Koposov}, S.~E. and {Kordopatis}, G. and {Koubsky}, P. and {Kowalczyk}, A. and {Krone-Martins}, A. and {Kudryashova}, M. and {Kull}, I. and {Bachchan}, R.~K. and {Lacoste-Seris}, F. and {Lanza}, A.~F. and {Lavigne}, J. -B. and {Le Poncin-Lafitte}, C. and {Lebreton}, Y. and {Lebzelter}, T. and {Leccia}, S. and {Leclerc}, N. and {Lecoeur-Taibi}, I. and {Lemaitre}, V. and {Lenhardt}, H. and {Leroux}, F. and {Liao}, S. and {Licata}, E. and {Lindstr{\o}m}, H.~E.~P. and {Lister}, T.~A. and {Livanou}, E. and {Lobel}, A. and {L{\"o}ffler}, W. and {L{\'o}pez}, M. and {Lopez-Lozano}, A. and {Lorenz}, D. and {Loureiro}, T. and {MacDonald}, I. and {Magalh{\~a}es Fernandes}, T. and {Managau}, S. and {Mann}, R.~G. and {Mantelet}, G. and {Marchal}, O. and {Marchant}, J.~M. and {Marconi}, M. and {Marie}, J. and {Marinoni}, S. and {Marrese}, P.~M. and {Marschalk{\'o}}, G. and {Marshall}, D.~J. and {Mart{\'\i}n-Fleitas}, J.~M. and {Martino}, M. and {Mary}, N. and {Matijevi{\v{c}}}, G. and {Mazeh}, T. and {McMillan}, P.~J. and {Messina}, S. and {Mestre}, A. and {Michalik}, D. and {Millar}, N.~R. and {Miranda}, B.~M.~H. and {Molina}, D. and {Molinaro}, R. and {Molinaro}, M. and {Moln{\'a}r}, L. and {Moniez}, M. and {Montegriffo}, P. and {Monteiro}, D. and {Mor}, R. and {Mora}, A. and {Morbidelli}, R. and {Morel}, T. and {Morgenthaler}, S. and {Morley}, T. and {Morris}, D. and {Mulone}, A.~F. and {Muraveva}, T. and {Musella}, I. and {Narbonne}, J. and {Nelemans}, G. and {Nicastro}, L. and {Noval}, L. and {Ord{\'e}novic}, C. and {Ordieres-Mer{\'e}}, J. and {Osborne}, P. and {Pagani}, C. and {Pagano}, I. and {Pailler}, F. and {Palacin}, H. and {Palaversa}, L. and {Parsons}, P. and {Paulsen}, T. and {Pecoraro}, M. and {Pedrosa}, R. and {Pentik{\"a}inen}, H. and {Pereira}, J. and {Pichon}, B. and {Piersimoni}, A.~M. and {Pineau}, F. -X. and {Plachy}, E. and {Plum}, G. and {Poujoulet}, E. and {Pr{\v{s}}a}, A. and {Pulone}, L. and {Ragaini}, S. and {Rago}, S. and {Rambaux}, N. and {Ramos-Lerate}, M. and {Ranalli}, P. and {Rauw}, G. and {Read}, A. and {Regibo}, S. and {Renk}, F. and {Reyl{\'e}}, C. and {Ribeiro}, R.~A. and {Rimoldini}, L. and {Ripepi}, V. and {Riva}, A. and {Rixon}, G. and {Roelens}, M. and {Romero-G{\'o}mez}, M. and {Rowell}, N. and {Royer}, F. and {Rudolph}, A. and {Ruiz-Dern}, L. and {Sadowski}, G. and {Sagrist{\`a} Sell{\'e}s}, T. and {Sahlmann}, J. and {Salgado}, J. and {Salguero}, E. and {Sarasso}, M. and {Savietto}, H. and {Schnorhk}, A. and {Schultheis}, M. and {Sciacca}, E. and {Segol}, M. and {Segovia}, J.~C. and {Segransan}, D. and {Serpell}, E. and {Shih}, I. -C. and {Smareglia}, R. and {Smart}, R.~L. and {Smith}, C. and {Solano}, E. and {Solitro}, F. and {Sordo}, R. and {Soria Nieto}, S. and {Souchay}, J. and {Spagna}, A. and {Spoto}, F. and {Stampa}, U. and {Steele}, I.~A. and {Steidelm{\"u}ller}, H. and {Stephenson}, C.~A. and {Stoev}, H. and {Suess}, F.~F. and {S{\"u}veges}, M. and {Surdej}, J. and {Szabados}, L. and {Szegedi-Elek}, E. and {Tapiador}, D. and {Taris}, F. and {Tauran}, G. and {Taylor}, M.~B. and {Teixeira}, R. and {Terrett}, D. and {Tingley}, B. and {Trager}, S.~C. and {Turon}, C. and {Ulla}, A. and {Utrilla}, E. and {Valentini}, G. and {van Elteren}, A. and {Van Hemelryck}, E. and {van Leeuwen}, M. and {Varadi}, M. and {Vecchiato}, A. and {Veljanoski}, J. and {Via}, T. and {Vicente}, D. and {Vogt}, S. and {Voss}, H. and {Votruba}, V. and {Voutsinas}, S. and {Walmsley}, G. and {Weiler}, M. and {Weingrill}, K. and {Werner}, D. and {Wevers}, T. and {Whitehead}, G. and {Wyrzykowski}, {\L}. and {Yoldas}, A. and {{\v{Z}}erjal}, M. and {Zucker}, S. and {Zurbach}, C. and {Zwitter}, T. and {Alecu}, A. and {Allen}, M. and {Allende Prieto}, C. and {Amorim}, A. and {Anglada-Escud{\'e}}, G. and {Arsenijevic}, V. and {Azaz}, S. and {Balm}, P. and {Beck}, M. and {Bernstein}, H. -H. and {Bigot}, L. and {Bijaoui}, A. and {Blasco}, C. and {Bonfigli}, M. and {Bono}, G. and {Boudreault}, S. and {Bressan}, A. and {Brown}, S. and {Brunet}, P. -M. and {Bunclark}, P. and {Buonanno}, R. and {Butkevich}, A.~G. and {Carret}, C. and {Carrion}, C. and {Chemin}, L. and {Ch{\'e}reau}, F. and {Corcione}, L. and {Darmigny}, E. and {de Boer}, K.~S. and {de Teodoro}, P. and {de Zeeuw}, P.~T. and {Delle Luche}, C. and {Domingues}, C.~D. and {Dubath}, P. and {Fodor}, F. and {Fr{\'e}zouls}, B. and {Fries}, A. and {Fustes}, D. and {Fyfe}, D. and {Gallardo}, E. and {Gallegos}, J. and {Gardiol}, D. and {Gebran}, M. and {Gomboc}, A. and {G{\'o}mez}, A. and {Grux}, E. and {Gueguen}, A. and {Heyrovsky}, A. and {Hoar}, J. and {Iannicola}, G. and {Isasi Parache}, Y. and {Janotto}, A. -M. and {Joliet}, E. and {Jonckheere}, A. and {Keil}, R. and {Kim}, D. -W. and {Klagyivik}, P. and {Klar}, J. and {Knude}, J. and {Kochukhov}, O. and {Kolka}, I. and {Kos}, J. and {Kutka}, A. and {Lainey}, V. and {LeBouquin}, D. and {Liu}, C. and {Loreggia}, D. and {Makarov}, V.~V. and {Marseille}, M.~G. and {Martayan}, C. and {Martinez-Rubi}, O. and {Massart}, B. and {Meynadier}, F. and {Mignot}, S. and {Munari}, U. and {Nguyen}, A. -T. and {Nordlander}, T. and {Ocvirk}, P. and {O'Flaherty}, K.~S. and {Olias Sanz}, A. and {Ortiz}, P. and {Osorio}, J. and {Oszkiewicz}, D. and {Ouzounis}, A. and {Palmer}, M. and {Park}, P. and {Pasquato}, E. and {Peltzer}, C. and {Peralta}, J. and {P{\'e}turaud}, F. and {Pieniluoma}, T. and {Pigozzi}, E. and {Poels}, J. and {Prat}, G. and {Prod'homme}, T. and {Raison}, F. and {Rebordao}, J.~M. and {Risquez}, D. and {Rocca-Volmerange}, B. and {Rosen}, S. and {Ruiz-Fuertes}, M.~I. and {Russo}, F. and {Sembay}, S. and {Serraller Vizcaino}, I. and {Short}, A. and {Siebert}, A. and {Silva}, H. and {Sinachopoulos}, D. and {Slezak}, E. and {Soffel}, M. and {Sosnowska}, D. and {Strai{\v{z}}ys}, V. and {ter Linden}, M. and {Terrell}, D. and {Theil}, S. and {Tiede}, C. and {Troisi}, L. and {Tsalmantza}, P. and {Tur}, D. and {Vaccari}, M. and {Vachier}, F. and {Valles}, P. and {Van Hamme}, W. and {Veltz}, L. and {Virtanen}, J. and {Wallut}, J. -M. and {Wichmann}, R. and {Wilkinson}, M.~I. and {Ziaeepour}, H. and {Zschocke}, S.},
        title = "{The Gaia mission}",
      journal = {\aap},
         year = 2016,
        month = nov,
       volume = {595},
          eid = {A1},
        pages = {A1},
          doi = {10.1051/0004-6361/201629272},
archivePrefix = {arXiv},
       eprint = {1609.04153},
 primaryClass = {astro-ph.IM},
       adsurl = {https://ui.adsabs.harvard.edu/abs/2016A&A...595A...1G}
}

@ARTICLE{Gaia_2018,
       author = {{Gaia Collaboration} and {Helmi}, A. and {van Leeuwen}, F. and {McMillan}, P.~J. and {Massari}, D. and {Antoja}, T. and {Robin}, A.~C. and {Lindegren}, L. and {Bastian}, U. and {Arenou}, F. and {Babusiaux}, C. and {Biermann}, M. and {Breddels}, M.~A. and {Hobbs}, D. and {Jordi}, C. and {Pancino}, E. and {Reyl{\'e}}, C. and {Veljanoski}, J. and {Brown}, A.~G.~A. and {Vallenari}, A. and {Prusti}, T. and {de Bruijne}, J.~H.~J. and {Bailer-Jones}, C.~A.~L. and {Evans}, D.~W. and {Eyer}, L. and {Jansen}, F. and {Klioner}, S.~A. and {Lammers}, U. and {Luri}, X. and {Mignard}, F. and {Panem}, C. and {Pourbaix}, D. and {Randich}, S. and {Sartoretti}, P. and {Siddiqui}, H.~I. and {Soubiran}, C. and {Walton}, N.~A. and {Cropper}, M. and {Drimmel}, R. and {Katz}, D. and {Lattanzi}, M.~G. and {Bakker}, J. and {Cacciari}, C. and {Casta{\~n}eda}, J. and {Chaoul}, L. and {Cheek}, N. and {De Angeli}, F. and {Fabricius}, C. and {Guerra}, R. and {Holl}, B. and {Masana}, E. and {Messineo}, R. and {Mowlavi}, N. and {Nienartowicz}, K. and {Panuzzo}, P. and {Portell}, J. and {Riello}, M. and {Seabroke}, G.~M. and {Tanga}, P. and {Th{\'e}venin}, F. and {Gracia-Abril}, G. and {Comoretto}, G. and {Garcia-Reinaldos}, M. and {Teyssier}, D. and {Altmann}, M. and {Andrae}, R. and {Audard}, M. and {Bellas-Velidis}, I. and {Benson}, K. and {Berthier}, J. and {Blomme}, R. and {Burgess}, P. and {Busso}, G. and {Carry}, B. and {Cellino}, A. and {Clementini}, G. and {Clotet}, M. and {Creevey}, O. and {Davidson}, M. and {De Ridder}, J. and {Delchambre}, L. and {Dell'Oro}, A. and {Ducourant}, C. and {Fern{\'a}ndez-Hern{\'a}ndez}, J. and {Fouesneau}, M. and {Fr{\'e}mat}, Y. and {Galluccio}, L. and {Garc{\'\i}a-Torres}, M. and {Gonz{\'a}lez-N{\'u}{\~n}ez}, J. and {Gonz{\'a}lez-Vidal}, J.~J. and {Gosset}, E. and {Guy}, L.~P. and {Halbwachs}, J. -L. and {Hambly}, N.~C. and {Harrison}, D.~L. and {Hern{\'a}ndez}, J. and {Hestroffer}, D. and {Hodgkin}, S.~T. and {Hutton}, A. and {Jasniewicz}, G. and {Jean-Antoine-Piccolo}, A. and {Jordan}, S. and {Korn}, A.~J. and {Krone-Martins}, A. and {Lanzafame}, A.~C. and {Lebzelter}, T. and {L{\"o}ffler}, W. and {Manteiga}, M. and {Marrese}, P.~M. and {Mart{\'\i}n-Fleitas}, J.~M. and {Moitinho}, A. and {Mora}, A. and {Muinonen}, K. and {Osinde}, J. and {Pauwels}, T. and {Petit}, J. -M. and {Recio-Blanco}, A. and {Richards}, P.~J. and {Rimoldini}, L. and {Sarro}, L.~M. and {Siopis}, C. and {Smith}, M. and {Sozzetti}, A. and {S{\"u}veges}, M. and {Torra}, J. and {van Reeven}, W. and {Abbas}, U. and {Abreu Aramburu}, A. and {Accart}, S. and {Aerts}, C. and {Altavilla}, G. and {{\'A}lvarez}, M.~A. and {Alvarez}, R. and {Alves}, J. and {Anderson}, R.~I. and {Andrei}, A.~H. and {Anglada Varela}, E. and {Antiche}, E. and {Arcay}, B. and {Astraatmadja}, T.~L. and {Bach}, N. and {Baker}, S.~G. and {Balaguer-N{\'u}{\~n}ez}, L. and {Balm}, P. and {Barache}, C. and {Barata}, C. and {Barbato}, D. and {Barblan}, F. and {Barklem}, P.~S. and {Barrado}, D. and {Barros}, M. and {Barstow}, M.~A. and {Bartholom{\'e} Mu{\~n}oz}, S. and {Bassilana}, J. -L. and {Becciani}, U. and {Bellazzini}, M. and {Berihuete}, A. and {Bertone}, S. and {Bianchi}, L. and {Bienaym{\'e}}, O. and {Blanco-Cuaresma}, S. and {Boch}, T. and {Boeche}, C. and {Bombrun}, A. and {Borrachero}, R. and {Bossini}, D. and {Bouquillon}, S. and {Bourda}, G. and {Bragaglia}, A. and {Bramante}, L. and {Bressan}, A. and {Brouillet}, N. and {Br{\"u}semeister}, T. and {Brugaletta}, E. and {Bucciarelli}, B. and {Burlacu}, A. and {Busonero}, D. and {Butkevich}, A.~G. and {Buzzi}, R. and {Caffau}, E. and {Cancelliere}, R. and {Cannizzaro}, G. and {Cantat-Gaudin}, T. and {Carballo}, R. and {Carlucci}, T. and {Carrasco}, J.~M. and {Casamiquela}, L. and {Castellani}, M. and {Castro-Ginard}, A. and {Charlot}, P. and {Chemin}, L. and {Chiavassa}, A. and {Cocozza}, G. and {Costigan}, G. and {Cowell}, S. and {Crifo}, F. and {Crosta}, M. and {Crowley}, C. and {Cuypers}, J. and {Dafonte}, C. and {Damerdji}, Y. and {Dapergolas}, A. and {David}, P. and {David}, M. and {de Laverny}, P. and {De Luise}, F. and {De March}, R. and {de Martino}, D. and {de Souza}, R. and {de Torres}, A. and {Debosscher}, J. and {del Pozo}, E. and {Delbo}, M. and {Delgado}, A. and {Delgado}, H.~E. and {Di Matteo}, P. and {Diakite}, S. and {Diener}, C. and {Distefano}, E. and {Dolding}, C. and {Drazinos}, P. and {Dur{\'a}n}, J. and {Edvardsson}, B. and {Enke}, H. and {Eriksson}, K. and {Esquej}, P. and {Eynard Bontemps}, G. and {Fabre}, C. and {Fabrizio}, M. and {Faigler}, S. and {Falc{\~a}o}, A.~J. and {Farr{\`a}s Casas}, M. and {Federici}, L. and {Fedorets}, G. and {Fernique}, P. and {Figueras}, F. and {Filippi}, F. and {Findeisen}, K. and {Fonti}, A. and {Fraile}, E. and {Fraser}, M. and {Fr{\'e}zouls}, B. and {Gai}, M. and {Galleti}, S. and {Garabato}, D. and {Garc{\'\i}a-Sedano}, F. and {Garofalo}, A. and {Garralda}, N. and {Gavel}, A. and {Gavras}, P. and {Gerssen}, J. and {Geyer}, R. and {Giacobbe}, P. and {Gilmore}, G. and {Girona}, S. and {Giuffrida}, G. and {Glass}, F. and {Gomes}, M. and {Granvik}, M. and {Gueguen}, A. and {Guerrier}, A. and {Guiraud}, J. and {Guti{\'e}rrez-S{\'a}nchez}, R. and {Hofmann}, W. and {Holland}, G. and {Huckle}, H.~E. and {Hypki}, A. and {Icardi}, V. and {Jan{\ss}en}, K. and {Jevardat de Fombelle}, G. and {Jonker}, P.~G. and {Juh{\'a}sz}, {\'A}. L. and {Julbe}, F. and {Karampelas}, A. and {Kewley}, A. and {Klar}, J. and {Kochoska}, A. and {Kohley}, R. and {Kolenberg}, K. and {Kontizas}, M. and {Kontizas}, E. and {Koposov}, S.~E. and {Kordopatis}, G. and {Kostrzewa-Rutkowska}, Z. and {Koubsky}, P. and {Lambert}, S. and {Lanza}, A.~F. and {Lasne}, Y. and {Lavigne}, J. -B. and {Le Fustec}, Y. and {Le Poncin-Lafitte}, C. and {Lebreton}, Y. and {Leccia}, S. and {Leclerc}, N. and {Lecoeur-Taibi}, I. and {Lenhardt}, H. and {Leroux}, F. and {Liao}, S. and {Licata}, E. and {Lindstr{\o}m}, H.~E.~P. and {Lister}, T.~A. and {Livanou}, E. and {Lobel}, A. and {L{\'o}pez}, M. and {Managau}, S. and {Mann}, R.~G. and {Mantelet}, G. and {Marchal}, O. and {Marchant}, J.~M. and {Marconi}, M. and {Marinoni}, S. and {Marschalk{\'o}}, G. and {Marshall}, D.~J. and {Martino}, M. and {Marton}, G. and {Mary}, N. and {Matijevi{\v{c}}}, G. and {Mazeh}, T. and {Messina}, S. and {Michalik}, D. and {Millar}, N.~R. and {Molina}, D. and {Molinaro}, R. and {Moln{\'a}r}, L. and {Montegriffo}, P. and {Mor}, R. and {Morbidelli}, R. and {Morel}, T. and {Morris}, D. and {Mulone}, A.~F. and {Muraveva}, T. and {Musella}, I. and {Nelemans}, G. and {Nicastro}, L. and {Noval}, L. and {O'Mullane}, W. and {Ord{\'e}novic}, C. and {Ord{\'o}{\~n}ez-Blanco}, D. and {Osborne}, P. and {Pagani}, C. and {Pagano}, I. and {Pailler}, F. and {Palacin}, H. and {Palaversa}, L. and {Panahi}, A. and {Pawlak}, M. and {Piersimoni}, A.~M. and {Pineau}, F. -X. and {Plachy}, E. and {Plum}, G. and {Poggio}, E. and {Poujoulet}, E. and {Pr{\v{s}}a}, A. and {Pulone}, L. and {Racero}, E. and {Ragaini}, S. and {Rambaux}, N. and {Ramos-Lerate}, M. and {Regibo}, S. and {Riclet}, F. and {Ripepi}, V. and {Riva}, A. and {Rivard}, A. and {Rixon}, G. and {Roegiers}, T. and {Roelens}, M. and {Romero-G{\'o}mez}, M. and {Rowell}, N. and {Royer}, F. and {Ruiz-Dern}, L. and {Sadowski}, G. and {Sagrist{\`a} Sell{\'e}s}, T. and {Sahlmann}, J. and {Salgado}, J. and {Salguero}, E. and {Sanna}, N. and {Santana-Ros}, T. and {Sarasso}, M. and {Savietto}, H. and {Schultheis}, M. and {Sciacca}, E. and {Segol}, M. and {Segovia}, J.~C. and {S{\'e}gransan}, D. and {Shih}, I. -C. and {Siltala}, L. and {Silva}, A.~F. and {Smart}, R.~L. and {Smith}, K.~W. and {Solano}, E. and {Solitro}, F. and {Sordo}, R. and {Soria Nieto}, S. and {Souchay}, J. and {Spagna}, A. and {Spoto}, F. and {Stampa}, U. and {Steele}, I.~A. and {Steidelm{\"u}ller}, H. and {Stephenson}, C.~A. and {Stoev}, H. and {Suess}, F.~F. and {Surdej}, J. and {Szabados}, L. and {Szegedi-Elek}, E. and {Tapiador}, D. and {Taris}, F. and {Tauran}, G. and {Taylor}, M.~B. and {Teixeira}, R. and {Terrett}, D. and {Teyssandier}, P. and {Thuillot}, W. and {Titarenko}, A. and {Torra Clotet}, F. and {Turon}, C. and {Ulla}, A. and {Utrilla}, E. and {Uzzi}, S. and {Vaillant}, M. and {Valentini}, G. and {Valette}, V. and {van Elteren}, A. and {Van Hemelryck}, E. and {van Leeuwen}, M. and {Vaschetto}, M. and {Vecchiato}, A. and {Viala}, Y. and {Vicente}, D. and {Vogt}, S. and {von Essen}, C. and {Voss}, H. and {Votruba}, V. and {Voutsinas}, S. and {Walmsley}, G. and {Weiler}, M. and {Wertz}, O. and {Wevems}, T. and {Wyrzykowski}, {\L}. and {Yoldas}, A. and {{\v{Z}}erjal}, M. and {Ziaeepour}, H. and {Zorec}, J. and {Zschocke}, S. and {Zucker}, S. and {Zurbach}, C. and {Zwitter}, T.},
        title = "{Gaia Data Release 2. Kinematics of globular clusters and dwarf galaxies around the Milky Way}",
      journal = {\aap},
         year = 2018,
        month = aug,
       volume = {616},
          eid = {A12},
        pages = {A12},
          doi = {10.1051/0004-6361/201832698},
archivePrefix = {arXiv},
       eprint = {1804.09381},
 primaryClass = {astro-ph.GA},
       adsurl = {https://ui.adsabs.harvard.edu/abs/2018A&A...616A..12G}
}

@ARTICLE{Massari_2018,
       author = {{Massari}, D. and {Breddels}, M.~A. and {Helmi}, A. and {Posti}, L. and {Brown}, A.~G.~A. and {Tolstoy}, E.},
        title = "{Three-dimensional motions in the Sculptor dwarf galaxy as a glimpse of a new era}",
      journal = {Nature Astronomy},
         year = 2018,
        month = nov,
       volume = {2},
        pages = {156-161},
          doi = {10.1038/s41550-017-0322-y},
archivePrefix = {arXiv},
       eprint = {1711.08945},
 primaryClass = {astro-ph.GA},
       adsurl = {https://ui.adsabs.harvard.edu/abs/2018NatAs...2..156M}
}

@MISC{Anderson_2004,
       author = {{Anderson}, Jay and {King}, Ivan R.},
        title = "{Multi-filter PSFs and Distortion Corrections for the HRC}",
 howpublished = {Instrument Science Report ACS 2004-15, 51 pages},
         year = 2004,
        month = aug,
        pages = {15},
       adsurl = {https://ui.adsabs.harvard.edu/abs/2004acs..rept...15A}
}

@MISC{Anderson_2006,
       author = {{Anderson}, Jay and {King}, Ivan R.},
        title = "{PSFs, Photometry, and Astronomy for the ACS/WFC}",
 howpublished = {Instrument Science Report ACS 2006-01, 34 pages},
         year = 2006,
        month = feb,
        pages = {1},
       adsurl = {https://ui.adsabs.harvard.edu/abs/2006acs..rept....1A}
}

@ARTICLE{Dey_2023,
       author = {{Dey}, Arjun and {Najita}, Joan and {Filion}, Carrie and {Han}, Jiwon Jesse and {Pearson}, Sarah and {Wyse}, Rosemary and {Thob}, Adrien C.~R. and {Anguiano}, Borja and {Apfel}, Miranda and {Arnaboldi}, Magda and {Bell}, Eric F. and {Beraldo e Silva}, Leandro and {Besla}, Gurtina and {Bhattacharya}, Aparajito and {Bhattacharya}, Souradeep and {Chandra}, Vedant and {Choi}, Yumi and {Collins}, Michelle L.~M. and {Cunningham}, Emily C. and {Dalcanton}, Julianne J. and {Escala}, Ivanna and {Foote}, Hayden R. and {Ferguson}, Annette M.~N. and {Gibson}, Benjamin J. and {Gnedin}, Oleg Y. and {Guhathakurta}, Puragra and {Hawkins}, Keith and {Horta}, Danny and {Ibata}, Rodrigo and {Kallivayalil}, Nitya and {Koch}, Eric W. and {Koposov}, Sergey and {Lewis}, Geraint F. and {Macri}, Lucas and {McKinnon}, Kevin A. and {Nidever}, David L. and {Olsen}, Knut A.~G. and {Patel}, Ekta and {Petersen}, Michael S. and {Petric}, Andreea and {Price-Whelan}, Adrian M. and {Rich}, R. Michael and {Riley}, Alexander H. and {Saha}, Abhijit and {Sanderson}, Robyn E. and {Sharma}, Sanjib and {Sohn}, Sangmo Tony and {Soraisam}, Monika D. and {Steinmetz}, Matthias and {Valluri}, Monica and {Vivas}, A. Katherina and {Williams}, Benjamin F. and {Wojno}, J. Leigh},
        title = "{RomAndromeda: The Roman Survey of the Andromeda Halo}",
      journal = {arXiv e-prints},
         year = 2023,
        month = jun,
          eid = {arXiv:2306.12302},
        pages = {arXiv:2306.12302},
          doi = {10.48550/arXiv.2306.12302},
archivePrefix = {arXiv},
       eprint = {2306.12302},
 primaryClass = {astro-ph.GA},
       adsurl = {https://ui.adsabs.harvard.edu/abs/2023arXiv230612302D}
}

@article{Rao_1947,
author = {Rao, C.},
year = {1947},
month = {04},
pages = {280 - 283},
title = {Minimum variance and the estimation of several parameters},
volume = {43},
journal = {Mathematical Proceedings of the Cambridge Philosophical Society},
doi = {10.1017/S0305004100023471}
}

@book{Cramer_1946,
author = {Cram\'er, H.},
ISBN = {9780691005478},
URL = {http://www.jstor.org/stable/j.ctt1bpm9r4},
publisher = {Princeton University Press},
year = {1946},
month = {01},
pages = {},
title = {Mathematical Methods Of Statistics},
volume = {9},
journal = {Princeton}
}

@ARTICLE{Anderson_2000,
       author = {{Anderson}, Jay and {King}, Ivan R.},
        title = "{Toward High-Precision Astrometry with WFPC2. I. Deriving an Accurate Point-Spread Function}",
      journal = {\pasp},
         year = 2000,
        month = oct,
       volume = {112},
       number = {776},
        pages = {1360-1382},
          doi = {10.1086/316632},
archivePrefix = {arXiv},
       eprint = {astro-ph/0006325},
 primaryClass = {astro-ph},
       adsurl = {https://ui.adsabs.harvard.edu/abs/2000PASP..112.1360A}
}

@ARTICLE{Savino_2024,
       author = {{Savino}, Alessandro and {Gennaro}, Mario and {Dolphin}, Andrew E. and {Weisz}, Daniel R. and {Correnti}, Matteo and {Anderson}, Jay and {Beaton}, Rachael and {Boyer}, Martha L. and {Cohen}, Roger E. and {Cole}, Andrew A. and {Durbin}, Meredith J. and {Garling}, Christopher T. and {Geha}, Marla C. and {Gilbert}, Karoline M. and {Kalirai}, Jason and {Kallivayalil}, Nitya and {McQuinn}, Kristen B.~W. and {Newman}, Max J.~B. and {Richstein}, Hannah and {Skillman}, Evan D. and {Warfield}, Jack T. and {Williams}, Benjamin F.},
        title = "{The JWST Resolved Stellar Populations Early Release Science Program. VII. Stress Testing the NIRCam Exposure Time Calculator}",
      journal = {\apj},
         year = 2024,
        month = jul,
       volume = {970},
       number = {1},
          eid = {36},
        pages = {36},
          doi = {10.3847/1538-4357/ad4e2f},
archivePrefix = {arXiv},
       eprint = {2405.17547},
 primaryClass = {astro-ph.GA},
       adsurl = {https://ui.adsabs.harvard.edu/abs/2024ApJ...970...36S}
}

@ARTICLE{Bedin_2025,
       author = {{Bedin}, Luigi ''Rolly''},
        title = "{The case for an astrometric mission extension of Euclid: Extending Gaia by six magnitudes with Euclid covering one third of the sky}",
      journal = {\aap},
         year = 2025,
        month = dec,
       volume = {704},
          eid = {A193},
        pages = {A193},
          doi = {10.1051/0004-6361/202557407},
archivePrefix = {arXiv},
       eprint = {2510.23694},
 primaryClass = {astro-ph.IM},
       adsurl = {https://ui.adsabs.harvard.edu/abs/2025A&A...704A.193B}
}

@ARTICLE{Han_2023,
       author = {{Han}, Jiwon Jesse and {Dey}, Arjun and {Price-Whelan}, Adrian M. and {Najita}, Joan and {Schlafly}, Edward F. and {Saydjari}, Andrew and {Wechsler}, Risa H. and {Bonaca}, Ana and {Schlegel}, David J and {Conroy}, Charlie and {Raichoor}, Anand and {Drlica-Wagner}, Alex and {Kollmeier}, Juna A. and {Koposov}, Sergey E. and {Besla}, Gurtina and {Rix}, Hans-Walter and {Goodman}, Alyssa and {Finkbeiner}, Douglas and {Anand}, Abhijeet and {Ashby}, Matthew and {Bahr-Kalus}, Benedict and {Beaton}, Rachel and {Behera}, Jayashree and {Bell}, Eric F. and {Bellm}, Eric C and {BenZvi}, Segev and {Beraldo e Silva}, Leandro and {Birrer}, Simon and {Blanton}, Michael R. and {Bock}, Jamie and {Broekgaarden}, Floor and {Brout}, Dillon and {Brown}, Warren and {Brown}, Anthony G.~A. and {Bulbul}, Esra and {Calderon}, Rodrigo and {Carlin}, Jeffrey L and {Carrillo}, Andreia and {Castander}, Francisco Javier and {Chakraborty}, Priyanka and {Chandra}, Vedant and {Chiang}, Yi-Kuan and {Choi}, Yumi and {Clark}, Susan E. and {Clarkson}, William I. and {Cooper}, Andrew and {Crill}, Brendan and {Cunha}, Katia and {Cunningham}, Emily and {Dalcanton}, Julianne and {Danieli}, Shany and {Daylan}, Tansu and {de Jong}, Roelof S. and {DeRose}, Joseph and {Dey}, Biprateep and {Dickinson}, Mark and {Dominguez}, Mariano and {Dong}, Dillon and {Eifler}, Tim and {El-Badry}, Kareem and {Erkal}, Denis and {Escala}, Ivanna and {Fazio}, Giovanni and {Ferguson}, Annette M.~N. and {Ferraro}, Simone and {Filion}, Carrie and {Forero-Romero}, Jaime E. and {Fu}, Shenming and {Galbany}, Llu{\'\i}s and {Garavito-Camargo}, Nicolas and {Gawiser}, Eric and {Geha}, Marla and {Gnedin}, Oleg Y. and {Gomez}, Sebastian and {Greene}, Jenny and {Guy}, Julien and {Hadzhiyska}, Boryana and {Hawkins}, Keith and {Heinrich}, Chen and {Hernquist}, Lars and {Hirata}, Christopher and {Hora}, Joseph and {Horowitz}, Benjamin and {Horta}, Danny and {Huang}, Caroline and {Huang}, Xiaosheng and {Huanyuan}, Shan and {Hunt}, Jason A.~S. and {Ibata}, Rodrigo and {Jannuzi}, Buell and {Johnston}, Kathryn V. and {Jones}, Michael G. and {Juneau}, Stephanie and {Kado-Fong}, Erin and {Kalari}, Venu and {Kallivayalil}, Nitya and {Karim}, Tanveer and {Keeley}, Ryan and {Khoperskov}, Sergey and {Kim}, Bokyoung and {Kov{\'a}cs}, Andr{\'a}s and {Krause}, Elisabeth and {Kremer}, Kyle and {Kremin}, Anthony and {Krolewski}, Alex and {Kulkarni}, S.~R. and {Kuna}, Marine and {L'Huillier}, Benjamin and {Lacy}, Mark and {Lan}, Ting-Wen and {Lang}, Dustin and {Leahy}, Denis and {Li}, Jiaxuan and {Lim}, Seunghwan and {L{\'o}pez-Morales}, Mercedes and {Macri}, Lucas and {Marc}, Manera and {Mau}, Sidney and {McCarthy}, Patrick J and {McDonald}, Eithne and {McQuinn}, Kristen and {Meisner}, Aaron and {Melnick}, Gary and {Merloni}, Andrea and {Millard}, Cl{\'e}a and {Millon}, Martin and {Minchev}, Ivan and {Montero-Camacho}, Paulo and {Morales-Gutierrez}, Catalina and {Morrell}, Nidia and {Moustakas}, John and {Moustakas}, Leonidas and {Murray}, Zachary and {Mutlu-Pakdil}, Burcin and {Myeong}, GyuChul and {Myers}, Adam D. and {Nadler}, Ethan and {Navarete}, Felipe and {Ness}, Melissa and {Nidever}, David L. and {Nikutta}, Robert and {Nushkia}, Chamba and {Olsen}, Knut and {Pace}, Andrew B. and {Pacucci}, Fabio and {Padmanabhan}, Nikhil and {Parkinson}, David and {Pearson}, Sarah and {Peng}, Eric W. and {Petric}, Andreea O. and {Petric}, Andreea and {Ratcliffe}, Bridget and {Razieh}, Emami and {Reiprich}, Thomas and {Rezaie}, Mehdi and {Ricci}, Marina and {Rich}, R. Michael and {Richstein}, Hannah and {Riley}, Alexander H. and {Rockosi}, Constance and {Rossi}, Graziano and {Salvato}, Mara and {Samushia}, Lado and {Sanchez}, Javier and {Sand}, David J and {E Sanderson}, Robyn and {{\v{S}}ar{\v{c}}evi{\'c}}, Nikolina and {Sarkar}, Arnab and {Savino}, Alessandro and {Schweizer}, Francois and {Shafieloo}, Arman and {Shengqi}, Yang and {Shields}, Joseph and {Shipp}, Nora and {Simon}, Josh and {Siudek}, Malgorzata and {Siwei}, Zou and {Slepian}, Zachary and {Smith}, Verne and {Sobeck}, Jennifer and {Sohn}, Sangmo Tony and {Som}, Debopam and {Speagle}, Joshua S. and {Spergel}, David and {Szabo}, Robert and {Tan}, Ting and {Theissen}, Christopher and {Tollerud}, Erik and {Tolls}, Volker and {Tran}, Kim-Vy and {Tsiane}, Kabelo and {Vacca}, William D. and {Valluri}, Monica and {Verberi}, TonyLouis and {Warfield}, Jack and {Weaverdyck}, Noah and {Weiner}, Benjamin and {Weisz}, Daniel and {Wetzel}, Andrew and {White}, Martin},
        title = "{NANCY: Next-generation All-sky Near-infrared Community surveY}",
      journal = {arXiv e-prints},
         year = 2023,
        month = jun,
          eid = {arXiv:2306.11784},
        pages = {arXiv:2306.11784},
          doi = {10.48550/arXiv.2306.11784},
archivePrefix = {arXiv},
       eprint = {2306.11784},
 primaryClass = {astro-ph.IM},
       adsurl = {https://ui.adsabs.harvard.edu/abs/2023arXiv230611784H}
}

@ARTICLE{Spergel_2015,
       author = {{Spergel}, D. and {Gehrels}, N. and {Baltay}, C. and {Bennett}, D. and {Breckinridge}, J. and {Donahue}, M. and {Dressler}, A. and {Gaudi}, B.~S. and {Greene}, T. and {Guyon}, O. and {Hirata}, C. and {Kalirai}, J. and {Kasdin}, N.~J. and {Macintosh}, B. and {Moos}, W. and {Perlmutter}, S. and {Postman}, M. and {Rauscher}, B. and {Rhodes}, J. and {Wang}, Y. and {Weinberg}, D. and {Benford}, D. and {Hudson}, M. and {Jeong}, W. -S. and {Mellier}, Y. and {Traub}, W. and {Yamada}, T. and {Capak}, P. and {Colbert}, J. and {Masters}, D. and {Penny}, M. and {Savransky}, D. and {Stern}, D. and {Zimmerman}, N. and {Barry}, R. and {Bartusek}, L. and {Carpenter}, K. and {Cheng}, E. and {Content}, D. and {Dekens}, F. and {Demers}, R. and {Grady}, K. and {Jackson}, C. and {Kuan}, G. and {Kruk}, J. and {Melton}, M. and {Nemati}, B. and {Parvin}, B. and {Poberezhskiy}, I. and {Peddie}, C. and {Ruffa}, J. and {Wallace}, J.~K. and {Whipple}, A. and {Wollack}, E. and {Zhao}, F.},
        title = "{Wide-Field InfrarRed Survey Telescope-Astrophysics Focused Telescope Assets WFIRST-AFTA 2015 Report}",
      journal = {arXiv e-prints},
         year = 2015,
        month = mar,
          eid = {arXiv:1503.03757},
        pages = {arXiv:1503.03757},
          doi = {10.48550/arXiv.1503.03757},
archivePrefix = {arXiv},
       eprint = {1503.03757},
 primaryClass = {astro-ph.IM},
       adsurl = {https://ui.adsabs.harvard.edu/abs/2015arXiv150303757S}
}

@ARTICLE{Akeson_2019,
       author = {{Akeson}, Rachel and {Armus}, Lee and {Bachelet}, Etienne and {Bailey}, Vanessa and {Bartusek}, Lisa and {Bellini}, Andrea and {Benford}, Dominic and {Bennett}, David and {Bhattacharya}, Aparna and {Bohlin}, Ralph and {Boyer}, Martha and {Bozza}, Valerio and {Bryden}, Geoffrey and {Calchi Novati}, Sebastiano and {Carpenter}, Kenneth and {Casertano}, Stefano and {Choi}, Ami and {Content}, David and {Dayal}, Pratika and {Dressler}, Alan and {Dor{\'e}}, Olivier and {Fall}, S. Michael and {Fan}, Xiaohui and {Fang}, Xiao and {Filippenko}, Alexei and {Finkelstein}, Steven and {Foley}, Ryan and {Furlanetto}, Steven and {Kalirai}, Jason and {Gaudi}, B. Scott and {Gilbert}, Karoline and {Girard}, Julien and {Grady}, Kevin and {Greene}, Jenny and {Guhathakurta}, Puragra and {Heinrich}, Chen and {Hemmati}, Shoubaneh and {Hendel}, David and {Henderson}, Calen and {Henning}, Thomas and {Hirata}, Christopher and {Ho}, Shirley and {Huff}, Eric and {Hutter}, Anne and {Jansen}, Rolf and {Jha}, Saurabh and {Johnson}, Samson and {Jones}, David and {Kasdin}, Jeremy and {Kelly}, Patrick and {Kirshner}, Robert and {Koekemoer}, Anton and {Kruk}, Jeffrey and {Lewis}, Nikole and {Macintosh}, Bruce and {Madau}, Piero and {Malhotra}, Sangeeta and {Mandel}, Kaisey and {Massara}, Elena and {Masters}, Daniel and {McEnery}, Julie and {McQuinn}, Kristen and {Melchior}, Peter and {Melton}, Mark and {Mennesson}, Bertrand and {Peeples}, Molly and {Penny}, Matthew and {Perlmutter}, Saul and {Pisani}, Alice and {Plazas}, Andr{\'e}s and {Poleski}, Radek and {Postman}, Marc and {Ranc}, Cl{\'e}ment and {Rauscher}, Bernard and {Rest}, Armin and {Roberge}, Aki and {Robertson}, Brant and {Rodney}, Steven and {Rhoads}, James and {Rhodes}, Jason and {Ryan}, Jr., Russell and {Sahu}, Kailash and {Sand}, David and {Scolnic}, Dan and {Seth}, Anil and {Shvartzvald}, Yossi and {Siellez}, Karelle and {Smith}, Arfon and {Spergel}, David and {Stassun}, Keivan and {Street}, Rachel and {Strolger}, Louis-Gregory and {Szalay}, Alexander and {Trauger}, John and {Troxel}, M.~A. and {Turnbull}, Margaret and {van der Marel}, Roeland and {von der Linden}, Anja and {Wang}, Yun and {Weinberg}, David and {Williams}, Benjamin and {Windhorst}, Rogier and {Wollack}, Edward and {Wu}, Hao-Yi and {Yee}, Jennifer and {Zimmerman}, Neil},
        title = "{The Wide Field Infrared Survey Telescope: 100 Hubbles for the 2020s}",
      journal = {arXiv e-prints},
         year = 2019,
        month = feb,
          eid = {arXiv:1902.05569},
        pages = {arXiv:1902.05569},
          doi = {10.48550/arXiv.1902.05569},
archivePrefix = {arXiv},
       eprint = {1902.05569},
 primaryClass = {astro-ph.IM},
       adsurl = {https://ui.adsabs.harvard.edu/abs/2019arXiv190205569A}
}

@ARTICLE{Ivezi_2019,
       author = {{Ivezi{\'c}}, {\v{Z}}eljko and {Kahn}, Steven M. and {Tyson}, J. Anthony and {Abel}, Bob and {Acosta}, Emily and {Allsman}, Robyn and {Alonso}, David and {AlSayyad}, Yusra and {Anderson}, Scott F. and {Andrew}, John and {Angel}, James Roger P. and {Angeli}, George Z. and {Ansari}, Reza and {Antilogus}, Pierre and {Araujo}, Constanza and {Armstrong}, Robert and {Arndt}, Kirk T. and {Astier}, Pierre and {Aubourg}, {\'E}ric and {Auza}, Nicole and {Axelrod}, Tim S. and {Bard}, Deborah J. and {Barr}, Jeff D. and {Barrau}, Aurelian and {Bartlett}, James G. and {Bauer}, Amanda E. and {Bauman}, Brian J. and {Baumont}, Sylvain and {Bechtol}, Ellen and {Bechtol}, Keith and {Becker}, Andrew C. and {Becla}, Jacek and {Beldica}, Cristina and {Bellavia}, Steve and {Bianco}, Federica B. and {Biswas}, Rahul and {Blanc}, Guillaume and {Blazek}, Jonathan and {Blandford}, Roger D. and {Bloom}, Josh S. and {Bogart}, Joanne and {Bond}, Tim W. and {Booth}, Michael T. and {Borgland}, Anders W. and {Borne}, Kirk and {Bosch}, James F. and {Boutigny}, Dominique and {Brackett}, Craig A. and {Bradshaw}, Andrew and {Brandt}, William Nielsen and {Brown}, Michael E. and {Bullock}, James S. and {Burchat}, Patricia and {Burke}, David L. and {Cagnoli}, Gianpietro and {Calabrese}, Daniel and {Callahan}, Shawn and {Callen}, Alice L. and {Carlin}, Jeffrey L. and {Carlson}, Erin L. and {Chandrasekharan}, Srinivasan and {Charles-Emerson}, Glenaver and {Chesley}, Steve and {Cheu}, Elliott C. and {Chiang}, Hsin-Fang and {Chiang}, James and {Chirino}, Carol and {Chow}, Derek and {Ciardi}, David R. and {Claver}, Charles F. and {Cohen-Tanugi}, Johann and {Cockrum}, Joseph J. and {Coles}, Rebecca and {Connolly}, Andrew J. and {Cook}, Kem H. and {Cooray}, Asantha and {Covey}, Kevin R. and {Cribbs}, Chris and {Cui}, Wei and {Cutri}, Roc and {Daly}, Philip N. and {Daniel}, Scott F. and {Daruich}, Felipe and {Daubard}, Guillaume and {Daues}, Greg and {Dawson}, William and {Delgado}, Francisco and {Dellapenna}, Alfred and {de Peyster}, Robert and {de Val-Borro}, Miguel and {Digel}, Seth W. and {Doherty}, Peter and {Dubois}, Richard and {Dubois-Felsmann}, Gregory P. and {Durech}, Josef and {Economou}, Frossie and {Eifler}, Tim and {Eracleous}, Michael and {Emmons}, Benjamin L. and {Fausti Neto}, Angelo and {Ferguson}, Henry and {Figueroa}, Enrique and {Fisher-Levine}, Merlin and {Focke}, Warren and {Foss}, Michael D. and {Frank}, James and {Freemon}, Michael D. and {Gangler}, Emmanuel and {Gawiser}, Eric and {Geary}, John C. and {Gee}, Perry and {Geha}, Marla and {Gessner}, Charles J.~B. and {Gibson}, Robert R. and {Gilmore}, D. Kirk and {Glanzman}, Thomas and {Glick}, William and {Goldina}, Tatiana and {Goldstein}, Daniel A. and {Goodenow}, Iain and {Graham}, Melissa L. and {Gressler}, William J. and {Gris}, Philippe and {Guy}, Leanne P. and {Guyonnet}, Augustin and {Haller}, Gunther and {Harris}, Ron and {Hascall}, Patrick A. and {Haupt}, Justine and {Hernandez}, Fabio and {Herrmann}, Sven and {Hileman}, Edward and {Hoblitt}, Joshua and {Hodgson}, John A. and {Hogan}, Craig and {Howard}, James D. and {Huang}, Dajun and {Huffer}, Michael E. and {Ingraham}, Patrick and {Innes}, Walter R. and {Jacoby}, Suzanne H. and {Jain}, Bhuvnesh and {Jammes}, Fabrice and {Jee}, M. James and {Jenness}, Tim and {Jernigan}, Garrett and {Jevremovi{\'c}}, Darko and {Johns}, Kenneth and {Johnson}, Anthony S. and {Johnson}, Margaret W.~G. and {Jones}, R. Lynne and {Juramy-Gilles}, Claire and {Juri{\'c}}, Mario and {Kalirai}, Jason S. and {Kallivayalil}, Nitya J. and {Kalmbach}, Bryce and {Kantor}, Jeffrey P. and {Karst}, Pierre and {Kasliwal}, Mansi M. and {Kelly}, Heather and {Kessler}, Richard and {Kinnison}, Veronica and {Kirkby}, David and {Knox}, Lloyd and {Kotov}, Ivan V. and {Krabbendam}, Victor L. and {Krughoff}, K. Simon and {Kub{\'a}nek}, Petr and {Kuczewski}, John and {Kulkarni}, Shri and {Ku}, John and {Kurita}, Nadine R. and {Lage}, Craig S. and {Lambert}, Ron and {Lange}, Travis and {Langton}, J. Brian and {Le Guillou}, Laurent and {Levine}, Deborah and {Liang}, Ming and {Lim}, Kian-Tat and {Lintott}, Chris J. and {Long}, Kevin E. and {Lopez}, Margaux and {Lotz}, Paul J. and {Lupton}, Robert H. and {Lust}, Nate B. and {MacArthur}, Lauren A. and {Mahabal}, Ashish and {Mandelbaum}, Rachel and {Markiewicz}, Thomas W. and {Marsh}, Darren S. and {Marshall}, Philip J. and {Marshall}, Stuart and {May}, Morgan and {McKercher}, Robert and {McQueen}, Michelle and {Meyers}, Joshua and {Migliore}, Myriam and {Miller}, Michelle and {Mills}, David J.},
        title = "{LSST: From Science Drivers to Reference Design and Anticipated Data Products}",
      journal = {\apj},
         year = 2019,
        month = mar,
       volume = {873},
       number = {2},
          eid = {111},
        pages = {111},
          doi = {10.3847/1538-4357/ab042c},
archivePrefix = {arXiv},
       eprint = {0805.2366},
 primaryClass = {astro-ph},
       adsurl = {https://ui.adsabs.harvard.edu/abs/2019ApJ...873..111I}
}

@ARTICLE{EuclidCollaboration_2025,
       author = {{Euclid Collaboration} and {Mellier}, Y. and {Abdurro'uf} and {Acevedo Barroso}, J.~A. and {Ach{\'u}carro}, A. and {Adamek}, J. and {Adam}, R. and {Addison}, G.~E. and {Aghanim}, N. and {Aguena}, M. and {Ajani}, V. and {Akrami}, Y. and {Al-Bahlawan}, A. and {Alavi}, A. and {Albuquerque}, I.~S. and {Alestas}, G. and {Alguero}, G. and {Allaoui}, A. and {Allen}, S.~W. and {Allevato}, V. and {Alonso-Tetilla}, A.~V. and {Altieri}, B. and {Alvarez-Candal}, A. and {Alvi}, S. and {Amara}, A. and {Amendola}, L. and {Amiaux}, J. and {Andika}, I.~T. and {Andreon}, S. and {Andrews}, A. and {Angora}, G. and {Angulo}, R.~E. and {Annibali}, F. and {Anselmi}, A. and {Anselmi}, S. and {Arcari}, S. and {Archidiacono}, M. and {Aric{\`o}}, G. and {Arnaud}, M. and {Arnouts}, S. and {Asgari}, M. and {Asorey}, J. and {Atayde}, L. and {Atek}, H. and {Atrio-Barandela}, F. and {Aubert}, M. and {Aubourg}, E. and {Auphan}, T. and {Auricchio}, N. and {Aussel}, B. and {Aussel}, H. and {Avelino}, P.~P. and {Avgoustidis}, A. and {Avila}, S. and {Awan}, S. and {Azzollini}, R. and {Baccigalupi}, C. and {Bachelet}, E. and {Bacon}, D. and {Baes}, M. and {Bagley}, M.~B. and {Bahr-Kalus}, B. and {Balaguera-Antolinez}, A. and {Balbinot}, E. and {Balcells}, M. and {Baldi}, M. and {Baldry}, I. and {Balestra}, A. and {Ballardini}, M. and {Ballester}, O. and {Balogh}, M. and {Ba{\~n}ados}, E. and {Barbier}, R. and {Bardelli}, S. and {Baron}, M. and {Barreiro}, T. and {Barrena}, R. and {Barriere}, J.-C. and {Barros}, B.~J. and {Barthelemy}, A. and {Bartolo}, N. and {Basset}, A. and {Battaglia}, P. and {Battisti}, A.~J. and {Baugh}, C.~M. and {Baumont}, L. and {Bazzanini}, L. and {Beaulieu}, J.-P. and {Beckmann}, V. and {Belikov}, A.~N. and {Bel}, J. and {Bellagamba}, F. and {Bella}, M. and {Bellini}, E. and {Benabed}, K. and {Bender}, R. and {Benevento}, G. and {Bennett}, C.~L. and {Benson}, K. and {Bergamini}, P. and {Bermejo-Climent}, J.~R. and {Bernardeau}, F. and {Bertacca}, D. and {Berthe}, M. and {Berthier}, J. and {Bethermin}, M. and {Beutler}, F. and {Bevillon}, C. and {Bhargava}, S. and {Bhatawdekar}, R. and {Bianchi}, D. and {Bisigello}, L. and {Biviano}, A. and {Blake}, R.~P. and {Blanchard}, A. and {Blazek}, J. and {Blot}, L. and {Bosco}, A. and {Bodendorf}, C. and {Boenke}, T. and {B{\"o}hringer}, H. and {Boldrini}, P. and {Bolzonella}, M. and {Bonchi}, A. and {Bonici}, M. and {Bonino}, D. and {Bonino}, L. and {Bonvin}, C. and {Bon}, W. and {Booth}, J.~T. and {Borgani}, S. and {Borlaff}, A.~S. and {Borsato}, E. and {Bose}, B. and {Botticella}, M.~T. and {Boucaud}, A. and {Bouche}, F. and {Boucher}, J.~S. and {Boutigny}, D. and {Bouvard}, T. and {Bouwens}, R. and {Bouy}, H. and {Bowler}, R.~A.~A. and {Bozza}, V. and {Bozzo}, E. and {Branchini}, E. and {Brando}, G. and {Brau-Nogue}, S. and {Brekke}, P. and {Bremer}, M.~N. and {Brescia}, M. and {Breton}, M.-A. and {Brinchmann}, J. and {Brinckmann}, T. and {Brockley-Blatt}, C. and {Brodwin}, M. and {Brouard}, L. and {Brown}, M.~L. and {Bruton}, S. and {Bucko}, J. and {Buddelmeijer}, H. and {Buenadicha}, G. and {Buitrago}, F. and {Burger}, P. and {Burigana}, C. and {Busillo}, V. and {Busonero}, D. and {Cabanac}, R. and {Cabayol-Garcia}, L. and {Cagliari}, M.~S. and {Caillat}, A. and {Caillat}, L. and {Calabrese}, M. and {Calabro}, A. and {Calderone}, G. and {Calura}, F. and {Camacho Quevedo}, B. and {Camera}, S. and {Campos}, L. and {Ca{\~n}as-Herrera}, G. and {Candini}, G.~P. and {Cantiello}, M. and {Capobianco}, V. and {Cappellaro}, E. and {Cappelluti}, N. and {Cappi}, A. and {Caputi}, K.~I. and {Cara}, C. and {Carbone}, C. and {Cardone}, V.~F. and {Carella}, E. and {Carlberg}, R.~G. and {Carle}, M. and {Carminati}, L. and {Caro}, F. and {Carrasco}, J.~M. and {Carretero}, J. and {Carrilho}, P. and {Carron Duque}, J. and {Carry}, B.},
        title = "{Euclid: I. Overview of the Euclid mission}",
      journal = {\aap},
         year = 2025,
        month = may,
       volume = {697},
          eid = {A1},
        pages = {A1},
          doi = {10.1051/0004-6361/202450810},
archivePrefix = {arXiv},
       eprint = {2405.13491},
 primaryClass = {astro-ph.CO},
       adsurl = {https://ui.adsabs.harvard.edu/abs/2025A&A...697A...1E}
}

@ARTICLE{Antoja_2018,
       author = {{Antoja}, T. and {Helmi}, A. and {Romero-G{\'o}mez}, M. and {Katz}, D. and {Babusiaux}, C. and {Drimmel}, R. and {Evans}, D.~W. and {Figueras}, F. and {Poggio}, E. and {Reyl{\'e}}, C. and {Robin}, A.~C. and {Seabroke}, G. and {Soubiran}, C.},
        title = "{A dynamically young and perturbed Milky Way disk}",
      journal = {\nat},
         year = 2018,
        month = sep,
       volume = {561},
       number = {7723},
        pages = {360-362},
          doi = {10.1038/s41586-018-0510-7},
archivePrefix = {arXiv},
       eprint = {1804.10196},
 primaryClass = {astro-ph.GA},
       adsurl = {https://ui.adsabs.harvard.edu/abs/2018Natur.561..360A}
}

@ARTICLE{Brandt_2024,
       author = {{Brandt}, Timothy D.},
        title = "{Optimal Fitting and Debiasing for Detectors Read Out Up-the-Ramp}",
      journal = {\pasp},
         year = 2024,
        month = apr,
       volume = {136},
       number = {4},
          eid = {045004},
        pages = {045004},
          doi = {10.1088/1538-3873/ad38d9},
archivePrefix = {arXiv},
       eprint = {2309.08753},
 primaryClass = {astro-ph.IM},
       adsurl = {https://ui.adsabs.harvard.edu/abs/2024PASP..136d5004B}
}

@ARTICLE{Offenberg_2001,
       author = {{Offenberg}, J.~D. and {Fixsen}, D.~J. and {Rauscher}, B.~J. and {Forrest}, W.~J. and {Hanisch}, R.~J. and {Mather}, J.~C. and {McKelvey}, M.~E. and {McMurray}, Jr., R.~E. and {Nieto-Santisteban}, M.~A. and {Pipher}, J.~L. and {Sengupta}, R. and {Stockman}, H.~S.},
        title = "{Validation of Up-the-Ramp Sampling with Cosmic-Ray Rejection on Infrared Detectors}",
      journal = {\pasp},
         year = 2001,
        month = feb,
       volume = {113},
       number = {780},
        pages = {240-254},
          doi = {10.1086/318615},
archivePrefix = {arXiv},
       eprint = {astro-ph/0008271},
 primaryClass = {astro-ph},
       adsurl = {https://ui.adsabs.harvard.edu/abs/2001PASP..113..240O}
}

@ARTICLE{Fixsen_2000,
       author = {{Fixsen}, D.~J. and {Offenberg}, J.~D. and {Hanisch}, R.~J. and {Mather}, J.~C. and {Nieto-Santisteban}, M.~A. and {Sengupta}, R. and {Stockman}, H.~S.},
        title = "{Cosmic-Ray Rejection and Readout Efficiency for Large-Area Arrays}",
      journal = {\pasp},
         year = 2000,
        month = oct,
       volume = {112},
       number = {776},
        pages = {1350-1359},
          doi = {10.1086/316626},
archivePrefix = {arXiv},
       eprint = {astro-ph/0005486},
 primaryClass = {astro-ph},
       adsurl = {https://ui.adsabs.harvard.edu/abs/2000PASP..112.1350F}
}

@ARTICLE{Vasiliev_2019,
       author = {{Vasiliev}, Eugene},
        title = "{Proper motions and dynamics of the Milky Way globular cluster system from Gaia DR2}",
      journal = {\mnras},
         year = 2019,
        month = apr,
       volume = {484},
       number = {2},
        pages = {2832-2850},
          doi = {10.1093/mnras/stz171},
archivePrefix = {arXiv},
       eprint = {1807.09775},
 primaryClass = {astro-ph.GA},
       adsurl = {https://ui.adsabs.harvard.edu/abs/2019MNRAS.484.2832V}
}

@ARTICLE{Rauscher_2007,
       author = {{Rauscher}, Bernard J. and {Fox}, Ori and {Ferruit}, Pierre and {Hill}, Robert J. and {Waczynski}, Augustyn and {Wen}, Yiting and {Xia-Serafino}, Wei and {Mott}, Brent and {Alexander}, David and {Brambora}, Clifford K. and {Derro}, Rebecca and {Engler}, Chuck and {Garrison}, Matthew B. and {Johnson}, Thomas and {Manthripragada}, Sridhar S. and {Marsh}, James M. and {Marshall}, Cheryl and {Martineau}, Robert J. and {Shakoorzadeh}, Kamdin B. and {Wilson}, Donna and {Roher}, Wayne D. and {Smith}, Miles and {Cabelli}, Craig and {Garnett}, James and {Loose}, Markus and {Wong-Anglin}, Selmer and {Zandian}, Majid and {Cheng}, Edward and {Ellis}, Timothy and {Howe}, Bryan and {Jurado}, Miriam and {Lee}, Ginn and {Nieznanski}, John and {Wallis}, Peter and {York}, James and {Regan}, Michael W. and {Hall}, Donald N.~B. and {Hodapp}, Klaus W. and {B{\"o}ker}, Torsten and {De Marchi}, Guido and {Jakobsen}, Peter and {Strada}, Paolo},
        title = "{Detectors for the James Webb Space Telescope Near-Infrared Spectrograph. I. Readout Mode, Noise Model, and Calibration Considerations}",
      journal = {\pasp},
         year = 2007,
        month = jul,
       volume = {119},
       number = {857},
        pages = {768-786},
          doi = {10.1086/520887},
archivePrefix = {arXiv},
       eprint = {0706.2344},
 primaryClass = {astro-ph},
       adsurl = {https://ui.adsabs.harvard.edu/abs/2007PASP..119..768R}
}

@software{Perrin_2025,
       author = {{Perrin}, Marshall and {Long}, Joseph and {Osborne}, Shannon and {Geda}, Robel and {Sappington}, Bradley and {Mel{\'e}ndez}, Marcio and {Lajoie}, Charles-Philippe and {Leisenring}, Jarron and {Zimmerman}, Neil and {Brooks}, Keira and {Otor}, O. Justin and {Kulp}, Trey and {Chambers}, Lauren and {Jurling}, Alden},
        title = "{STPSF}",
         year = 2025,
        month = jun,
          eid = {10.5281/zenodo.15747364},
          doi = {10.5281/zenodo.15747364},
      version = {2.1.0},
    publisher = {Zenodo},
       adsurl = {https://ui.adsabs.harvard.edu/abs/2025zndo..15747364P}
}

@INPROCEEDINGS{Pontoppidan_2016,
       author = {{Pontoppidan}, Klaus M. and {Pickering}, Timothy E. and {Laidler}, Victoria G. and {Gilbert}, Karoline and {Sontag}, Christopher D. and {Slocum}, Christine and {Sienkiewicz}, Mark J. and {Hanley}, Christopher and {Earl}, Nicholas M. and {Pueyo}, Laurent and {Ravindranath}, Swara and {Karakla}, Diane M. and {Robberto}, Massimo and {Noriega-Crespo}, Alberto and {Barker}, Elizabeth A.},
        title = "{Pandeia: a multi-mission exposure time calculator for JWST and WFIRST}",
    booktitle = {Observatory Operations: Strategies, Processes, and Systems VI},
         year = 2016,
       editor = {{Peck}, Alison B. and {Seaman}, Robert L. and {Benn}, Chris R.},
       series = {Society of Photo-Optical Instrumentation Engineers (SPIE) Conference Series},
       volume = {9910},
        month = jul,
          eid = {991016},
        pages = {991016},
          doi = {10.1117/12.2231768},
archivePrefix = {arXiv},
       eprint = {1707.02202},
 primaryClass = {astro-ph.IM},
       adsurl = {https://ui.adsabs.harvard.edu/abs/2016SPIE.9910E..16P}
}

@INPROCEEDINGS{Perrin_2012,
       author = {{Perrin}, Marshall D. and {Soummer}, R{\'e}mi and {Elliott}, Erin M. and {Lallo}, Matthew D. and {Sivaramakrishnan}, Anand},
        title = "{Simulating point spread functions for the James Webb Space Telescope with WebbPSF}",
    booktitle = {Space Telescopes and Instrumentation 2012: Optical, Infrared, and Millimeter Wave},
         year = 2012,
       editor = {{Clampin}, Mark C. and {Fazio}, Giovanni G. and {MacEwen}, Howard A. and {Oschmann}, Jr., Jacobus M.},
       series = {Society of Photo-Optical Instrumentation Engineers (SPIE) Conference Series},
       volume = {8442},
        month = sep,
          eid = {84423D},
        pages = {84423D},
          doi = {10.1117/12.925230},
       adsurl = {https://ui.adsabs.harvard.edu/abs/2012SPIE.8442E..3DP}
}

@MISC{Anderson_2022,
       author = {{Anderson}, Jay},
        title = "{One-Pass HST Photometry with hst1pass}",
 howpublished = {Instrument Science Report WFC3 2022-5, 55 pages},
         year = 2022,
        month = jul,
        pages = {5},
       adsurl = {https://ui.adsabs.harvard.edu/abs/2022wfc..rept....5A}
}

@ARTICLE{Vitral_2024,
       author = {{Vitral}, Eduardo and {van der Marel}, Roeland P. and {Sohn}, Sangmo Tony and {Libralato}, Mattia and {del Pino}, Andr{\'e}s and {Watkins}, Laura L. and {Bellini}, Andrea and {Walker}, Matthew G. and {Besla}, Gurtina and {Pawlowski}, Marcel S. and {Mamon}, Gary A.},
        title = "{HSTPROMO Internal Proper-motion Kinematics of Dwarf Spheroidal Galaxies. I. Velocity Anisotropy and Dark Matter Cusp Slope of Draco}",
      journal = {\apj},
         year = 2024,
        month = jul,
       volume = {970},
       number = {1},
          eid = {1},
        pages = {1},
          doi = {10.3847/1538-4357/ad571c},
archivePrefix = {arXiv},
       eprint = {2407.07769},
 primaryClass = {astro-ph.GA},
       adsurl = {https://ui.adsabs.harvard.edu/abs/2024ApJ...970....1V}
}

@ARTICLE{Libralato_2024b,
       author = {{Libralato}, M. and {Bedin}, L.~R. and {Griggio}, M. and {Massari}, D. and {Anderson}, J. and {Cuillandre}, J.-C. and {Ferguson}, A.~M.~N. and {Lan{\c{c}}on}, A. and {Larsen}, S.~S. and {Schirmer}, M. and {Annibali}, F. and {Balbinot}, E. and {Dalessandro}, E. and {Erkal}, D. and {Kuzma}, P.~B. and {Saifollahi}, T. and {Verdoes Kleijn}, G. and {K{\"u}mmel}, M. and {Nakajima}, R. and {Correnti}, M. and {Battaglia}, G. and {Altieri}, B. and {Amara}, A. and {Andreon}, S. and {Baccigalupi}, C. and {Baldi}, M. and {Balestra}, A. and {Bardelli}, S. and {Basset}, A. and {Battaglia}, P. and {Bonino}, D. and {Branchini}, E. and {Brescia}, M. and {Brinchmann}, J. and {Caillat}, A. and {Camera}, S. and {Capobianco}, V. and {Carbone}, C. and {Carretero}, J. and {Casas}, S. and {Castellano}, M. and {Castignani}, G. and {Cavuoti}, S. and {Cimatti}, A. and {Colodro-Conde}, C. and {Congedo}, G. and {Conselice}, C.~J. and {Conversi}, L. and {Copin}, Y. and {Courbin}, F. and {Courtois}, H.~M. and {Cropper}, M. and {Da Silva}, A. and {Degaudenzi}, H. and {De Lucia}, G. and {Dinis}, J. and {Dubath}, F. and {Dupac}, X. and {Dusini}, S. and {Fabricius}, M. and {Farina}, M. and {Farrens}, S. and {Faustini}, F. and {Ferriol}, S. and {Fosalba}, P. and {Frailis}, M. and {Franceschi}, E. and {Fumana}, M. and {Galeotta}, S. and {Garilli}, B. and {George}, K. and {Gillard}, W. and {Gillis}, B. and {Giocoli}, C. and {G{\'o}mez-Alvarez}, P. and {Grazian}, A. and {Grupp}, F. and {Guzzo}, L. and {Haugan}, S.~V.~H. and {Hoar}, J. and {Hoekstra}, H. and {Holmes}, W. and {Hormuth}, F. and {Hornstrup}, A. and {Hudelot}, P. and {Jahnke}, K. and {Jhabvala}, M. and {Keih{\"a}nen}, E. and {Kermiche}, S. and {Kiessling}, A. and {Kilbinger}, M. and {Kubik}, B. and {Kunz}, M. and {Kurki-Suonio}, H. and {Laureijs}, R. and {Le Mignant}, D. and {Ligori}, S. and {Lilje}, P.~B. and {Lindholm}, V. and {Lloro}, I. and {Maiorano}, E. and {Mansutti}, O. and {Marggraf}, O. and {Markovic}, K. and {Martinelli}, M. and {Martinet}, N. and {Marulli}, F. and {Massey}, R. and {Medinaceli}, E. and {Mei}, S. and {Melchior}, M. and {Mellier}, Y. and {Meneghetti}, M. and {Merlin}, E. and {Meylan}, G. and {Moresco}, M. and {Moscardini}, L. and {Neissner}, C. and {Nichol}, R.~C. and {Niemi}, S.-M. and {Nightingale}, J.~W. and {Padilla}, C. and {Paltani}, S. and {Pasian}, F. and {Pedersen}, K. and {Percival}, W.~J. and {Pettorino}, V. and {Pires}, S. and {Polenta}, G. and {Poncet}, M. and {Popa}, L.~A. and {Pozzetti}, L. and {Raison}, F. and {Rebolo}, R. and {Refregier}, A. and {Renzi}, A. and {Rhodes}, J. and {Riccio}, G. and {Romelli}, E. and {Roncarelli}, M. and {Rossetti}, E. and {Saglia}, R. and {Sakr}, Z. and {S{\'a}nchez}, A.~G. and {Sapone}, D. and {Sartoris}, B. and {Sauvage}, M. and {Schneider}, P. and {Schrabback}, T. and {Secroun}, A. and {Sefusatti}, E. and {Seidel}, G. and {Seiffert}, M. and {Serrano}, S. and {Sirignano}, C. and {Sirri}, G. and {Skottfelt}, J. and {Stanco}, L. and {Steinwagner}, J. and {Tallada-Cresp{\'\i}}, P. and {Taylor}, A.~N. and {Teplitz}, H.~I. and {Tereno}, I. and {Toledo-Moreo}, R. and {Torradeflot}, F. and {Tsyganov}, A. and {Tutusaus}, I. and {Valenziano}, L. and {Vassallo}, T. and {Veropalumbo}, A. and {Wang}, Y. and {Weller}, J. and {Zamorani}, G. and {Zucca}, E. and {Burigana}, C. and {Scottez}, V. and {Scott}, D. and {Smart}, R.~L.},
        title = "{Euclid: High-precision imaging astrometry and photometry from Early Release Observations: I. Internal kinematics of NGC6397 by combining Euclid and Gaia data}",
      journal = {\aap},
         year = 2024,
        month = dec,
       volume = {692},
          eid = {A96},
        pages = {A96},
          doi = {10.1051/0004-6361/202452295},
archivePrefix = {arXiv},
       eprint = {2411.02487},
 primaryClass = {astro-ph.SR},
       adsurl = {https://ui.adsabs.harvard.edu/abs/2024A&A...692A..96L}
}

@ARTICLE{Libralato_2024a,
       author = {{Libralato}, Mattia and {Argyriou}, Ioannis and {Dicken}, Dan and {Garc{\'\i}a Mar{\'\i}n}, Macarena and {Guillard}, Pierre and {Hines}, Dean C. and {Kavanagh}, Patrick J. and {Kendrew}, Sarah and {Law}, David R. and {Noriega-Crespo}, Alberto and {{\'A}lvarez-M{\'a}rquez}, Javier},
        title = "{High-precision Astrometry and Photometry with the JWST/MIRI Imager}",
      journal = {\pasp},
         year = 2024,
        month = mar,
       volume = {136},
       number = {3},
          eid = {034502},
        pages = {034502},
          doi = {10.1088/1538-3873/ad2551},
archivePrefix = {arXiv},
       eprint = {2311.12145},
 primaryClass = {astro-ph.IM},
       adsurl = {https://ui.adsabs.harvard.edu/abs/2024PASP..136c4502L}
}

@ARTICLE{Libralato_2023,
       author = {{Libralato}, Mattia and {Bellini}, Andrea and {van der Marel}, Roeland P. and {Anderson}, Jay and {Sohn}, Sangmo Tony and {Watkins}, Laura L. and {Alderson}, Lili and {Allen}, Natalie and {Clampin}, Mark and {Glidden}, Ana and {Goyal}, Jayesh and {Hoch}, Kielan and {Huang}, Jingcheng and {Kammerer}, Jens and {Lewis}, Nikole K. and {Lin}, Zifan and {Long}, Douglas and {Louie}, Dana and {MacDonald}, Ryan J. and {Mountain}, Matt and {Pe{\~n}a-Guerrero}, Maria and {Perrin}, Marshall D. and {Pueyo}, Laurent and {Rebollido}, Isabel and {Rickman}, Emily and {Seager}, Sara and {Stevenson}, Kevin B. and {Valenti}, Jeff A. and {Valentine}, Daniel and {Wakeford}, Hannah R.},
        title = "{JWST-TST Proper Motions. I. High-precision NIRISS Calibration and Large Magellanic Cloud Kinematics}",
      journal = {\apj},
         year = 2023,
        month = jun,
       volume = {950},
       number = {2},
          eid = {101},
        pages = {101},
          doi = {10.3847/1538-4357/acd04f},
archivePrefix = {arXiv},
       eprint = {2303.00009},
 primaryClass = {astro-ph.GA},
       adsurl = {https://ui.adsabs.harvard.edu/abs/2023ApJ...950..101L}
}

@ARTICLE{Bennet_2024,
       author = {{Bennet}, Paul and {Patel}, Ekta and {Sohn}, Sangmo Tony and {del Pino Molina}, Andr{\'e}s and {van der Marel}, Roeland P. and {Libralato}, Mattia and {Watkins}, Laura L. and {Aparicio}, Antonio and {Besla}, Gurtina and {Gallart}, Carme and {Fardal}, Mark A. and {Monelli}, Matteo and {Sacchi}, Elena and {Tollerud}, Erik and {Weisz}, Daniel R.},
        title = "{Proper Motions and Orbits of Distant Local Group Dwarf Galaxies from a Combination of Gaia and Hubble Data}",
      journal = {\apj},
         year = 2024,
        month = aug,
       volume = {971},
       number = {1},
          eid = {98},
        pages = {98},
          doi = {10.3847/1538-4357/ad5349},
archivePrefix = {arXiv},
       eprint = {2312.09276},
 primaryClass = {astro-ph.GA},
       adsurl = {https://ui.adsabs.harvard.edu/abs/2024ApJ...971...98B}
}

@ARTICLE{Warfield_2023,
       author = {{Warfield}, Jack T. and {Kallivayalil}, Nitya and {Zivick}, Paul and {Fritz}, Tobias and {Richstein}, Hannah and {Sohn}, Sangmo Tony and {del Pino}, Andr{\'e}s and {Savino}, Alessandro and {Weisz}, Daniel R.},
        title = "{HUBPUG: proper motions for local group dwarfs observed with HST utilizing Gaia as a reference frame}",
      journal = {\mnras},
         year = 2023,
        month = feb,
       volume = {519},
       number = {1},
        pages = {1189-1200},
          doi = {10.1093/mnras/stac3647},
archivePrefix = {arXiv},
       eprint = {2209.02751},
 primaryClass = {astro-ph.GA},
       adsurl = {https://ui.adsabs.harvard.edu/abs/2023MNRAS.519.1189W}
}

@ARTICLE{Sohn_2012,
       author = {{Sohn}, Sangmo Tony and {Anderson}, Jay and {van der Marel}, Roeland P.},
        title = "{The M31 Velocity Vector. I. Hubble Space Telescope Proper-motion Measurements}",
      journal = {\apj},
         year = 2012,
        month = jul,
       volume = {753},
       number = {1},
          eid = {7},
        pages = {7},
          doi = {10.1088/0004-637X/753/1/7},
archivePrefix = {arXiv},
       eprint = {1205.6863},
 primaryClass = {astro-ph.GA},
       adsurl = {https://ui.adsabs.harvard.edu/abs/2012ApJ...753....7S}
}

@ARTICLE{Sohn_2018,
       author = {{Sohn}, Sangmo Tony and {Watkins}, Laura L. and {Fardal}, Mark A. and {van der Marel}, Roeland P. and {Deason}, Alis J. and {Besla}, Gurtina and {Bellini}, Andrea},
        title = "{Absolute Hubble Space Telescope Proper Motion (HSTPROMO) of Distant Milky Way Globular Clusters: Galactocentric Space Velocities and the Milky Way Mass}",
      journal = {\apj},
         year = 2018,
        month = jul,
       volume = {862},
       number = {1},
          eid = {52},
        pages = {52},
          doi = {10.3847/1538-4357/aacd0b},
archivePrefix = {arXiv},
       eprint = {1804.01994},
 primaryClass = {astro-ph.GA},
       adsurl = {https://ui.adsabs.harvard.edu/abs/2018ApJ...862...52S}
}

@ARTICLE{vanderMarel_2012,
       author = {{van der Marel}, Roeland P. and {Fardal}, Mark and {Besla}, Gurtina and {Beaton}, Rachael L. and {Sohn}, Sangmo Tony and {Anderson}, Jay and {Brown}, Tom and {Guhathakurta}, Puragra},
        title = "{The M31 Velocity Vector. II. Radial Orbit toward the Milky Way and Implied Local Group Mass}",
      journal = {\apj},
         year = 2012,
        month = jul,
       volume = {753},
       number = {1},
          eid = {8},
        pages = {8},
          doi = {10.1088/0004-637X/753/1/8},
archivePrefix = {arXiv},
       eprint = {1205.6864},
 primaryClass = {astro-ph.GA},
       adsurl = {https://ui.adsabs.harvard.edu/abs/2012ApJ...753....8V}
}

@ARTICLE{Helmi_2018,
       author = {{Helmi}, Amina and {Babusiaux}, Carine and {Koppelman}, Helmer H. and {Massari}, Davide and {Veljanoski}, Jovan and {Brown}, Anthony G.~A.},
        title = "{The merger that led to the formation of the Milky Way's inner stellar halo and thick disk}",
      journal = {\nat},
         year = 2018,
        month = oct,
       volume = {563},
       number = {7729},
        pages = {85-88},
          doi = {10.1038/s41586-018-0625-x},
archivePrefix = {arXiv},
       eprint = {1806.06038},
 primaryClass = {astro-ph.GA},
       adsurl = {https://ui.adsabs.harvard.edu/abs/2018Natur.563...85H}
}

@ARTICLE{Belokurov_2018,
       author = {{Belokurov}, V. and {Erkal}, D. and {Evans}, N.~W. and {Koposov}, S.~E. and {Deason}, A.~J.},
        title = "{Co-formation of the disc and the stellar halo}",
      journal = {\mnras},
         year = 2018,
        month = jul,
       volume = {478},
       number = {1},
        pages = {611-619},
          doi = {10.1093/mnras/sty982},
archivePrefix = {arXiv},
       eprint = {1802.03414},
 primaryClass = {astro-ph.GA},
       adsurl = {https://ui.adsabs.harvard.edu/abs/2018MNRAS.478..611B}
}

@ARTICLE{Lindegren_2021,
       author = {{Lindegren}, L. and {Bastian}, U. and {Biermann}, M. and {Bombrun}, A. and {de Torres}, A. and {Gerlach}, E. and {Geyer}, R. and {Hern{\'a}ndez}, J. and {Hilger}, T. and {Hobbs}, D. and {Klioner}, S.~A. and {Lammers}, U. and {McMillan}, P.~J. and {Ramos-Lerate}, M. and {Steidelm{\"u}ller}, H. and {Stephenson}, C.~A. and {van Leeuwen}, F.},
        title = "{Gaia Early Data Release 3. Parallax bias versus magnitude, colour, and position}",
      journal = {\aap},
         year = 2021,
        month = may,
       volume = {649},
          eid = {A4},
        pages = {A4},
          doi = {10.1051/0004-6361/202039653},
archivePrefix = {arXiv},
       eprint = {2012.01742},
 primaryClass = {astro-ph.IM},
       adsurl = {https://ui.adsabs.harvard.edu/abs/2021A&A...649A...4L}
}
\bibliographystyle{aasjournal}

\appendix

\section{Line-of-Sight Velocity Impact on Astrometry} \label{sec:los_appendix}

In the absence of acceleration, the magnitude of a star's PM is defined to be 
$$|\mu| = \frac{|V_T|}{D}$$ where $|V_T|$ is the magnitude of the star's tangential velocity and $D$ is the star's distance. Taking a derivative with respect to time, we have:
$$\left| \frac{d\mu}{dt} \right| = \frac{|V_T|}{D^2} \cdot |V_{LOS}|$$ where the line-of-sight (LOS) velocity is defined by $V_{LOS} = \frac{dD}{dt}$. For the model presented in this work to hold, we require that the change in the PM magnitude over some time baseline $T$ is much smaller than the measured PM uncertainty,
$\sigma_{\mu} \gg \lvert T \cdot \frac{d\mu}{dt} \rvert$, which implies:
\begin{equation}
\begin{split}
     D &\gg \left(\frac{1}{\sigma_{\mu}} \cdot T\cdot |V_{LOS}|\cdot |{V_T}|\right)^{1/2} \\
     \left[\frac{D}{\mathrm{pc}}\right] &\gg \left(2.154\times10^{-4} \cdot \left[\frac{\mathrm{mas/yr}}{\sigma_{\mu}}\right] \cdot \left[\frac{T}{\mathrm{yr}}\right] \cdot\left[\frac{|V_{LOS}|}{\mathrm{km/s}}\right] \cdot\left[\frac{|{V_T}|}{\mathrm{km/s}}\right]\right)^{1/2}.
\end{split}
\end{equation}
Similarly, for parallax, we have $\varpi = \frac{1}{D}$, which ultimately leads to the following constraint:
\begin{equation}
\begin{split}
     D &\gg \left(\frac{1}{\sigma_{\varpi}} \cdot T\cdot |V_{LOS}|\right)^{1/2} \\
     \left[\frac{D}{\mathrm{pc}}\right] &\gg \left(1.022\times10^{-3} \cdot \left[\frac{\mathrm{mas}}{\sigma_{\mu}}\right] \cdot \left[\frac{T}{\mathrm{yr}}\right] \cdot\left[\frac{|V_{LOS}|}{\mathrm{km/s}}\right] \right)^{1/2}.
\end{split}
\end{equation}
Putting in some typical values -- such as $V_{LOS}= V_{T}= 100$~km/s, $T=10$~years, $\sigma_{\mu}=0.01$~mas/yr, and $\sigma_{\varpi}=0.01$~mas -- we find that the stellar distance must be much larger than $\sim46$~pc to be confident that the LOS velocity will not significantly change the parallax or PM over a 10 year baseline. \kam{For some of the highest-precision PMs and parallaxes expected from next-generation telescopes, astrometric uncertainties like $\sigma_{\mu}<1$~$\mu$as/yr and $\sigma_{\varpi}<1$~$\mu$as will be possible (e.g., Figure}~\ref{fig:GBTDS_astrometry_errs}\kam{). In these cases, the distance threshold would increase to $\sim 147$~pc (i.e. a factor of $\sqrt{10}$ times larger).}

\end{document}